\documentclass[iop]{emulateapj}

\begin{document}
\title{{\it Spitzer} Observations of Long Term Infrared Variability among Young Stellar Objects in Chamaeleon I}
\author{
Kevin M. Flaherty\altaffilmark{1},
Lindsay DeMarchi\altaffilmark{2,8},
James Muzerolle\altaffilmark{3},
Zoltan Balog\altaffilmark{4},
William Herbst\altaffilmark{1},
S. Thomas Megeath\altaffilmark{5},
Elise Furlan\altaffilmark{6},
Robert Gutermuth\altaffilmark{7}
}
\altaffiltext{1}{Van Vleck Observatory, Astronomy Department, Wesleyan University, 96 Foss Hill Drive, Middletown, CT 06459}
\altaffiltext{2}{Department of Physics and Astronomy, Colgate University, 13 Oak Drive, Hamilton, NY 13346}
\altaffiltext{3}{Space Telescope Science Institute, 3700 San Martin Drive, Baltimore, Maryland, 21218}
\altaffiltext{4}{Max-Planck-Institut fur Astronomie, Konigstuhl 17, 69117 Heidelberg, Germany}
\altaffiltext{5}{Ritter Astrophysical Research Center, Department of Physics and Astronomy, University of Toledo, Toledo, OH 43606}
\altaffiltext{6}{Infrared Processing and Analysis Center, California Institute of Technology, 770 S. Wilson Ave, Pasadena, CA 91125}
\altaffiltext{7}{Department of Astronomy, University of Massachusetts, Amherst, MA 01003}
\altaffiltext{8}{Department of Physics, Syracuse University, Syracuse, NY, 13244}

\begin{abstract}
Infrared variability is common among young stellar objects, with surveys finding daily to weekly fluctuations of a few tenths of a magnitude. Space-based observations can produce highly sampled infrared light curves, but are often limited to total baselines of about a month due to the orientation of the spacecraft. Here we present observations of the Chameleon I cluster whose low declination makes it observable by the {\it Spitzer} space telescope over a 200 day period. We observe 30 young stellar objects with a daily cadence to better sample variability on timescales of months. We find such variability is common, occurring in $\sim$80\% of the detected cluster members. The change in [3.6]-[4.5] color over 200 days for many of the sources falls between that expected for extinction and fluctuations in disk emission. With our high cadence and long baseline we can derive power spectral density curves covering two orders of magnitude in frequency and find significant power at low frequencies, up to the boundaries of our 200 day survey. Such long timescales are difficult to explain with variations driven by the interaction between the disk and stellar magnetic field, which has a dynamical timescale of days to weeks. The most likely explanation is either structural or temperature fluctuations spread throughout the inner $\sim$0.5 au of the disk, suggesting that the intrinsic dust structure is highly dynamic.
\end{abstract}

\section{Introduction}
In young stellar objects the inner disk ($\lesssim$1 au) stands at the intersection between gas, dust, and the star itself. The gas disk is truncated by its interaction with the stellar magnetic field ($\sim$0.05 au), while the dust disk inner edge occurs where the grains are heated to a high enough temperature to sublimate (0.1-1 au). The exact structure can be difficult to study given the small size of these regions, making resolved observations challenging, although not impossible \citep[e.g.][]{men15,mil16}. Unresolved observations have been successful in revealing the general structure of this region. Spectral line profiles trace the free-fall of gas onto the stellar surface along magnetic field lines after it is lifted away from the disk \citep[e.g.][]{lim10} as well as outflows \citep{edw13,cau15}. Infrared observations find strong emission from T$\sim$1500 K dust \citep[e.g.][]{mcc13a}, consistent with a curved inner wall \citep{ise05,flo16}. Recent studies have found evidence for multiple populations of dust grains with different inner radii \citep{mcc13b}, consistent with the expectation that small dust grains reach the sublimation temperature farther from the star because they more efficiently absorb the stellar irradiation. While here we focus on the dust, much has been learned about the gas structure within this region \citep[see reviews by ][]{bou07,dul10}.

Variability serves as an additional tool for studying the inner disk. Recent surveys have found that the vast majority of young stellar objects are variable in the mid-infrared on timescales of days to weeks \citep{cod14,reb14}, expanding on earlier ground-based near-infrared studies \citep[e.g.][]{car01}. These surveys have also revealed additional clues as to the structure of the inner disk. \citet{men16} use reverberation mapping to derive the location of the dust inner wall. \citet{esp11} demonstrate that the wavelength dependence of the variability in pre-transition disks, systems with an optically thick inner disk followed by an optically thin gap and an optically thick outer disk,  can be explained by varying the height of the inner disk, indicating that this region is not static. \citet{mcg15} find, based on quasi-periodic occultations of the stellar surface, that the height of the inner wall can vary by $\sim$10\% over short timescales even in full disks. \citet{bou03} find that the variability of AA Tau is consistent with periodic occultations by a warped inner disk. Such non-axisymmetric structure is difficult to study with spatially un-resolved observations, which often rely on fits to axisymmetric models. 

Most previous {\it Spitzer} time domain studies of YSOs were restricted to baselines of 1-2 months due to the duration of a single visibility window for most star forming regions, while ground-based near-infrared studies have found variability extending out to year long timescales \citep{wol13,par14}, although with a sparser cadence than is possible with space-based observatories. Long timescales push towards larger radii in the disk; if the timescale of the variability is proportional to the local Keplerian period then year-long fluctuations imply variability at $\sim$1 au from the central star. A detailed characterization of long-timescale variability can improve our understanding of this important planet-forming region of the disk.

Here we probe variability over weeks to months using {\it Spitzer} Space Telescope observations of the Chamaeleon I (Cha I) star-forming region \citep{luh08b}. Due to its low declination, Cha I can be observed by {\it Spitzer} over a 200 day window, much longer than the $\sim$40 day windows of other star forming regions. With photometry (Section~\ref{data}) taken roughly once-per-day for 200 days we produce well sampled light curves and select out a sample of stars with significant variability (Section~\ref{find_var}). We analyze the change in color (Section~\ref{color}) and the timescale of the variability (Section~\ref{timescale}) and relate this information to potential sources of the variability (Section~\ref{discussion}). We find that months-long variability is common and likely due to intrinsic instabilities in disk structure as far out as 0.5 au.

\section{Data\label{data}}
\begin{figure}
\includegraphics[scale=.3]{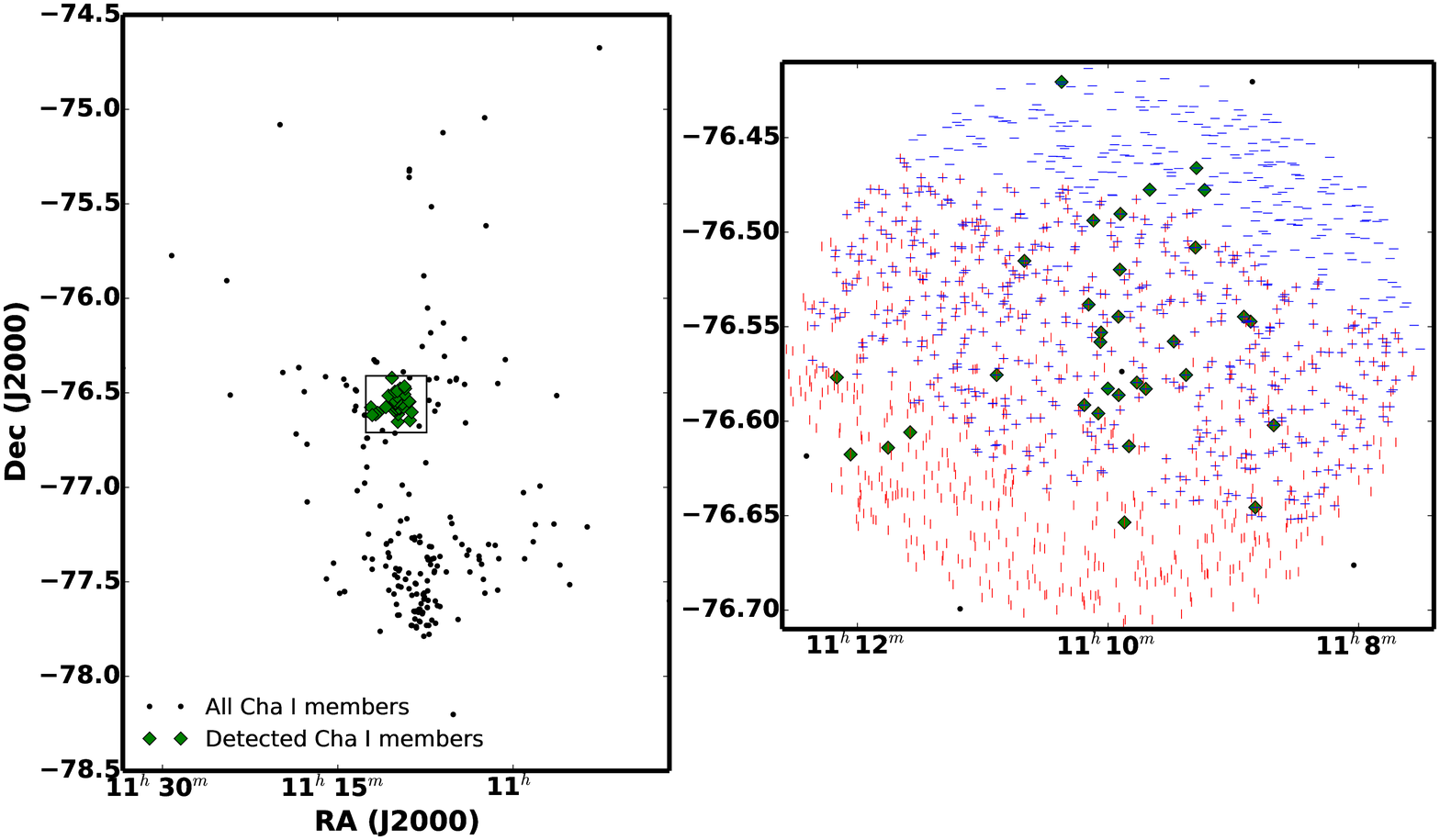}
\caption{Coverage map comparing the known Cha I cluster members (black dots) with those covered in our survey (green diamonds). The inset shows a close-up on our survey region, with every source detected on more than 20 days in [3.6] (vertical red ticks) and [4.5] (horizontal blue ticks). Our survey focused on region of Cha I with a relatively high density of young stellar objects. The rotation of the field of view during the visibility window, as well as the incomplete overlap between the [3.6] and [4.5] fields of view, leads to uneven [3.6] and [4.5] coverage of the cluster.\label{map}}
\end{figure}

During our survey we focused on a subset of the young, d$\sim$160 pc, Cha I low-mass star forming region \citep{luh08b}. Our field of view is located in the northern section of the region (Figure~\ref{map}), and encompasses over 30 known members, as defined by the samples of \citet{luh07} and \citet{luh08a}. Membership has been defined using a number of characteristics included infrared excess emission, spectroscopic properties and extinction \citep{luh07,luh08a}, and proper motions \citep{lop13}. This region was chosen because of its relatively high density of young stellar objects, allowing us to observe a modest sample in a small area while maximizing the time coverage. Stellar properties are listed in Table~\ref{stellar_params} with masses estimated using the measured T$_{\rm eff}$ and L$_*$ and the \citet{bar15} isochrones. Spectral types are taken from \citet{luh08b}, which agree with more recent surveys \citep{man16}. While the cluster members cover a wide range, from a B9 star down to a M9.5 star, the median star is effectively an early M type star, with T$_{\rm eff}$=3400, L$_*$=0.2L$_{\odot}$ and M$_*$=0.3M$_{\odot}$. While our sample is too small for a detailed comparison, the distribution of stellar masses is consistent with the shape of the initial mass function derived over the entire cluster \citep{luh07}. We draw information on the SED shape from the analysis of \citet{gut09}, and where a prior classification was not available we apply the \citet{gut09} classification scheme. Accordingly, our sample consists of three class I sources, 24 class II sources and five class III sources. None of the previously identified transition disks within Cha I \citep{kim09,man11} fall within our field of view.

\begin{figure*}
\center
\includegraphics[scale=.4]{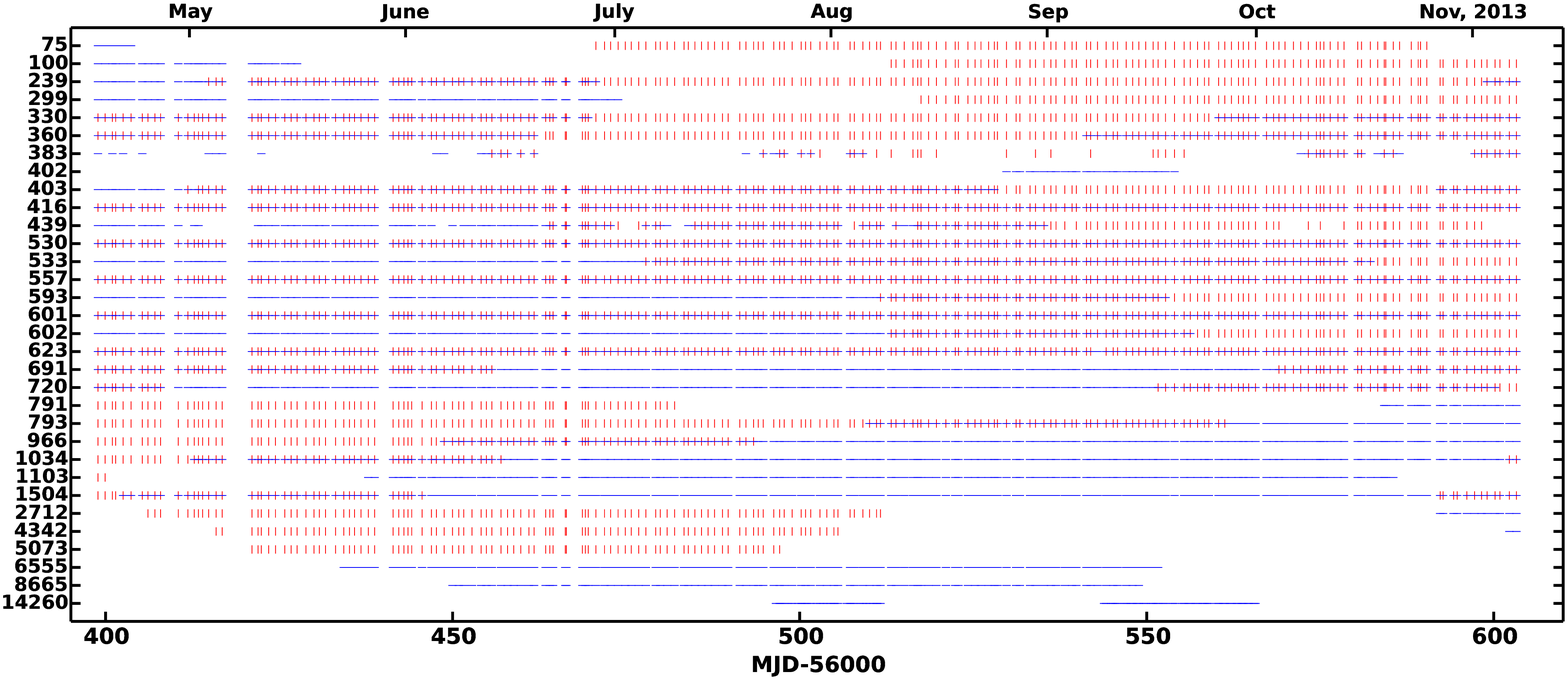}
\caption{Coverage of the 32 Cha I cluster members in our survey. Vertical red ticks mark observations at [3.6] while horizontal blue ticks mark observations at [4.5]. Due to the rotation of the field of view, and the incomplete overlap between the [3.6] and [4.5] fields of view, the time coverage varies between each star and each band. \label{coverage}}
\end{figure*}

The focus of our analysis is repeated observations using the IRAC instrument on the {\it Spitzer Space Telescope} (PID 90141). Both [3.6] and [4.5] photometry were obtained simultaneously on 200 epochs from 2013-4-16 to 2013-11-07. Each epoch consisted of 12s HDR exposures with three cycles at each position. The region of interest, centered at 11:10:00, -76:33:36,  was mapped using a sequence of four pointings, arranged in a 2 by 2 grid, with 260\arcsec\ between each pointing. The roughly once-per-day cadence was chosen to cover much of the 220 day long visibility while still tracking the light curve through fluctuations on daily and weekly timescales. Such a low cadence does leave us relatively insensitive to rapid features, such as short accretion bursts \citep{sta14} and 'dippers' \citep[e.g.][]{mcg15}. Our focus instead is on the long term variations that are common \citep{cod14} but difficult to characterize given the $\sim$40 day {\it Spitzer} observing windows of most previously surveyed regions \citep[e.g.][]{reb14}.

The photometry was derived using the {\it Spitzer} data reduction pipeline as described in \citet{gut09} with updates appropriate for the warm {\it Spitzer} mission \citep{mor12}. The uncertainties derived from the images do not include additional systematic uncertainties and we add systematic uncertainties, derived from the prescription of \citet{cod14}, in quadrature with statistical uncertainties. 
\begin{eqnarray}
\sigma_{\rm sys} = \left\{
\begin{array}{ll}
\exp^{(m-19.75)}+0.014 & \mbox{for [3.6]}\\
\exp^{(m-19.28)}+0.008 & \mbox{for [4.5]}\\
\end{array}
\right.\\
\sigma^2=\sigma_{\rm stat}^2+\sigma_{\rm sys}^2
\end{eqnarray}
Since most of cluster members are bright, the systematic uncertainties make up the dominant noise component. Selecting sources with $\sigma<$0.1 mag on at least one epoch in either [3.6] or [4.5], we detect a total of 32 cluster members within our field of view. In addition to these known cluster members, the {\it Spitzer} field of view includes $\sim$1000 other sources, most of which are background or foreground stars unrelated to the Cha I star forming region. We restrict our analysis to previously discovered cluster members, although additional targets of interest may exist within these data. While we observe the cluster over 200 epochs, we do not have have 200 epochs of [3.6] and [4.5] for every cluster member within our field of view (Figure~\ref{coverage}). Part of this is due to the incomplete overlap of the [3.6] and [4.5] fields of view, which results in many stars only being observed in one band at a time. Also, during the long visibility window the spacecraft itself rotates about its axis as it circles the Sun, causing the orientation of the field of view to change with time. The result is that while some stars have coverage from both bands over 200 epochs, many of the cluster members have only partial coverage. Much of the incomplete coverage of any particular source is due to the changing field of view; the stars rarely change flux dramatically enough to either saturate or become too faint for short periods. All but five stars (402, 5073, 6555, 8665, 14260) are detected in both bands, with the median number of detected epochs equal to $\sim$100. This incomplete coverage is accounted for in our analysis.


\section{Finding Variable Stars\label{find_var}}
\begin{figure*}
\center
\includegraphics[scale=.4]{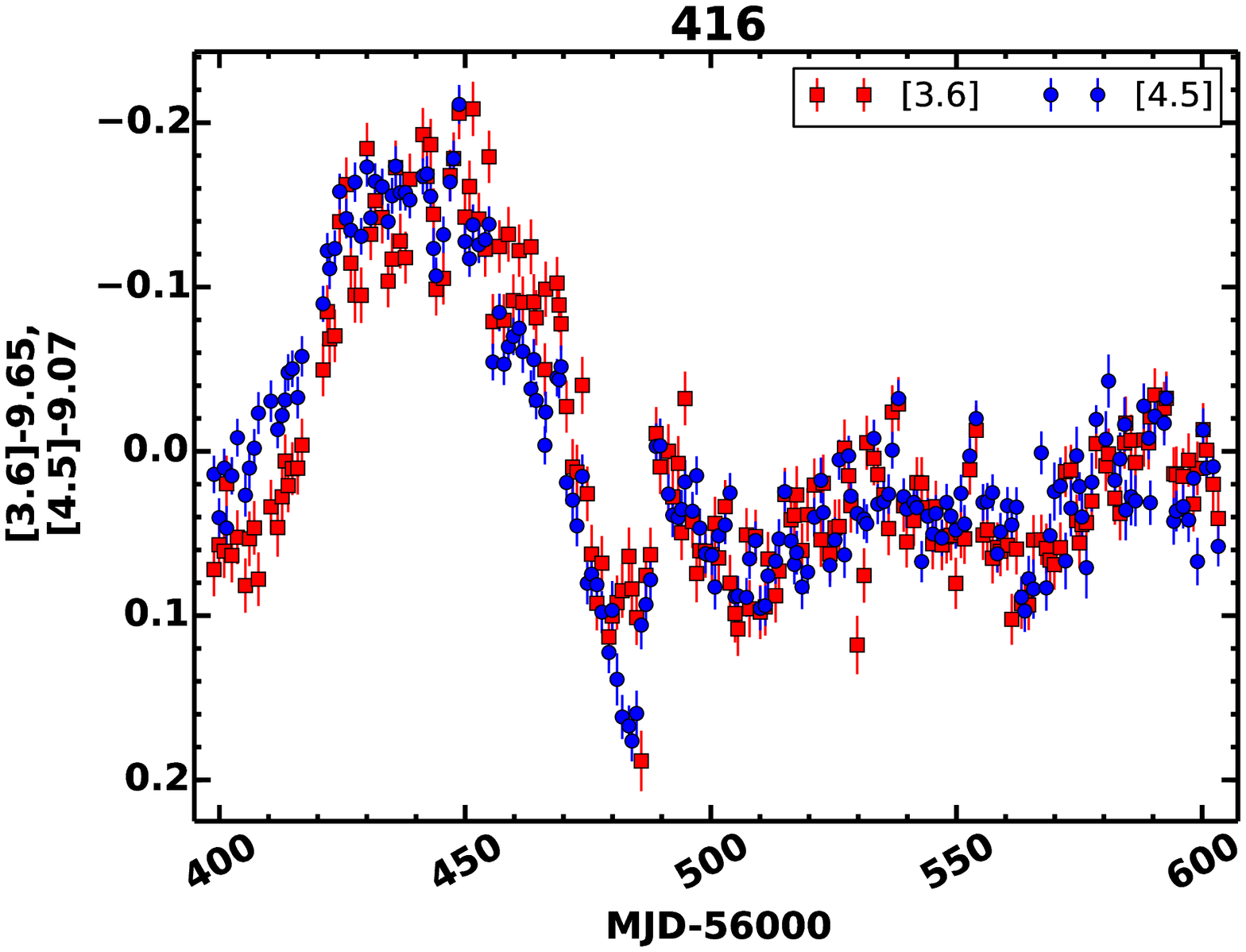}
\includegraphics[scale=.4]{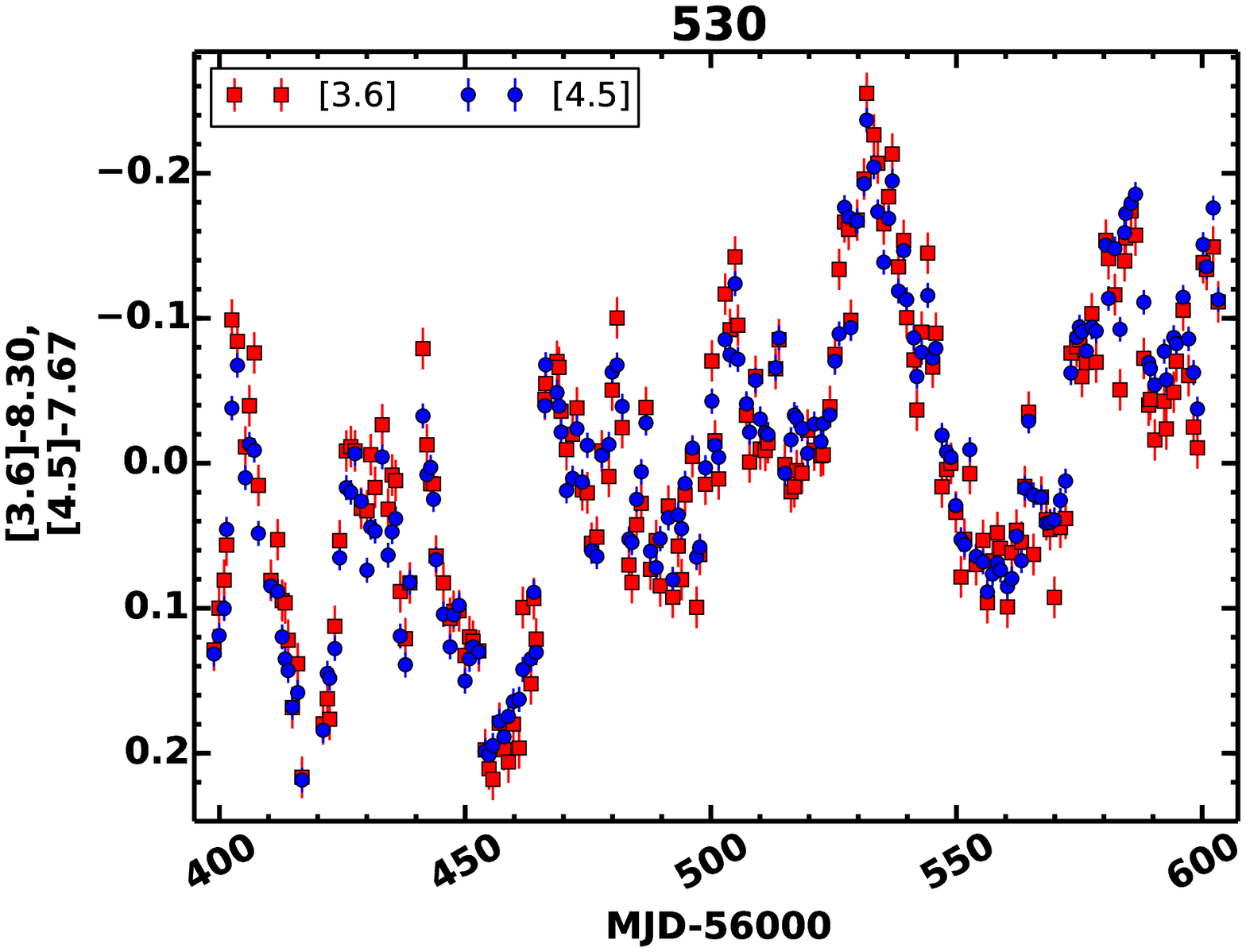}
\includegraphics[scale=.4]{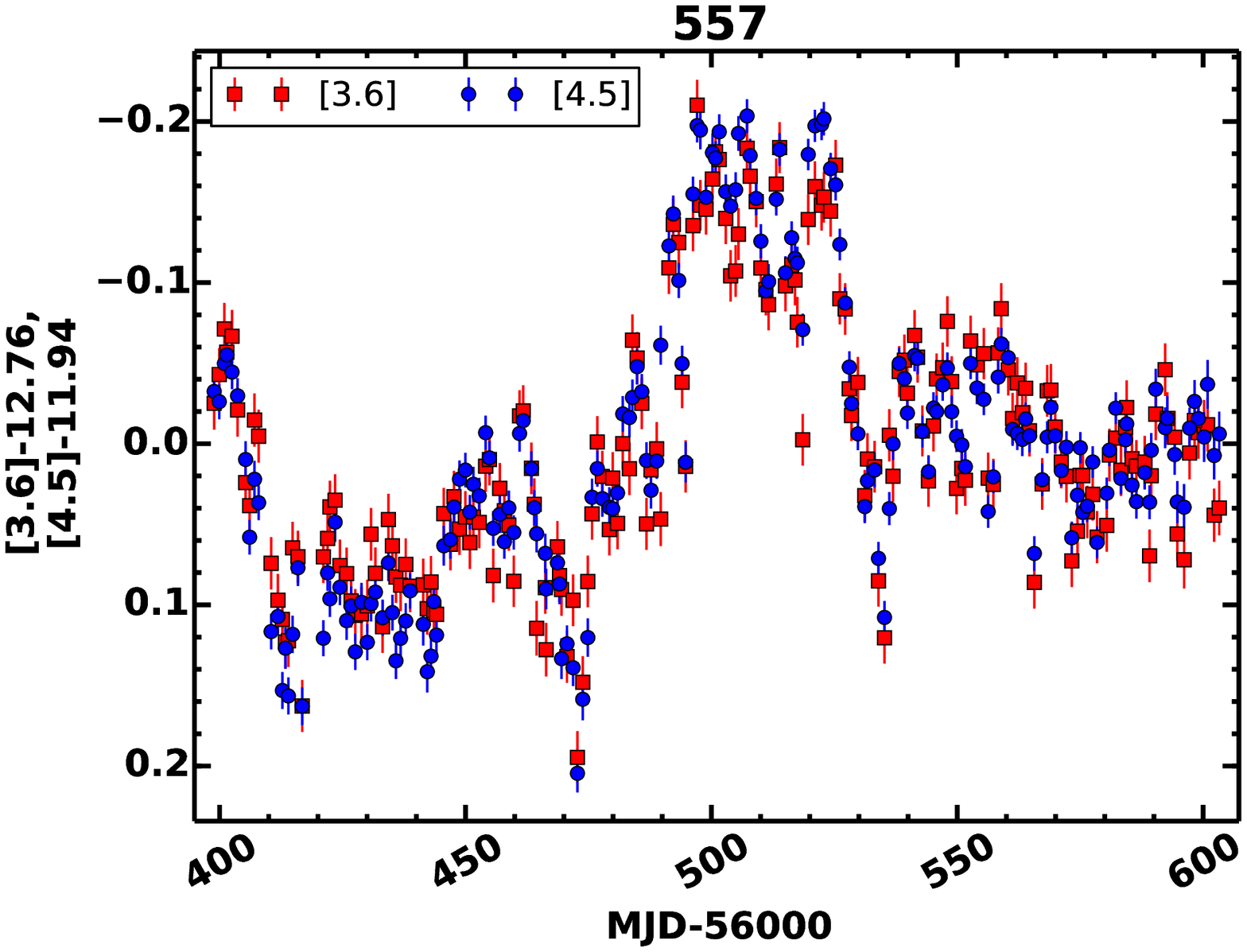}
\includegraphics[scale=.4]{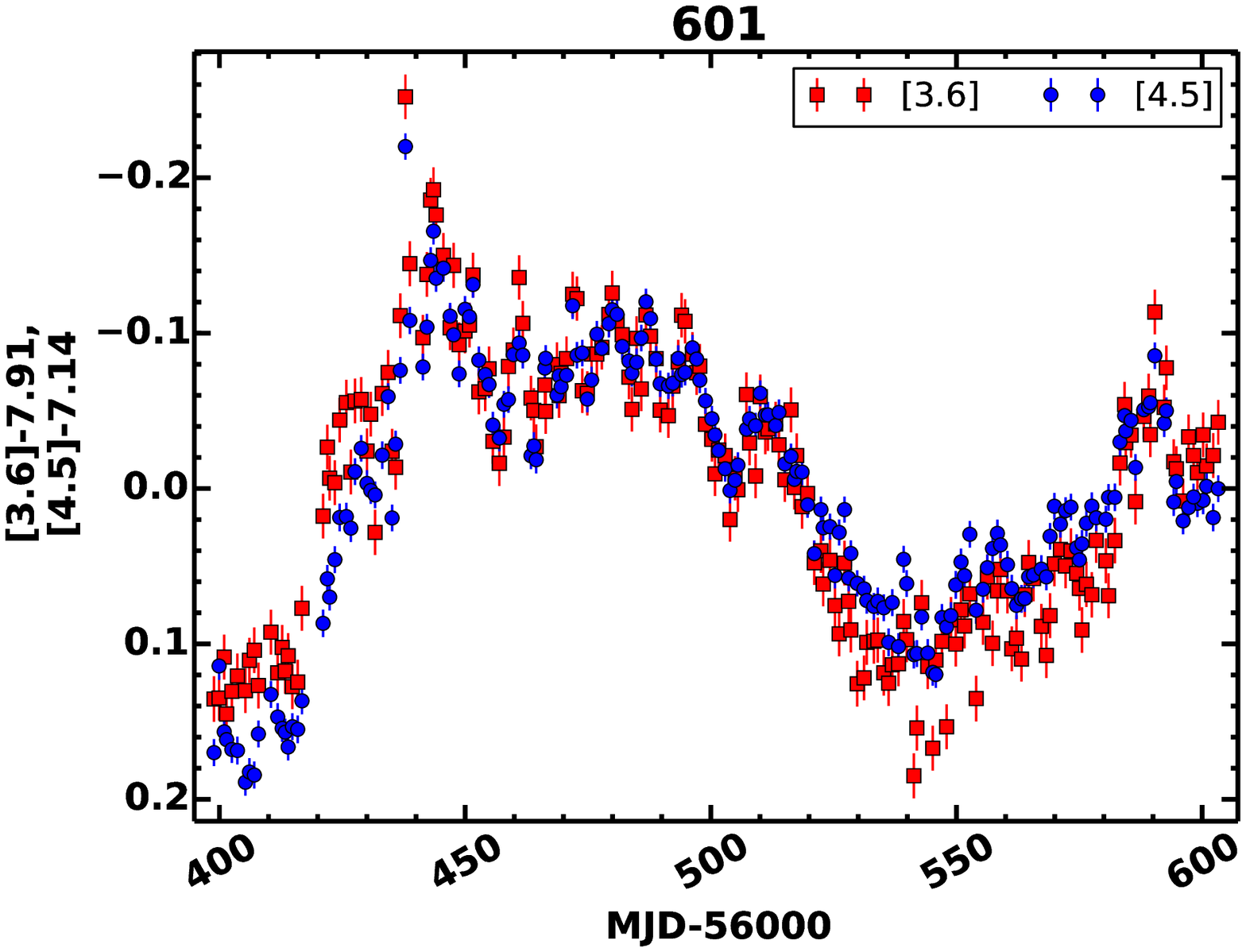}
\caption{Infrared light curves for select cluster members. [3.6] data is marked with red squares while [4.5] photometry is shown with blue circles. We see a wide variety of light curve shapes, with timescales covering the entire 200 day observing window. Light curves for all the detected cluster members are included in the online version of the journal. \label{light_curves}}
\end{figure*}

\begin{figure}
\center
\includegraphics[scale=.3]{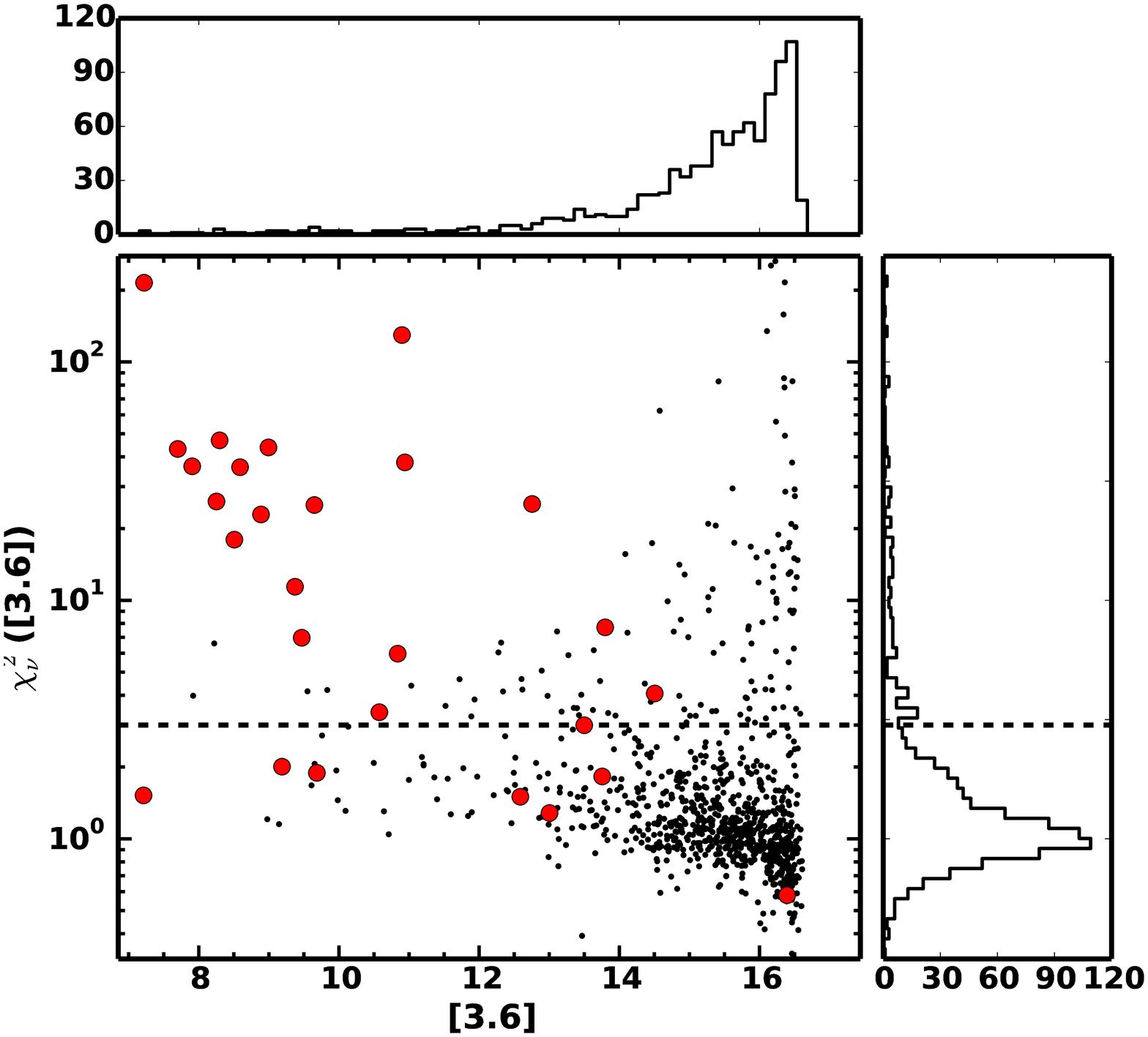}
\includegraphics[scale=.3]{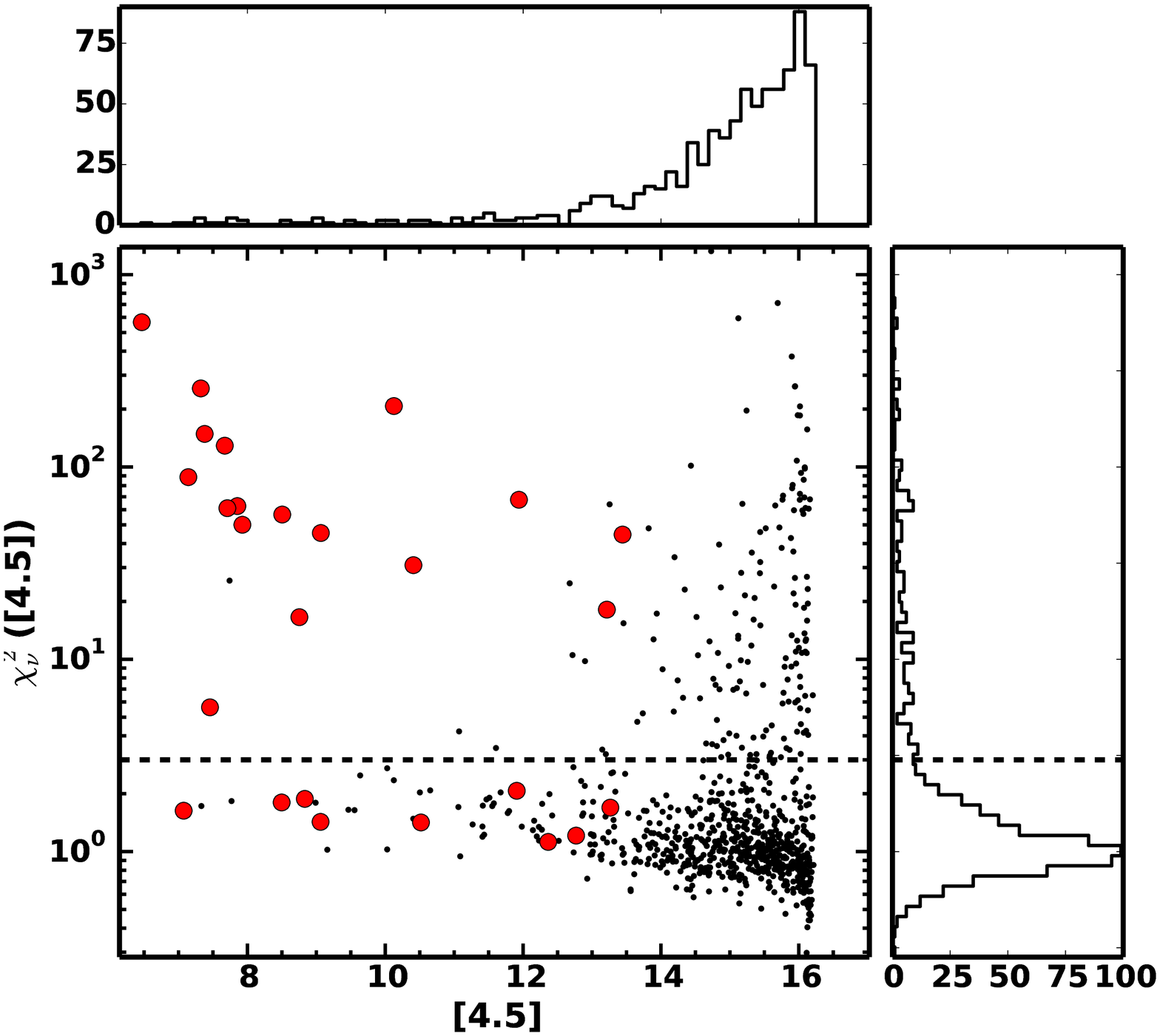}
\caption{Distribution of reduced chi-squared values and mean magnitudes for [3.6] (left) and [4.5] (right). All detected sources with more than 20 epochs in a particular band, predominately foreground and background stars, are shown with black dots, while known cluster members are marked by large red circles. The horizontal dashed line indicates a reduced chi-squared of three; points above this line are considered significantly variable.\label{chisq_mag}}
\end{figure}

The light curves show a wide range of behaviors (Figure~\ref{light_curves}) with many of the fluctuations appearing stochastic on both short and long timescales. Defining the size of the fluctuations as the difference between the 10th percentile magnitude and the 90th percentile magnitude ( $\Delta$mag in Table~\ref{invar_table}) we find fluctuations ranging from 0.05-0.5 magnitudes, with a mean of $\sim$0.15, similar to previous surveys (Table~\ref{cluster_comparison}). We use a combination of statistics to select out variable cluster members in a way that is independent of the exact light curve shape. We start with the reduced chi-squared, defined as: 
\begin{equation}
\chi^2_{\nu} = \frac{1}{n-1}\sum^n_{i=1}\frac{(m_i-\overline{m})^2}{\sigma_i^2},
\end{equation}
to test the hypothesis that the flux is consistent with a constant value defined by the mean. Here $n$ is the number of epochs, $m_i$ is the magnitude in either [3.6] or [4.5] at each epoch, $\overline{m}$ is the mean magnitude and $\sigma_i$ is the uncertainty at each epoch. Results for the cluster members are reported in Table~\ref{invar_table} and the distribution of chi-squared values, comparing the cluster members with all sources detected in our field of view, at both [3.6] and [4.5] is shown in Figure~\ref{chisq_mag}. The majority of non-cluster members are non-variable foreground or background stars whose $\chi^2_{\nu}$ peaks at one, consistent with no significant fluctuations. Among the non-variable sources the fluctuations are consistent with white noise, with no evidence of long-term trends or a frequency dependence for the noise over our 200 day observing window. As a boundary between variables and non-variables, we use $\chi^2_{\nu}=3$. Given the brightness of the cluster members, the uncertainties are mostly systematic rather than statistical, and are relatively uniform at $\sim$0.01 mag, which in turn implies that we are generally sensitive to variations larger than $\sim$0.03 mag in both bands. Based on the [3.6] photometry, 74$^{+15}_{-20}$\%\ (95\%\ confidence interval, 20/27) of the cluster members are variable, while the [4.5] photometry points to 68$^{+16}_{-20}$\%\ (19/28) of the cluster members as variable. There is a finite probability that a static source could have random fluctuations large enough to result in $\chi^2_{\nu}$, possibly biasing our statistics, and similar studies \citep[e.g.][]{reb14} have used a $\chi^2_{\nu}$ boundary of five instead of three to eliminate these false positives. Given our small sample size this does not substantially effect our results; only two stars at [3.6] and one at [4.5] fall within $3<\chi^2_{\nu}<5$. Removing these sources from our sample results in a decrease in the fraction of variable stars that is within our large Poisson errors. The difference between the fraction of variable stars between the two bands is due entirely to the difference in coverage of the two fields of view; no star is found to be variable in only one channel unless it was only observed in one channel. 

\begin{figure}
\center
\includegraphics[scale=.3]{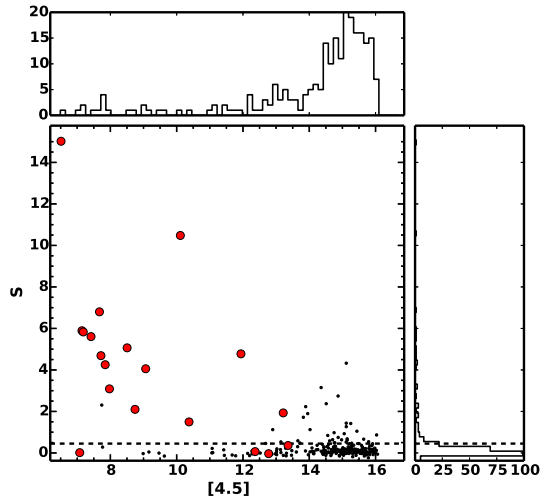}
\caption{Stetson index as a function of [4.5] magnitude. All detected sources, comprised mostly of background and foreground stars, with more than 20 simultaneous [3.6] and [4.5] detections are shown with black dots, while known cluster members are marked with large red circles. The horizontal dashed line indicates a Stetson index of 0.45, above which sources are included as variable. \label{stetson}}
\end{figure}

The [3.6] and [4.5] flux from young stars with disks traces a region of the spectral energy distribution (SED) whose emission is dominated by 1500 K blackbody emission from the inner wall \citep{mcc13a} and as a result we expect the flux in the two bands to be highly correlated. This is seen in previous surveys, with almost no sources exhibiting anti-correlated variability, and the most extreme cases showing a small change in [3.6]-[4.5] color \citep[e.g.][]{cod14,reb14}. This correlated behavior can be used to trace variability with the Stetson index \citep{ste96}, which is defined as:
\begin{equation}
S = \frac{\sum^n_{i=1}{\rm sgn}(P_i)\sqrt{|P_i|}}{N_{\rm pair}},
\end{equation}
where $N_{\rm pair}$ is the number of paired observations, which may or may not be equal to the number of total observations in each band $n$. The parameter $P_i$ is the product of the normalized residuals in the i-th epoch and is defined as:
\begin{equation}
P_i = \frac{N_{\rm pair}}{N_{\rm pair}-1}\frac{[3.6]_i-\overline{[3.6]}}{\sigma_{i,[3.6]}}\frac{[4.5]_i-\overline{[4.5]}}{\sigma_{i,[4.5]}}.
\end{equation}
Figure~\ref{stetson} shows the distribution of Stetson indices for the entire sample, with the cluster members marked. The length of the overlap between [3.6] and [4.5] ranges from zero to 200 days, with $\sim100$ days typical for our sample. We use a boundary of $S>0.45$ to define variables and find that 80$^{+14}_{-24}$\%\ (16/20) of the cluster members are variable. All of the sources found to be variable by the Stetson index are also variable based on the chi-squared, while only star 383 is variable based on the [3.6] and [4.5] chi-squared but does not meet our Stetson index threshold. Star 383 has a highly erratic light curve during which there is little overlap between the two bands, which may contribute to its mis-classification by the Stetson index. The consistency between the variability detected by the chi-squared and by the Stetson index indicates that the assumption of correlated [3.6] and [4.5] fluctuations does not substantially bias the Stetson index. Our choice of boundary also does not effect our results; similar studies \citep[e.g.][]{reb14} have used S$>$0.9, while our weakest variable has S=1.5.

The high fraction of variable stars is consistent with {\it Spitzer} surveys of infrared variability (Table~\ref{cluster_comparison}). In particular the YSOVAR survey, with its {\it Spitzer} observations of IRAS:20050+2720 \citep{pop15}, NGC 1333 \citep{reb15}, the ONC \citep{mor11}, L1688 \citep{gun14}, and GGD 12-15 \citep{wol15}, find that 60-85\%\ of young stellar objects with an infrared excess are variable. Studies of IC 348 \citep{fla13}, NGC 2264 \citep{cod14} and the Orion molecular cloud \citep{meg12} have found similar results. Breaking down the sample by SED class we find 2/3 class I objects, 18/24 class II objects and 2/5 class III objects are variable. Previous studies have found infrared variability to be more common among less evolved sources \citep[e.g.][]{meg12,fla13} and our results, while suffering from small number statistics, are marginally consistent with this trend.

\section{Characterizing the Variability}
Different physical models make different predictions for the properties of the variability. Hot or cold spots on the stellar surface will produce periodic behavior if the spots are long lived, with a timescale equal to the stellar rotation period, and create fluctuations whose size and color depend on the covering fraction and temperature of the spot. A highly-variable warp at the inner edge of the disk will produce quasi-periodic fluctuations that will cause extinction if the system is viewed close to edge-on, but will appear as fluctuating disk emission if the system is viewed closer to face-on. By quantifying the size, color (Section~\ref{color}) and timescale (Section~\ref{timescale}) of the fluctuations we can add to our understanding of the underlying physical source of the variability (Section~\ref{discussion}).

\subsection{Color Variations\label{color}}
\begin{figure*}
\center
\includegraphics[scale=.5]{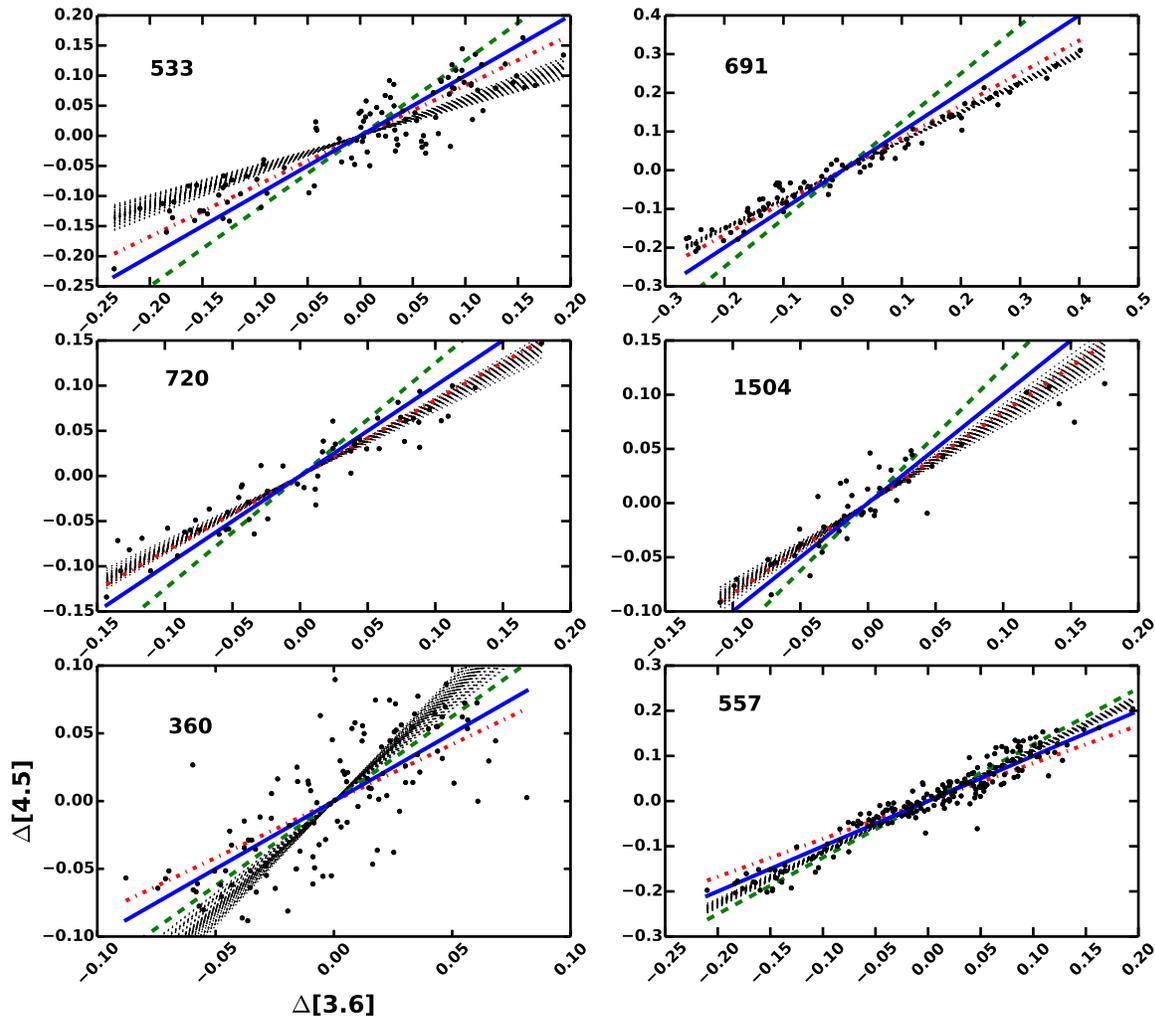}
\caption{Change in the [3.6] and [4.5] magnitude for the six sources with significant color variations. The blue solid line shows no change in color, the red dotted line shows extinction by A$_K$=1, and the green dashed line shows a change in dust emission. The measured slope, along with its uncertainty, is indicated by the distribution of dashed lines. The first four systems (533, 691, 720, 1504) exhibit reddening, while the last two systems (360, 557) experience bluening. \label{color_slopes1}}
\end{figure*}

\begin{figure*}
\center
\includegraphics[scale=.5]{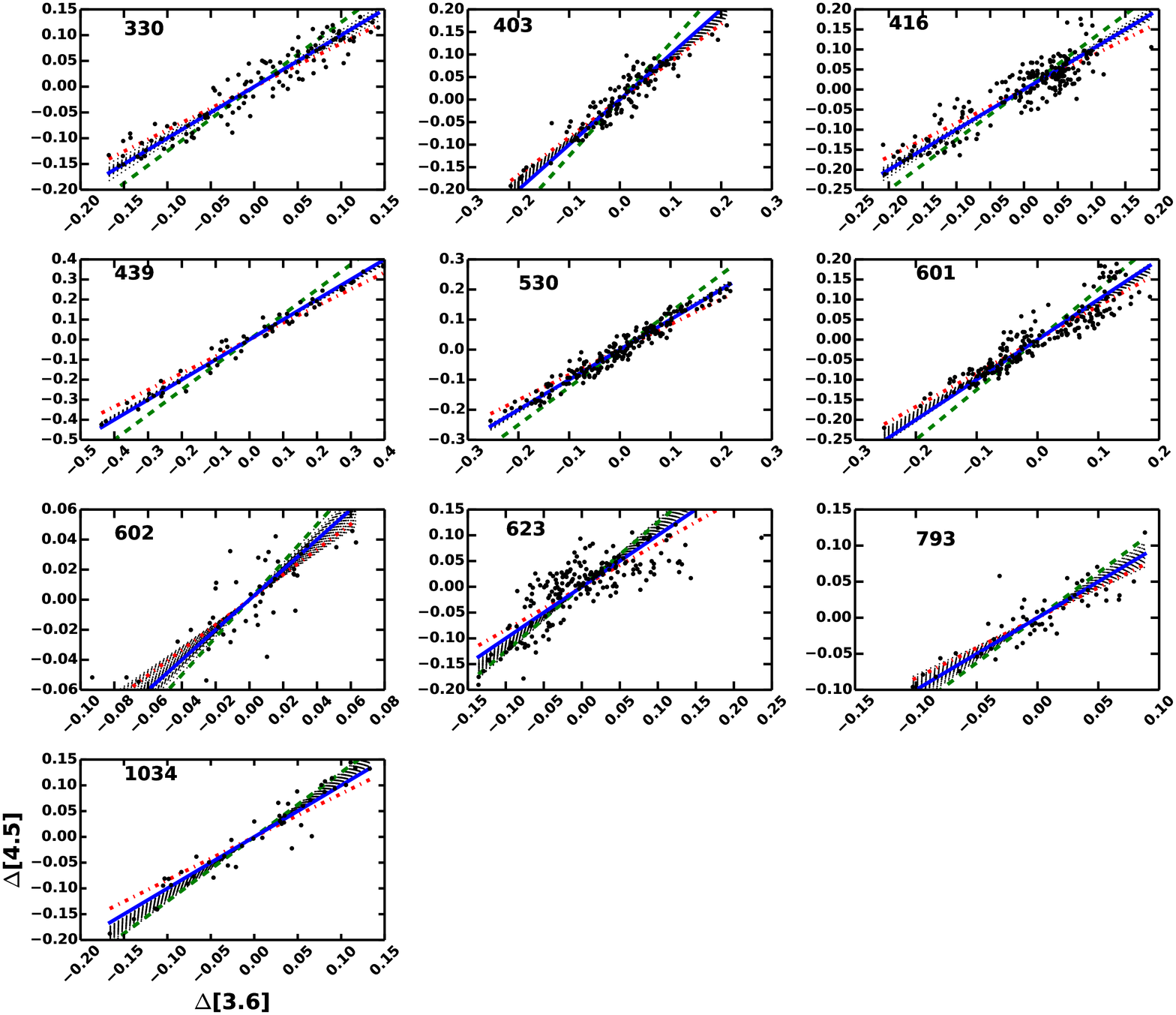}
\caption{Change in the [3.6] and [4.5] magnitude for the ten sources with substantial overlap in both bands but no significant average color variations. Lines and points are as in Figure~\ref{color_slopes1}. Even when the color does not vary on average, there are points that follow the extinction or disk fluctuation model predictions, indicating that a mix of different processes may be operating within these systems. \label{color_slopes2}}
\end{figure*}

To begin to understand the source of the variability, we look for changes over time in the [3.6]-[4.5] color. To quantify the change in color we measure the slope of the correlation between the [3.6] and [4.5] photometry, which translates into a direct measure of $\Delta$[4.5]/$\Delta$[3.6] ($\Delta$[3.6] = [3.6]-$\overline{[3.6]}$, $\Delta$[4.5] = [4.5]-$\overline{[4.5]}$), and in turn a measure of the change in [3.6]-[4.5] averaged over the observing window. Previous studies \citep[e.g.][]{pop15} have utilized color-magnitude diagrams, such as [3.6] vs [3.6]-[4.5], which contain the same information but are slightly biased by the correlated errors between the two axes. In our diagnostic, $\Delta$[4.5]/$\Delta$[3.6]=1 indicates no change in color, $\Delta$[4.5]/$\Delta$[3.6]$>$1 indicates a 'blueining' (the source gets bluer as it gets fainter, or similarly, redder as it gets brighter) and $\Delta$[4.5]/$\Delta$[3.6]$<$1 indicates a reddening (the source gets redder as it gets fainter, or similarly, bluer as it gets brighter). We measure the slope using the python Orthogonal Distance Regression method, which accounts for uncertainties in both axes, with results for those stars with substantial [3.6] and [4.5] overlap (N$_{\rm pair}>$20) listed in Table~\ref{color_table} and shown in Figures~\ref{color_slopes1} and~\ref{color_slopes2}. 

\begin{figure}
\center
\includegraphics[scale=.3]{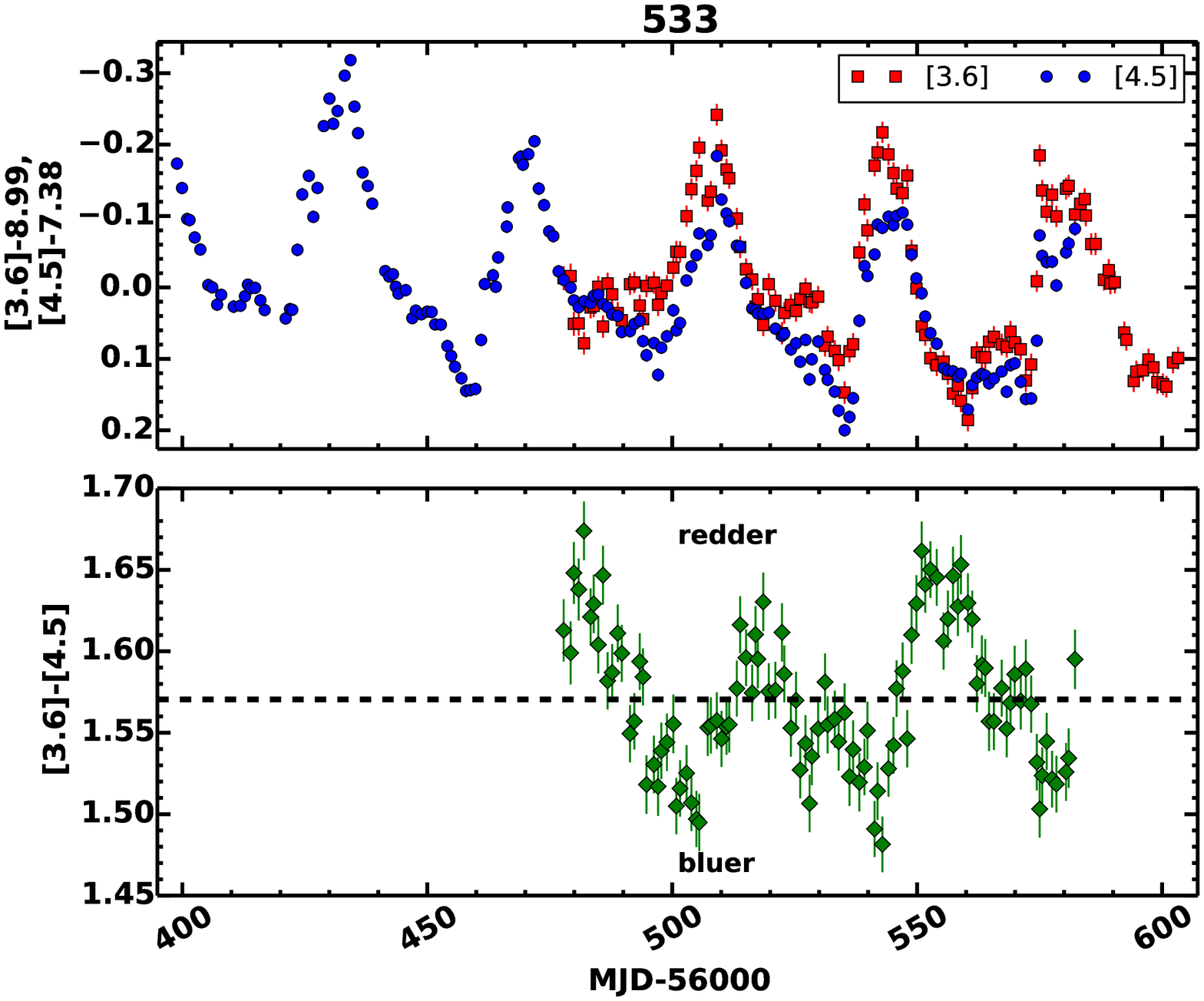}
\includegraphics[scale=.3]{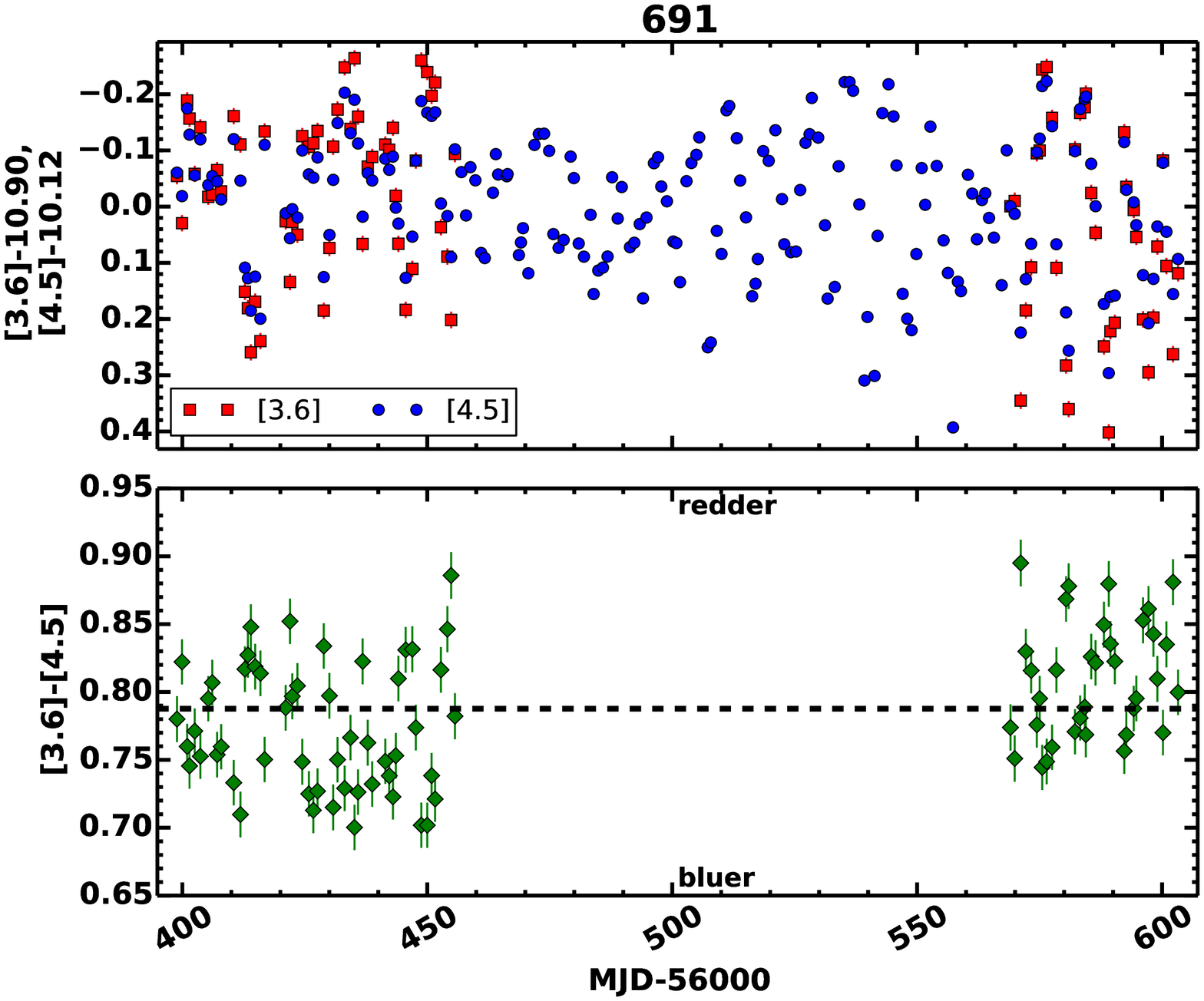}
\includegraphics[scale=.3]{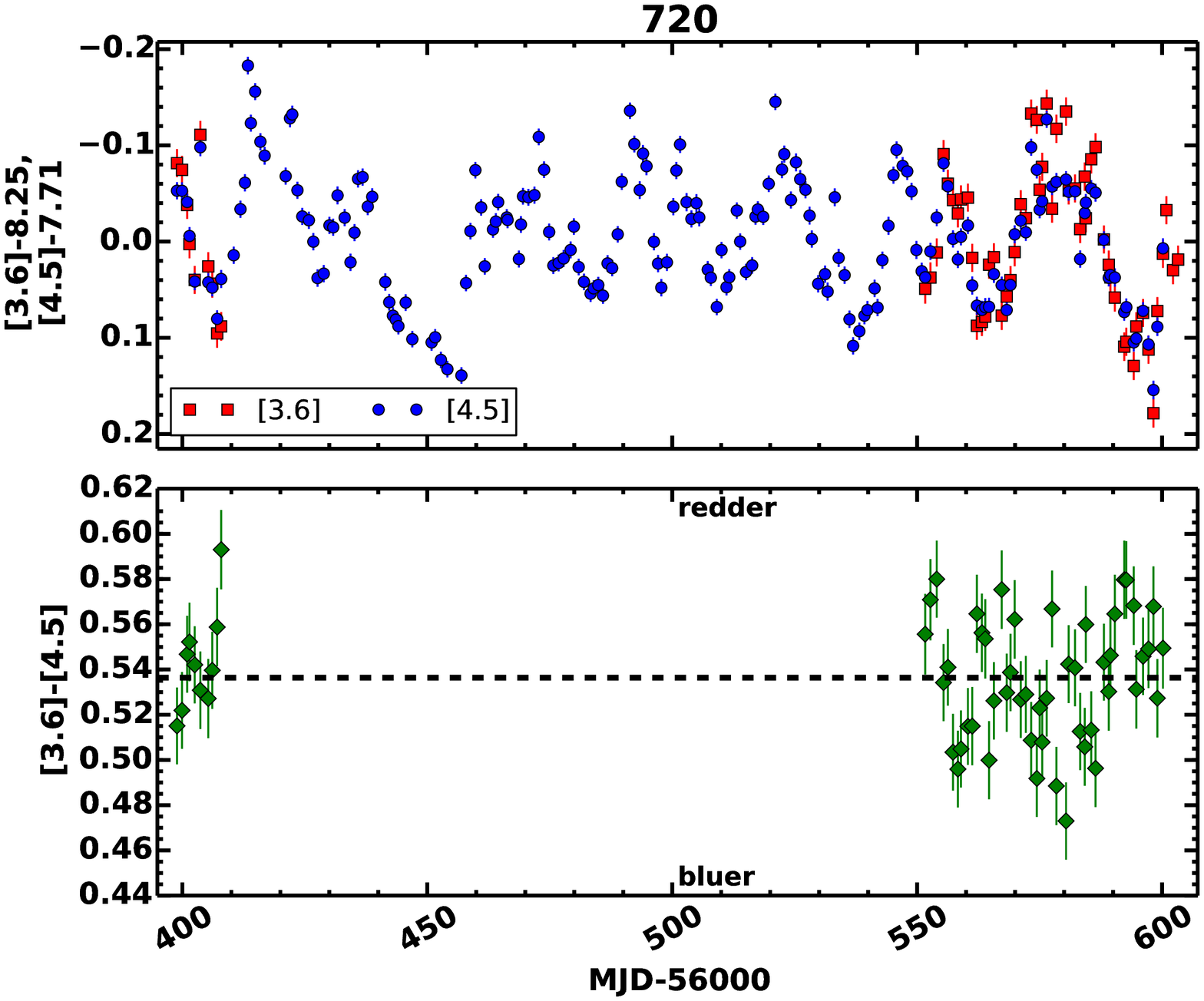}
\includegraphics[scale=.3]{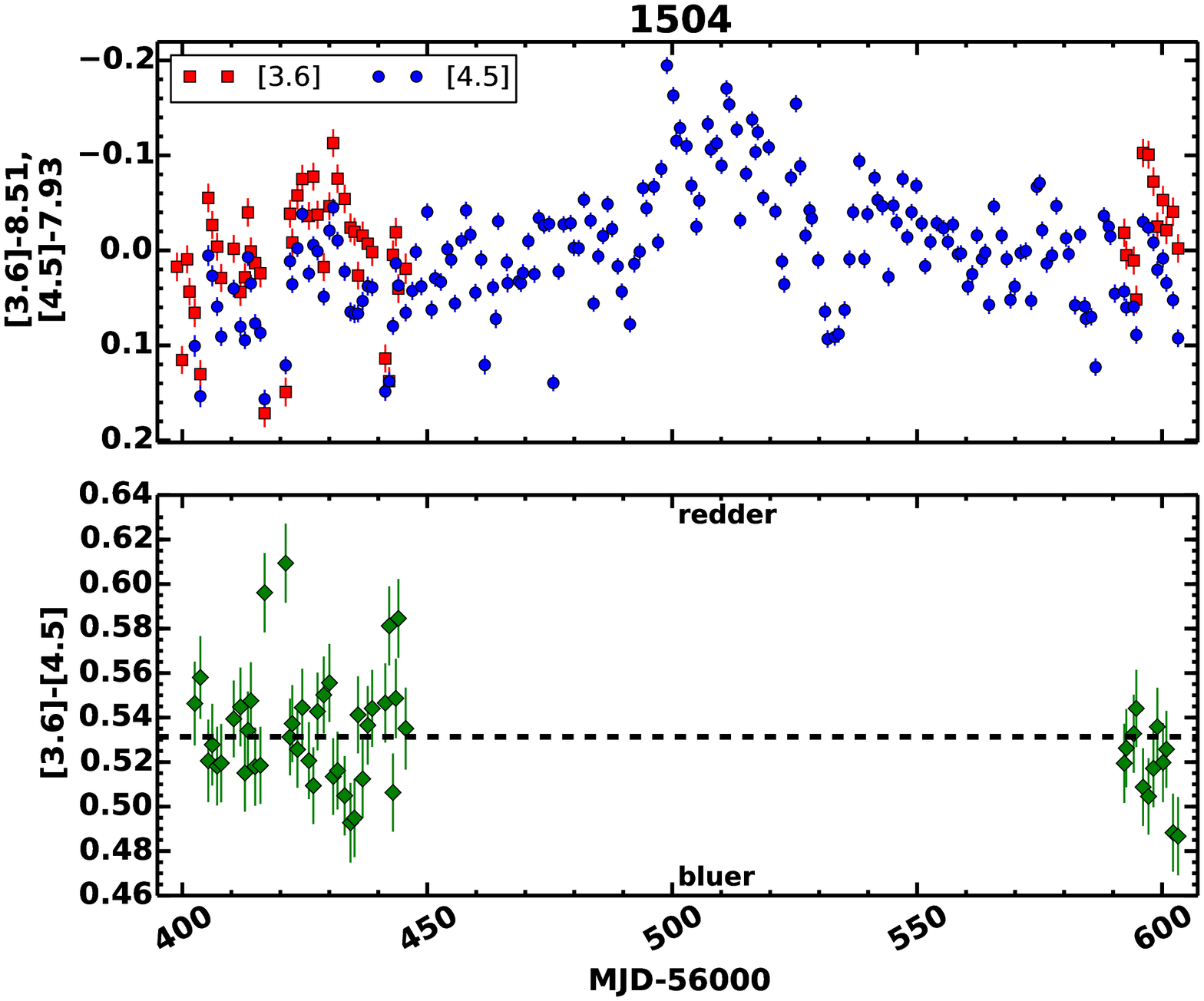}
\caption{Light curves of stars that get significantly redder as they get fainter ($\Delta$[4.5]/$\Delta$[3.6]$<$1). This reddening behavior may be a sign of extinction by small dust grains. \label{lc_redder}}
\end{figure}

\begin{figure}
\center
\includegraphics[scale=.3]{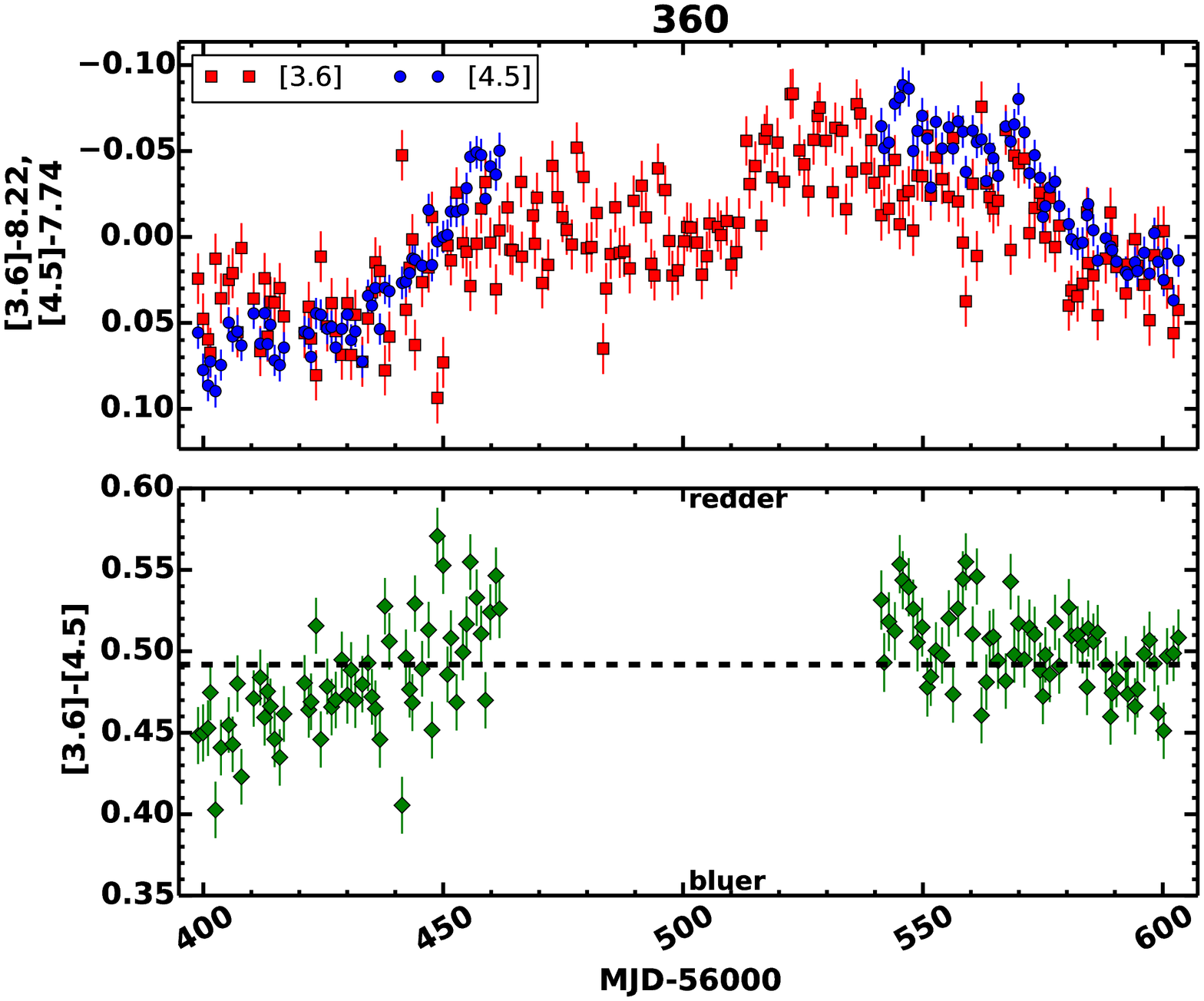}
\includegraphics[scale=.3]{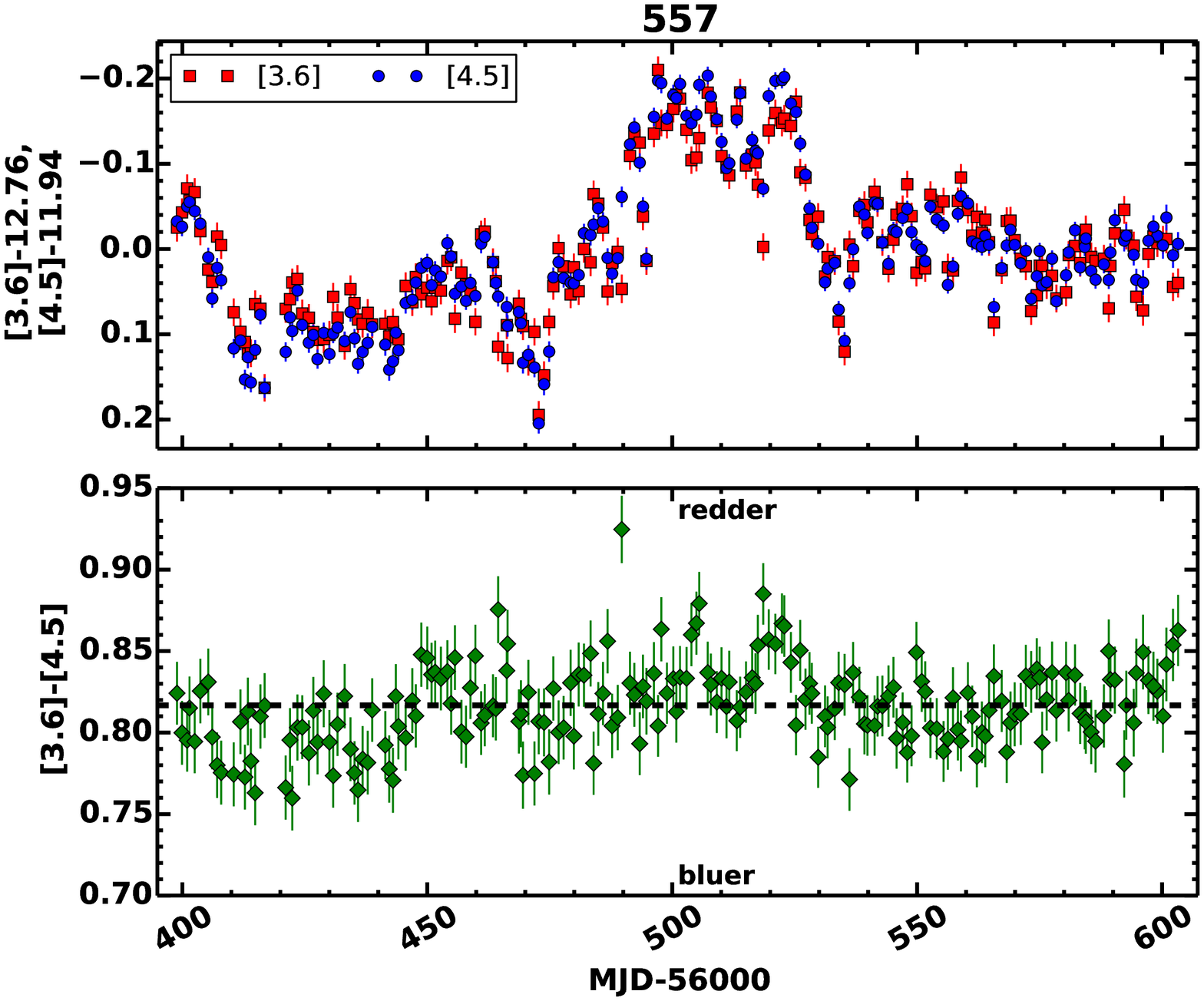}
\caption{Light curves of stars that get significantly bluer as they get fainter ($\Delta$[4.5]/$\Delta$[3.6]$>$1). This behavior may be a sign of variable dust emission.\label{lc_bluer}}
\end{figure}

The majority of the young stellar objects are consistent with having no significant change in color. Given our typical uncertainties this corresponds to 0.9$<\Delta$[4.5]/$\Delta$[3.6]$<$1.1, or a change in [3.5]-[4.5] smaller than 0.05 mag for every half magnitude change in [3.6]. We do find four sources (533,691,720,1504, Figure~\ref{lc_redder}) that become significantly redder as they get fainter ($\Delta$[4.5]/$\Delta$[3.6]$<$1) and two sources (360,557, Figure~\ref{lc_bluer}) that become significantly bluer as they become fainter ($\Delta$[4.5]/$\Delta$[3.6]$>$1). For these sources, the [3.6]-[4.5] color varies by 0.06-0.3 magnitudes for every half magnitude change in [3.6] magnitude. None of these color variations is large enough to change the evolutionary stage classification. \citet{reb14} find such events to be very rare, occurring in $<$0.02\% of infrared sources, although some cases have been found \citep{ric12}.

\begin{figure*}
\center
\includegraphics[scale=.3]{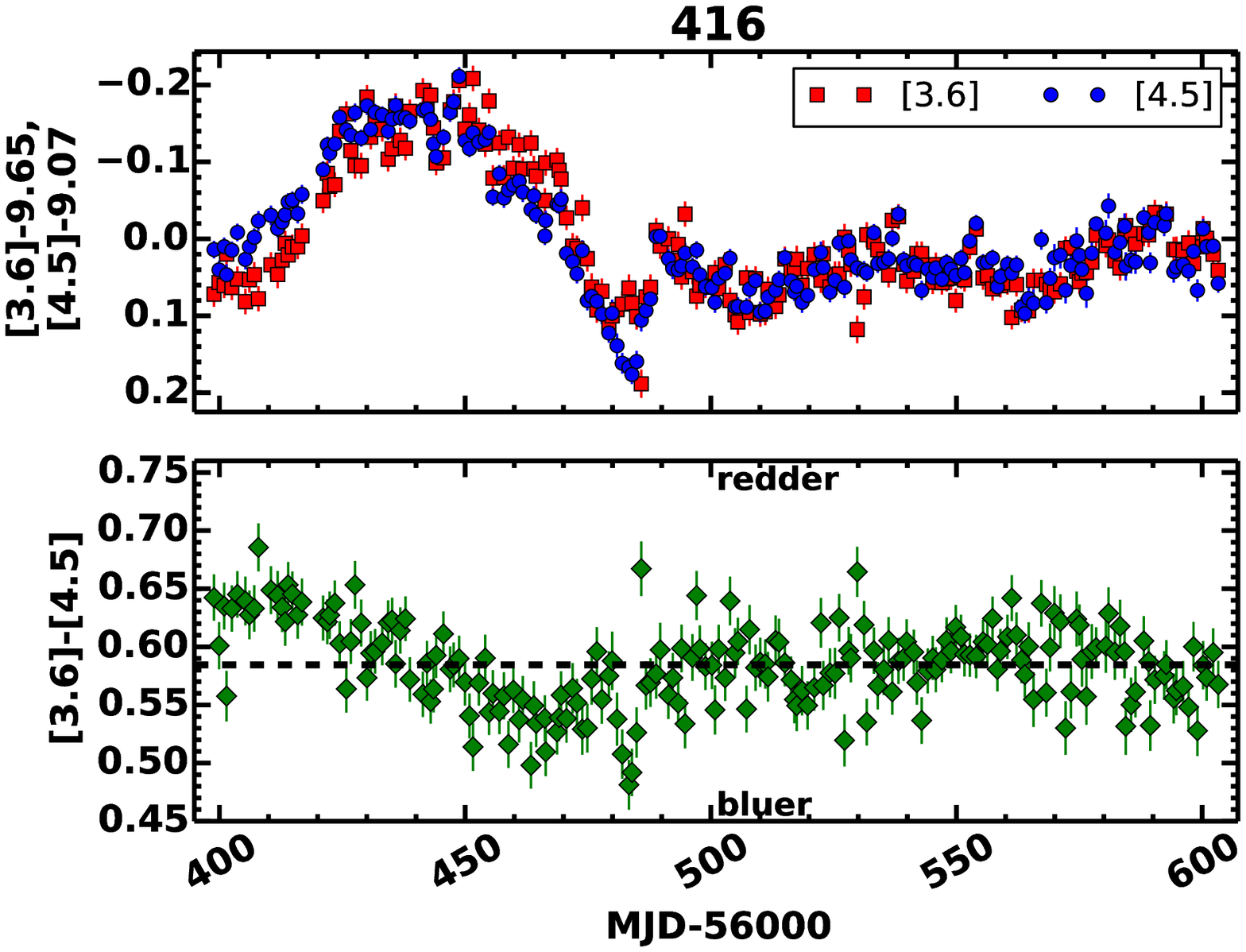}
\includegraphics[scale=.3]{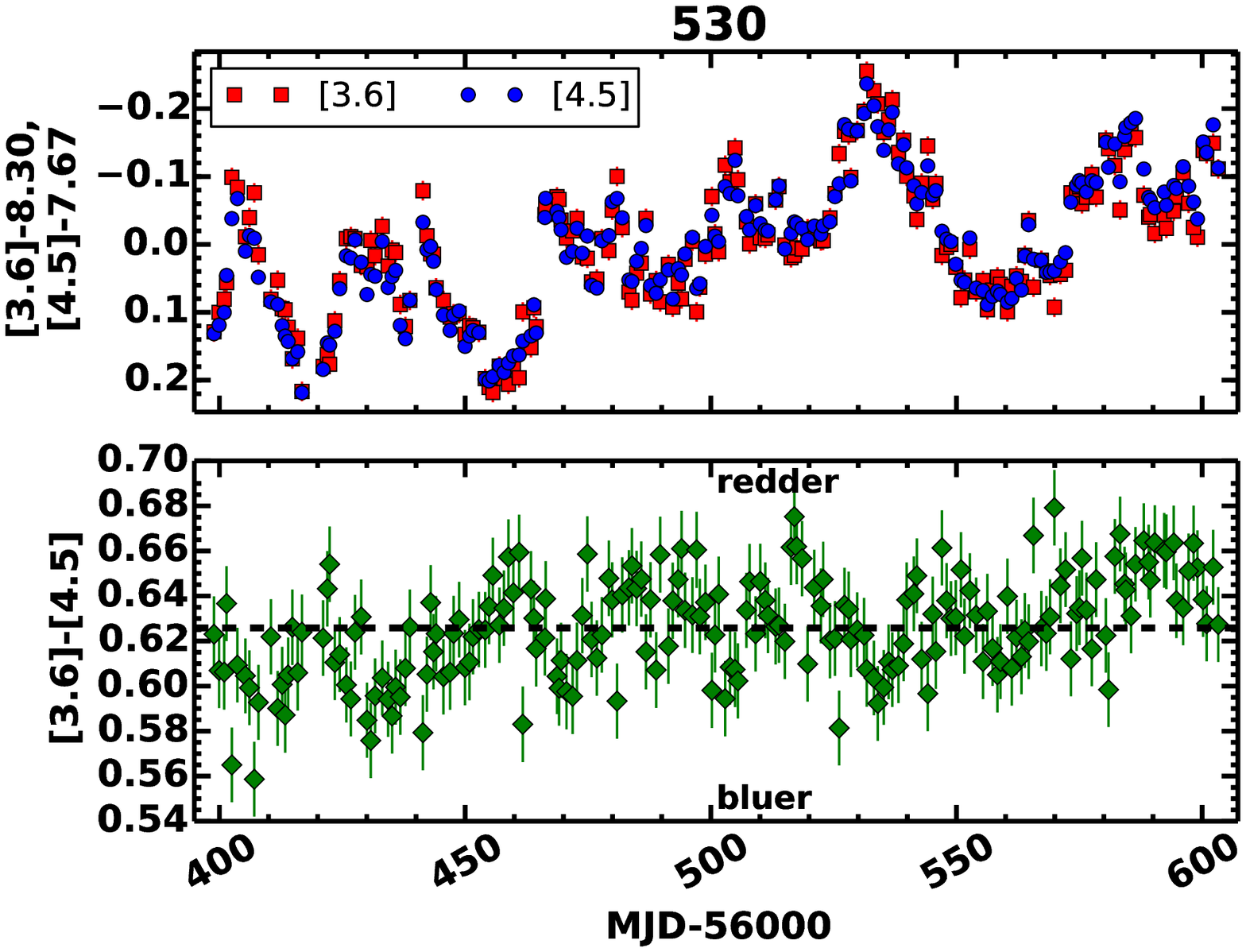}
\includegraphics[scale=.3]{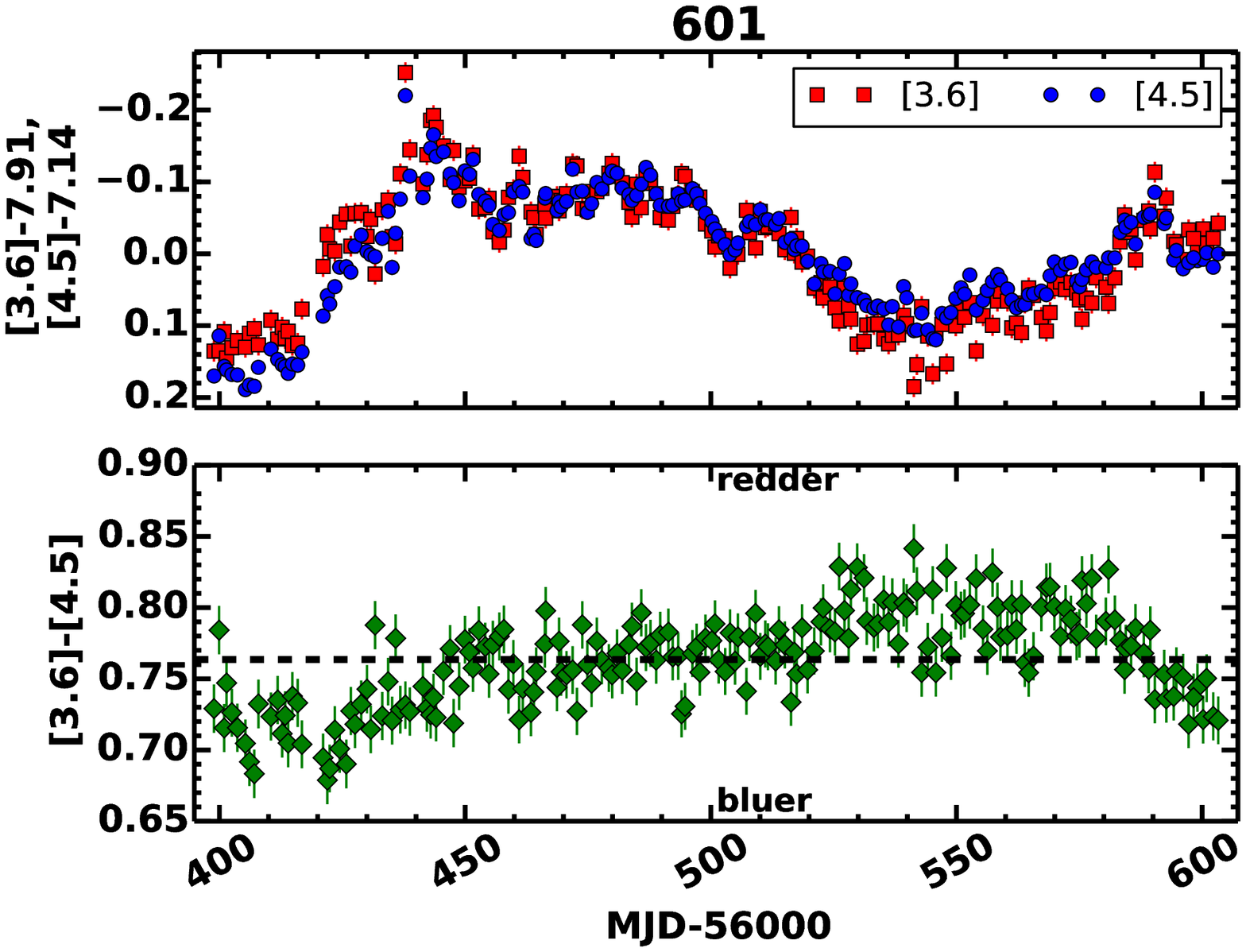}
\includegraphics[scale=.3]{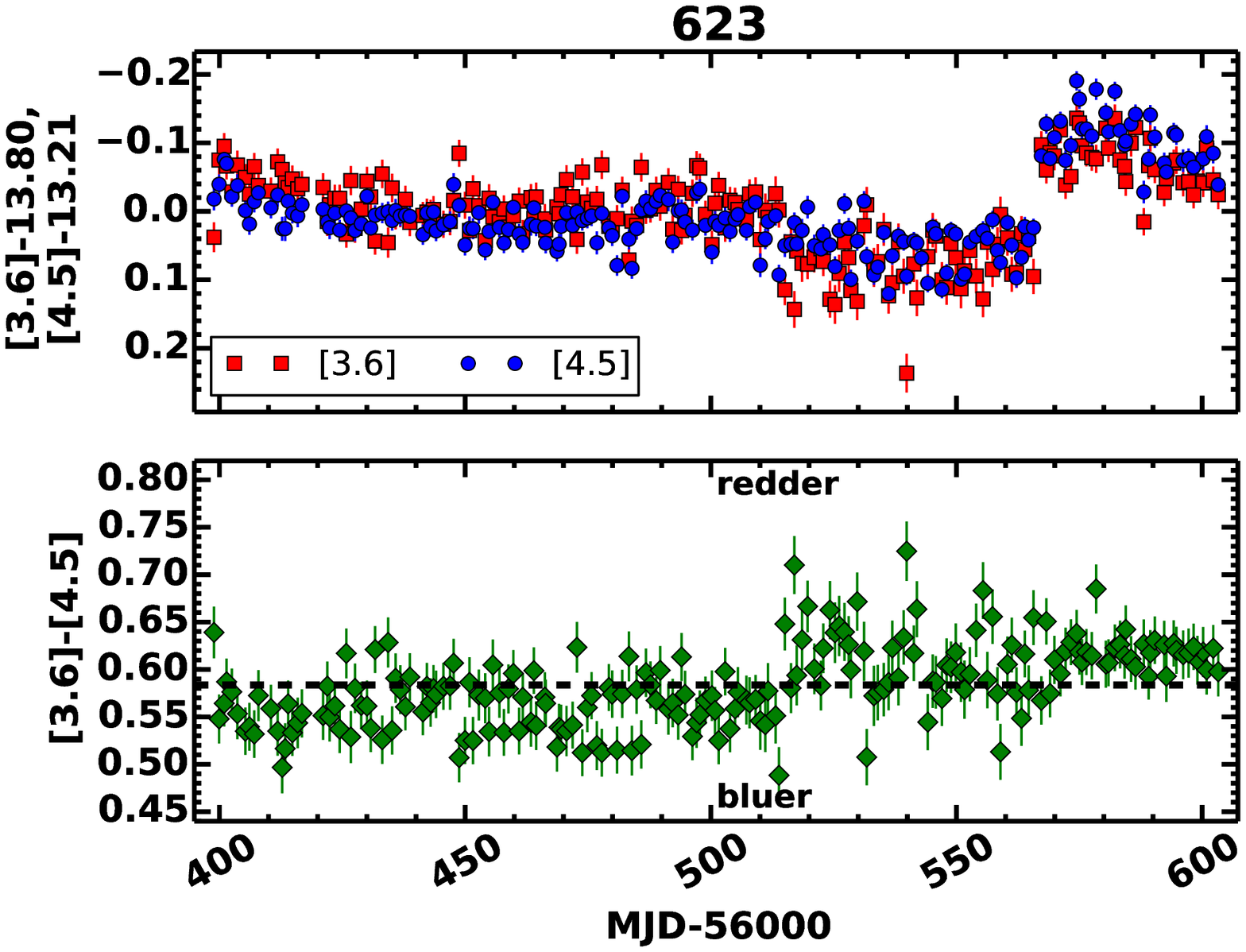}
\caption{Light curves of stars for which the average color does not vary, but the color varies on short timescales within the light curve. A mix of multiple processes (extinction, disk fluctuations, etc.) may cancel each other out producing effectively zero change in overall color over the course of the entire observing window.\label{lc_nocolor}}
\end{figure*}

By measuring $\Delta$[4.5]/$\Delta$[3.6] we are tracing the average change in color over the entire 200 day observing window. Multiple processes that produce opposite effects on the color can operate simultaneously, effectively canceling each other out in the long term, while still producing short term fluctuations in color. We see evidence for this in some of the light curves that show no average change in color (Figure~\ref{lc_nocolor}), as well as the distribution of $\Delta$[4.5]/$\Delta$[3.6] (Figure~\ref{color_slopes2}). Star 601 shows a long term trend where it gets redder over the course of the observing window, before turning bluer over the last 40 days. Star 530 oscillates in color, with the color light curve tending to become bluer during outbursts, although not consistently and strongly enough to produce a significant signal in $\Delta$[4.5]/$\Delta$[3.6]. Star 416 gets significantly bluer during the large initial outburst, but has a relatively constant color afterwards, while star 623 has a much shallower slope in $\Delta$[4.5]/$\Delta$[3.6] when it is fainter. The structure in the lightcurves suggests that there may be a timescale dependence to the color variations, which we examine in more detail in Section~\ref{timescale}.

Since the [3.6] and [4.6] flux is a mix of stellar and dust emission and stellar emission is hotter than the dust emission, changes to one or the other will have distinct effects on the observed color. We consider the variations in [3.6] and [4.5] for three scenarios (1) hot/cold spots, (2) extinction and (3) changes in the dust emission strength. The hot/cold spots are modeled as T$_{\rm spot}$=8000,2000 K spots covering a fixed fraction of a star with an effective temperature representative of the median of the cluster (3500 K). With covering fractions of 5\%\ and 30\%\ for the hot and cold spot we find $\Delta$[3.6] = 0.16 and 0.12 and $\Delta$[4.5]/$\Delta$[3.6] = 1.1 and 0.9 respectively. A dust cloud with a depth of A$_K$=1 creates $\Delta$[3.6]=0.6 with a $\Delta$[4.5]/$\Delta$[3.6]=0.84 based on the extinction law of \citet{fla07}, and $\Delta$[4.5]/$\Delta$[3.6] closer to one if the grains within the disk are larger than those in giant molecular clouds. To model variable dust emission we consider a simple model with a T$_{\rm eff}$=3500 K blackbody representing the stellar emission underneath a T$_{\rm dust}$=1500 K blackbody with varying strength. This model is meant to mimic a system in which the stellar contribution to the [3.6] and [4.5] emission is constant while the contribution from the disk itself changes with time, possibly due to changes in the emitting area (e.g. scale height) of the dust. This model produces fluctuations of a few tenths of a magnitude with $\Delta$[4.5]/$\Delta$[3.6]=1.25, with slight variations in this slope expected for stars of different T$_{\rm eff}$ and for different fiducial dust strengths. 

Comparing the predictions from these phenomenological models to the data (Table~\ref{color_table}, Figure~\ref{color_slopes1}), we find that fluctuations from hot/cold spots are inconsistent with the large observed fluctuations, and that the data are more consistent with either extinction or disk emission fluctuations. Dust emission fluctuations are more consistent with the sources that show 'bluening' (Figure~\ref{lc_bluer}: stars 330, 557), both of which exhibit class II SED shapes. Extinction is most consistent with those sources that show reddening of their infrared colors (Figure~\ref{lc_redder}: stars 533, 691, 720, 1504). Of these sources, 720 and 1504 have class II SED shapes, while 533 and 691 are more consistent with class I objects. We note that $\Delta$[4.5]/$\Delta$[3.6]$<$1 can be equivalently interpreted as meaning the system becomes redder as it becomes fainter, or that it becomes bluer as it becomes brighter. The former interpretation can be explained by dust extinction, while the later is more characteristic of a large increase in the hot accretion emission. Star 533, discussed in more detail below, may be better explained by an accretion outburst rather than obscuration by dust. 


Previous studies have found similar results in other clusters. \citet{pop15} find that color changes in the Spitzer bands of the YSOs in IRAS:250050+2720 are consistent with a mix of extinction and disk variability, with spots more prevalent among sources with heavily depleted dust emission. \citet{wol15} also find that extinction and disk emission changes are most consistent with the Spitzer variability in the cluster GGD 12-15. Near-infrared studies of Orion have also found similar results \citep{car01,par14}, with a larger role for spots due to the shorter wavelength near-infrared bands being more sensitive to stellar emission.

\subsection{Timescale of the Variability\label{timescale}}
Another property that can help to determine the nature of the variability is its timescale. Processes associated with the stellar surface, such as heating of the disk by variable hot spots \citep[e.g.][]{her94,nag15} will operate on a timescale similar to the stellar rotation period, which is typically 1-15 days \citep{reb04,reb06,cie06,cod10}. The timescales of 'dippers', in which the flux briefly decreases on quasi-periodic timescales of days, has been used to constrain the location of the obscuring material to close to the star \citep{sta15,mcg15}. Processes originating in the disk will operate on the local dynamical timescales, which can range from weeks to years. Periodic fluctuations indicate stable, long-lived effects, while aperiodic behavior indicates that the features responsible for the variability are quickly created and destroyed. When combined with the color information discussed above, the physical cause of the variability can be constrained. Here, we search for periodic fluctuations and also characterize the power spectrum of the aperiodic fluctuations that are seen among the majority of stars.

\subsubsection{Periodicity}
We look for periodic behavior using the NASA Exoplanet Archive Periodogram Service\footnote{http://exoplanetarchive.ipac.caltech.edu/cgi-bin/Pgram/nph-pgram}, which examines the light curves using the Lomb-Scargle periodogram \citep{sca82}, the Box-fitting Least Squares algorithm \citep{kov02} and the Plavchan Algorithm \citep{pla08}. Each light curve has been smoothed by a 50 day moving average; long-term trends in the data will enhance the power at low frequencies resulting in spurious period detections. We consider a star as periodic if one of the methods finds a peak with power $>$20, which roughly corresponds to a false-alarm probability $<$10$^{-5}$, this period is recovered in the other methods and in both wavelengths. We also confirm the variability based on the presence of well-phased features at the measured period. This conservative set of criteria results in only five periodic stars (Table~\ref{periodic_table}), for a periodic fraction of 16$^{+17}_{-11}$\%\ (5/32). Although the period search algorithms are sensitive to weaker variability than the $\chi^2_{\nu}$ or Stetson index, no periodicity was detected in a source for which the $\chi^2_{\nu}$ or Stetson index did not already indicate strong variability. 



\begin{figure*}
\center
\includegraphics[scale=.25]{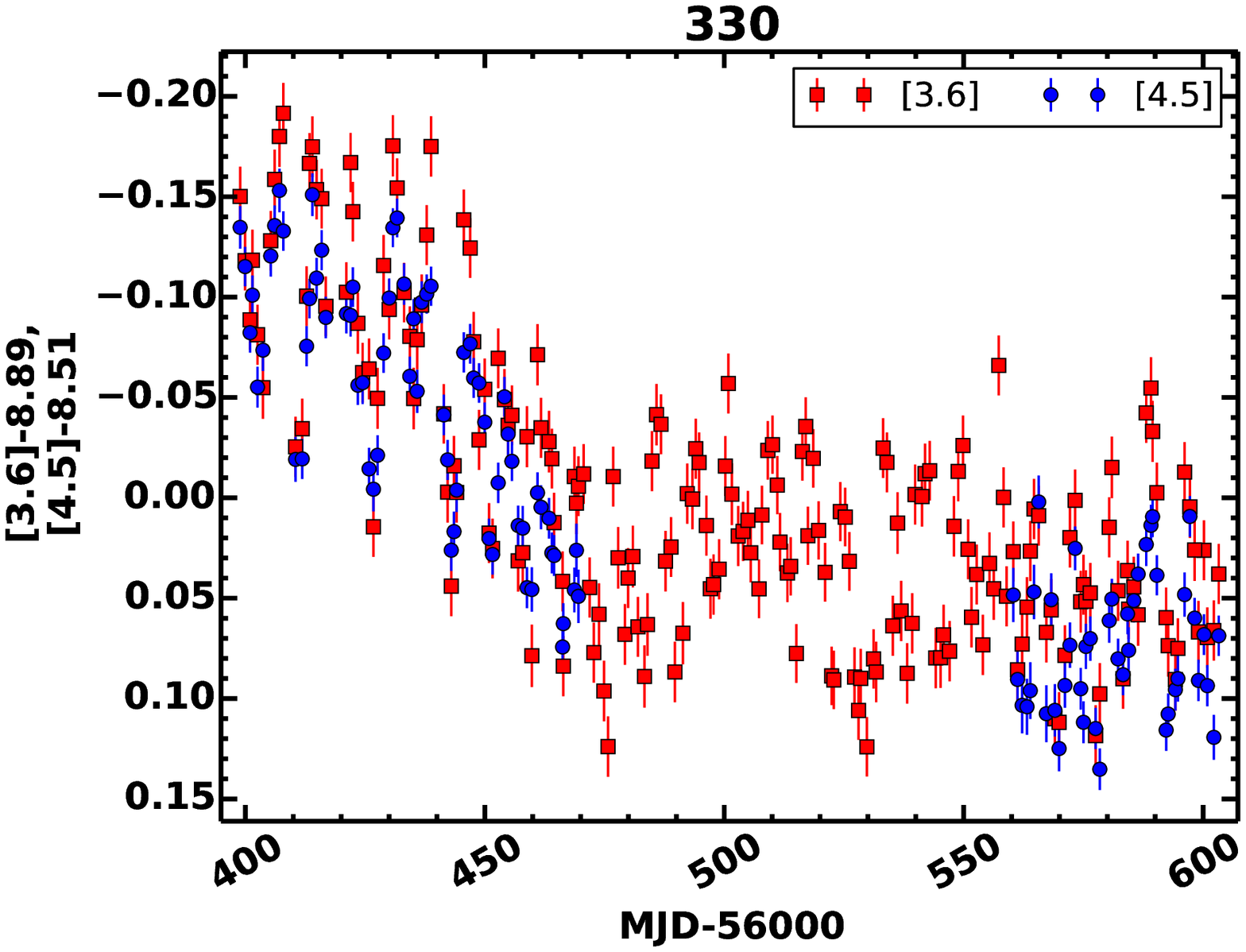}
\includegraphics[scale=.25]{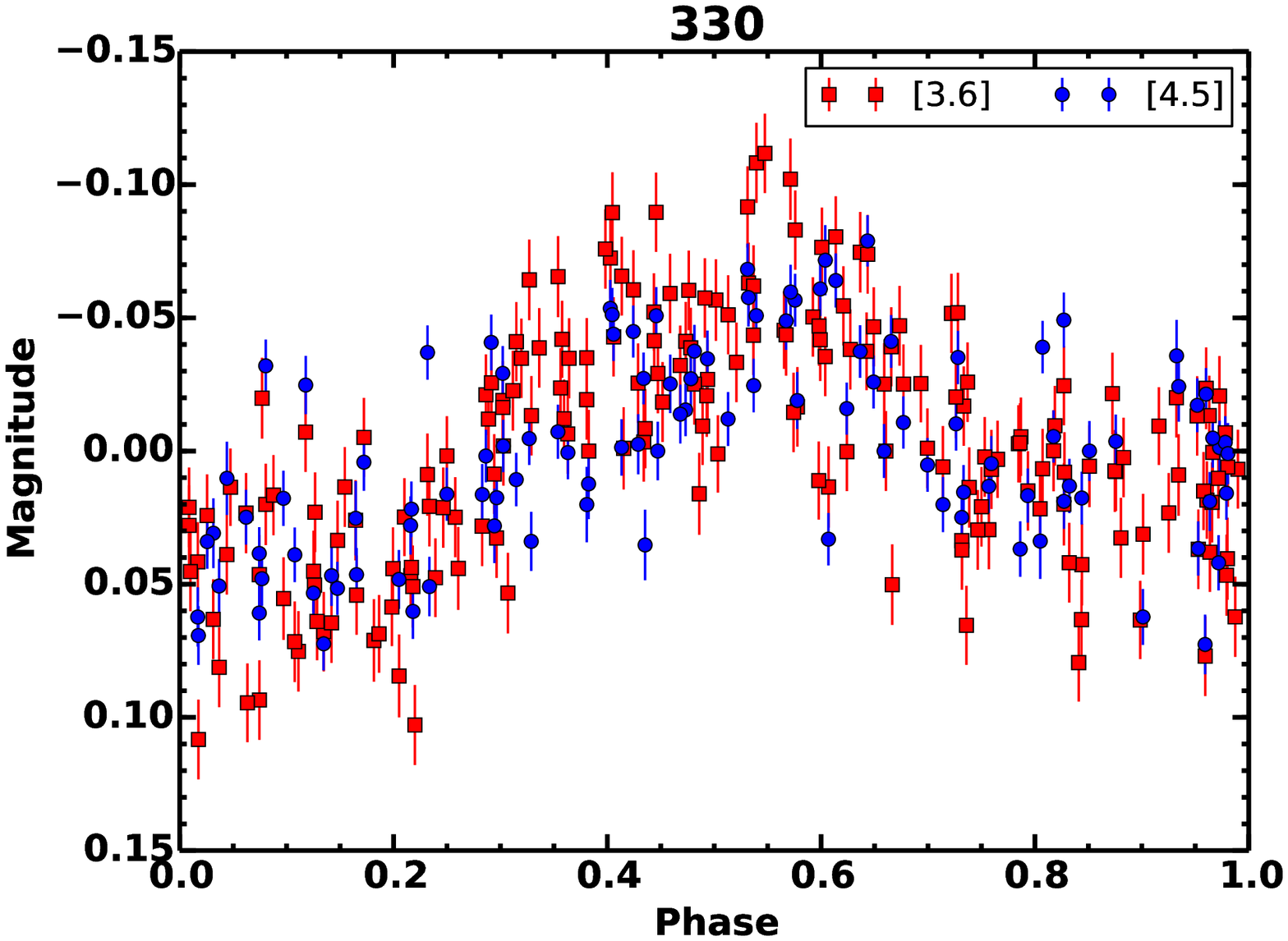}
\includegraphics[scale=.25]{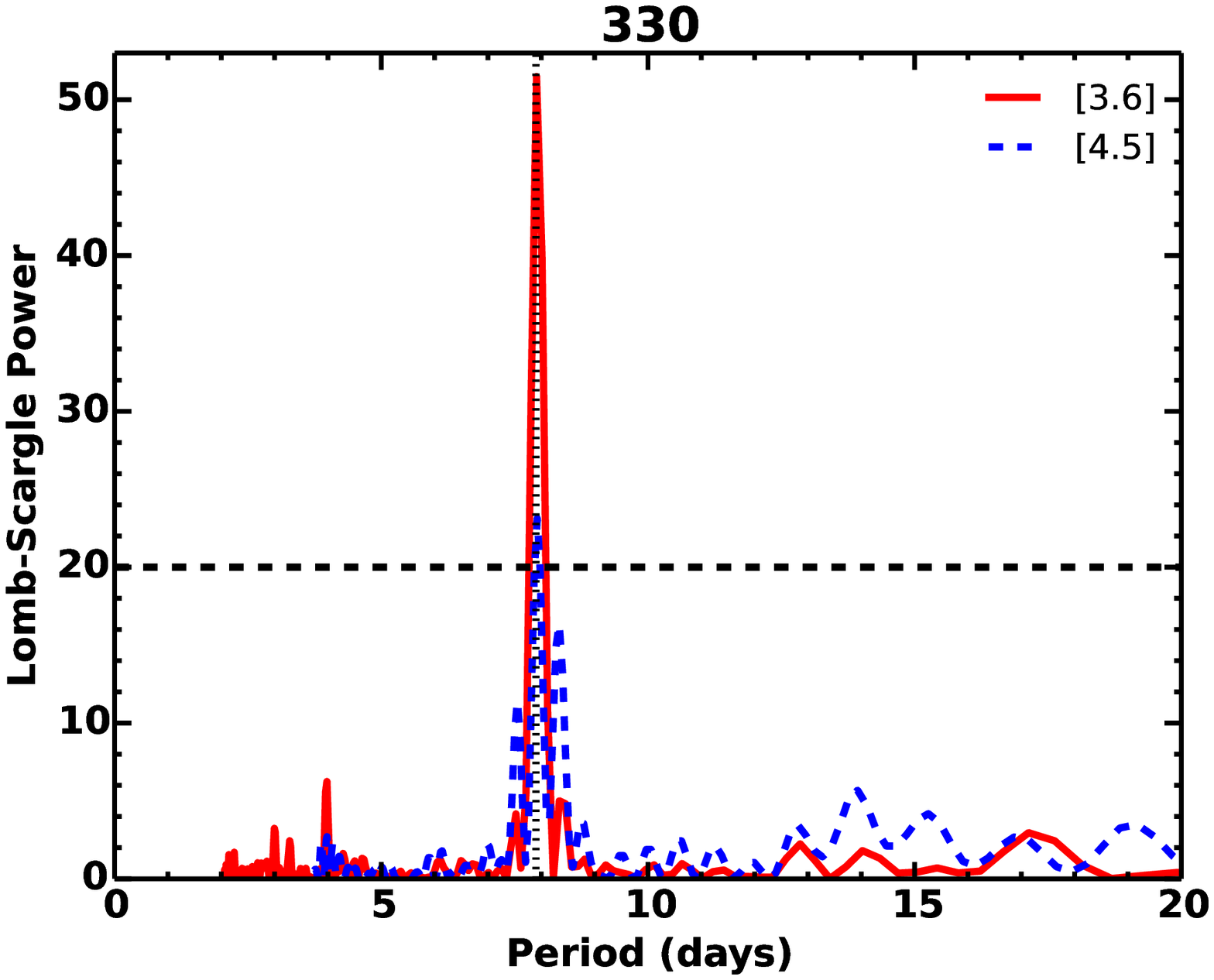}
\includegraphics[scale=.25]{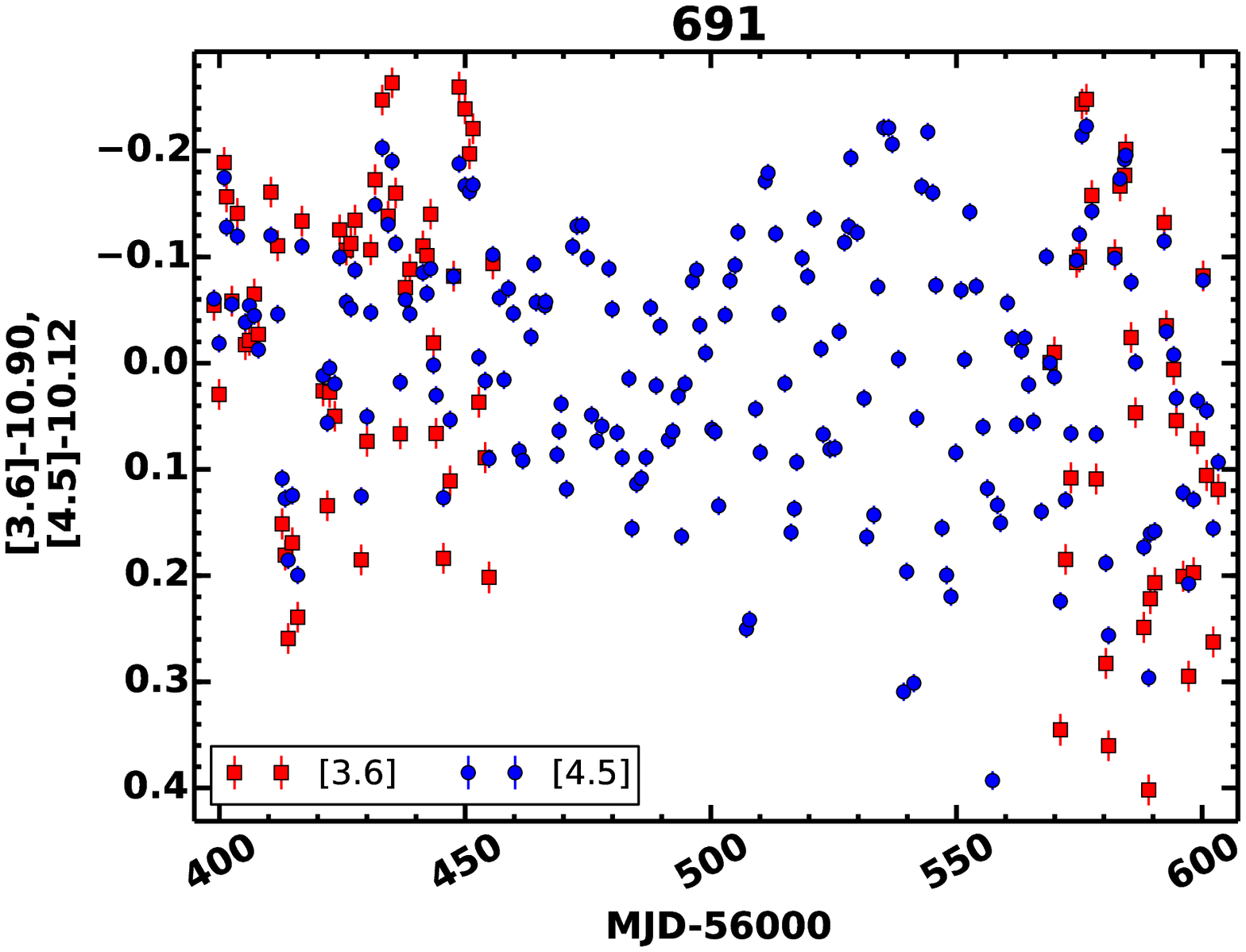}
\includegraphics[scale=.25]{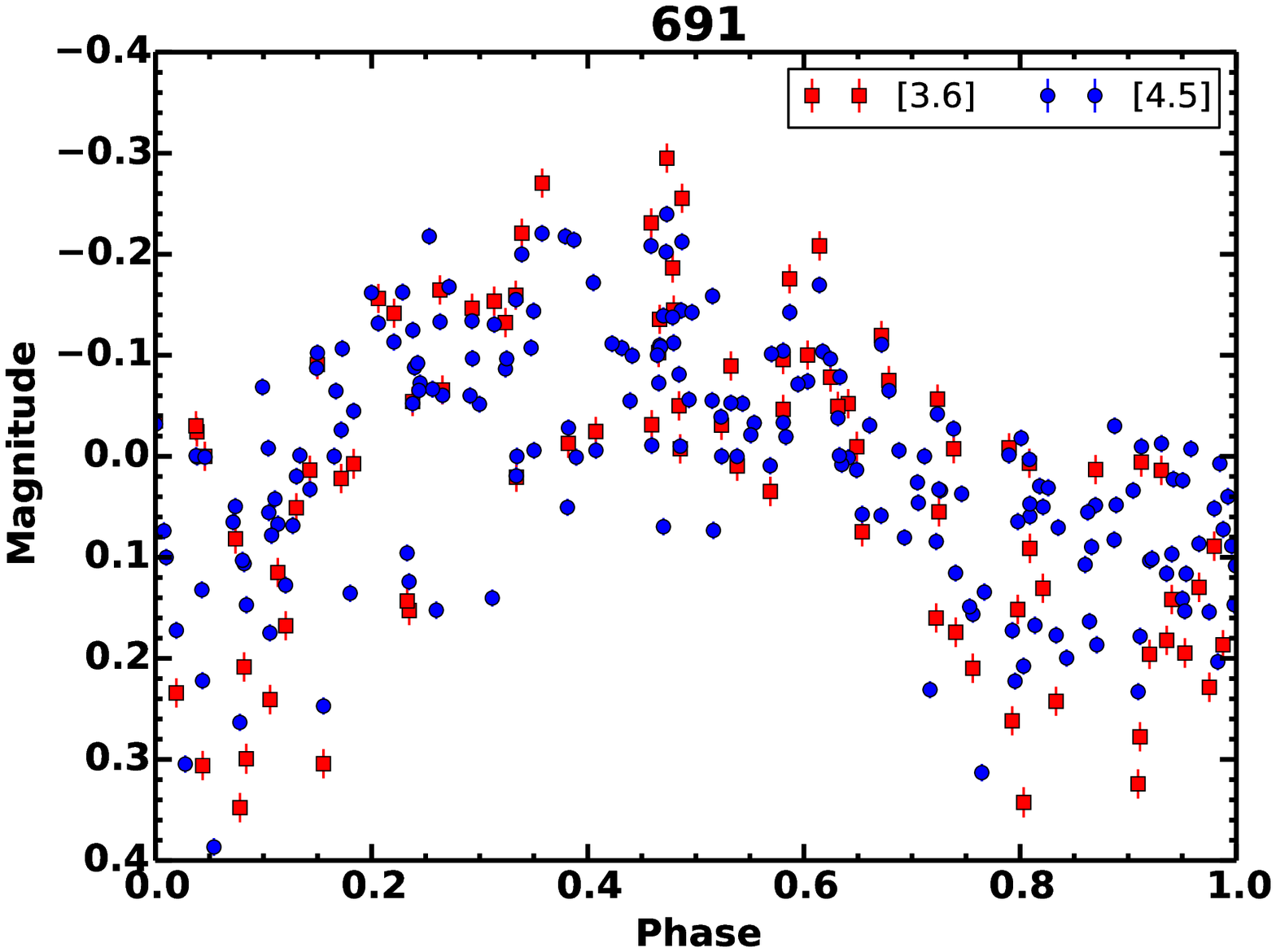}
\includegraphics[scale=.25]{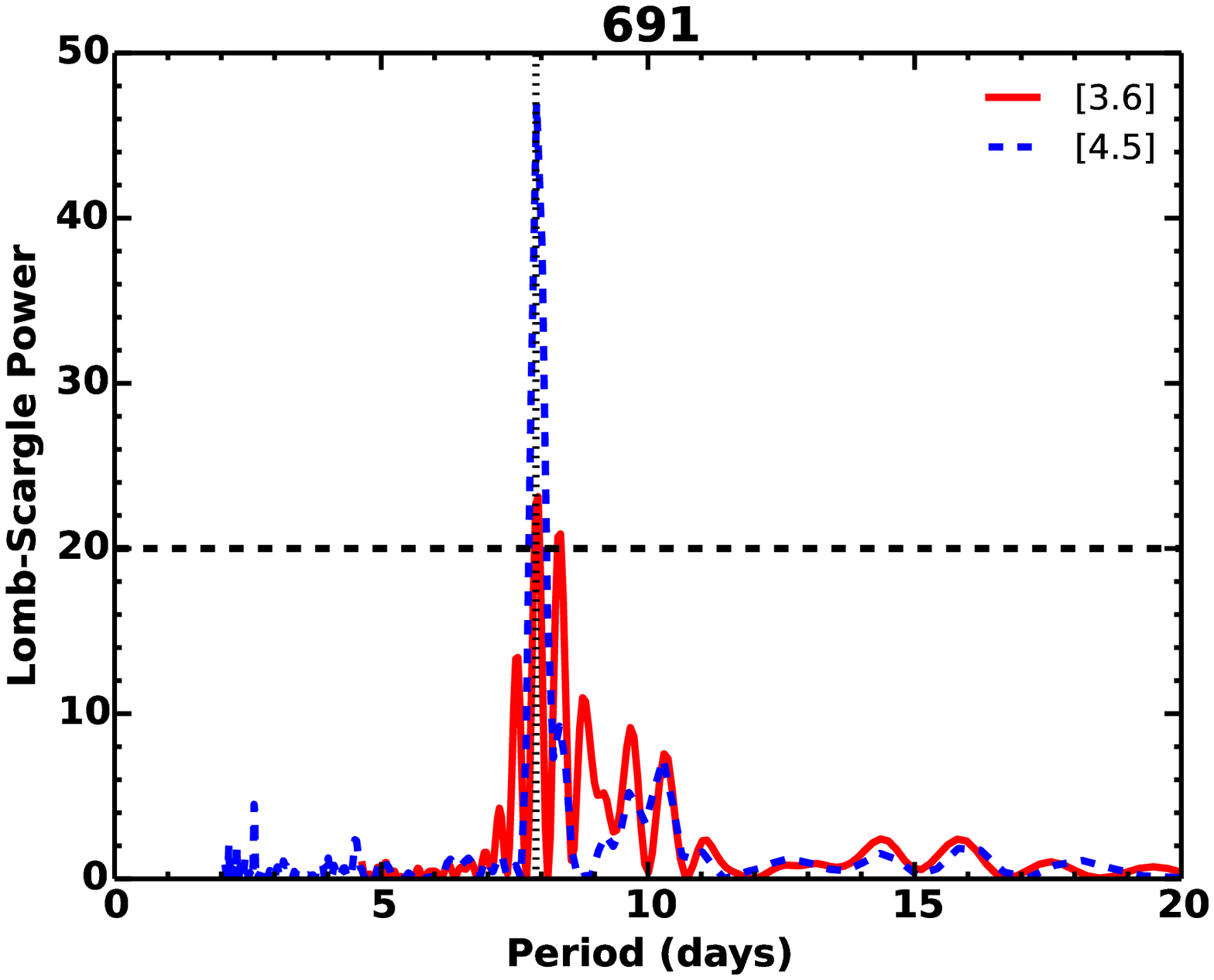}
\caption{Original (left) and phased (middle) light curves, along with the Lomb-Scargle periodogram (right), for the two sources that show rapid (P=7.9 day) periodic behavior. In the periodograms the horizontal line marks the boundary for a significant signal, while the vertical dotted line indicates a period of 7.9 days. The slow decrease in flux in star 330 has been removed to better highlight the rapid periodicity. \label{short_period}}
\end{figure*}

Two of the periodic stars (\# 330 and 691) exhibit short periods (P=7.9 days) with original and phased light curves for these two sources shown in Figure~\ref{short_period}. Both of these sources have M1-M2 spectral types, with 330 displaying a class II SED shape, while 691 has a class I SED shape. To test if the similarity in the periods of these two systems is an artifact of e.g. the time sampling, or is simply a coincidence, we repeat our period search on randomly shuffled versions of the light curves of stars 330 and 691. We do not find any significant (false-alarm probability$<$0.05) peaks at P=7.9 days, or any other period, in these shuffled light curves. Similarly, we examine the light curves of non-variable non-cluster members and again find no significant peaks near 7.9 days. This indicates that the similarity between the periods of these two sources is a coincidence.

While the stellar rotation period has not been directly measured for these two stars, the observed period is similar to what has been measured in young stellar objects of similar ages \citep{reb04}. In the optical, periodic variability is often associated with the rotation of cold spots across the visible stellar surface \citep{her94}, although as discussed above, the size and color of the fluctuations within these objects is not consistent with a spot interpretation. Instead we may be seeing periodic obscuration of the stellar surface by a large warp at a location in the disk close to the co-rotation radius, a scenario that is discussed in more detail in section~\ref{discussion}. 

\begin{figure*}
\center
\includegraphics[scale=.25]{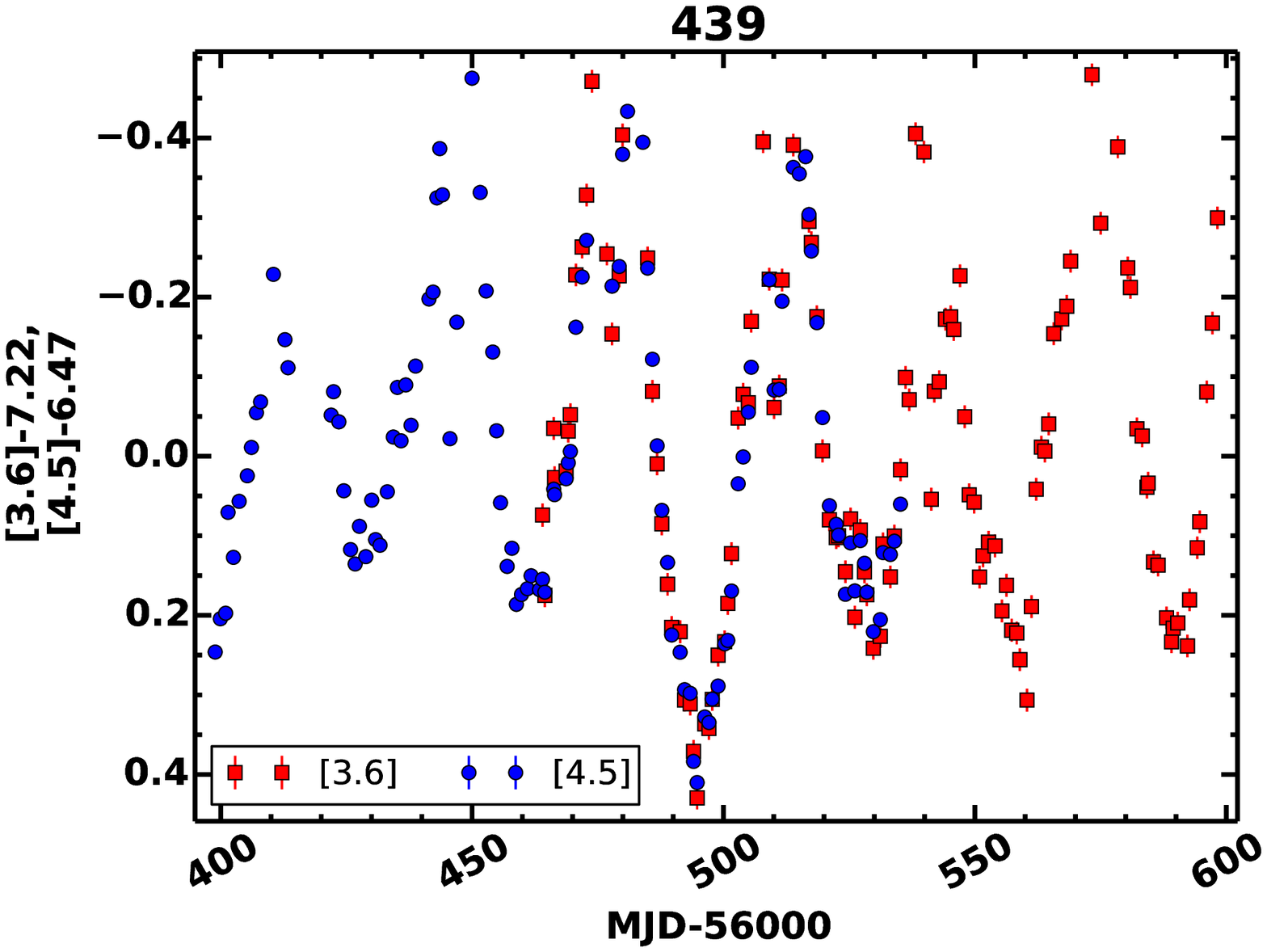}
\includegraphics[scale=.25]{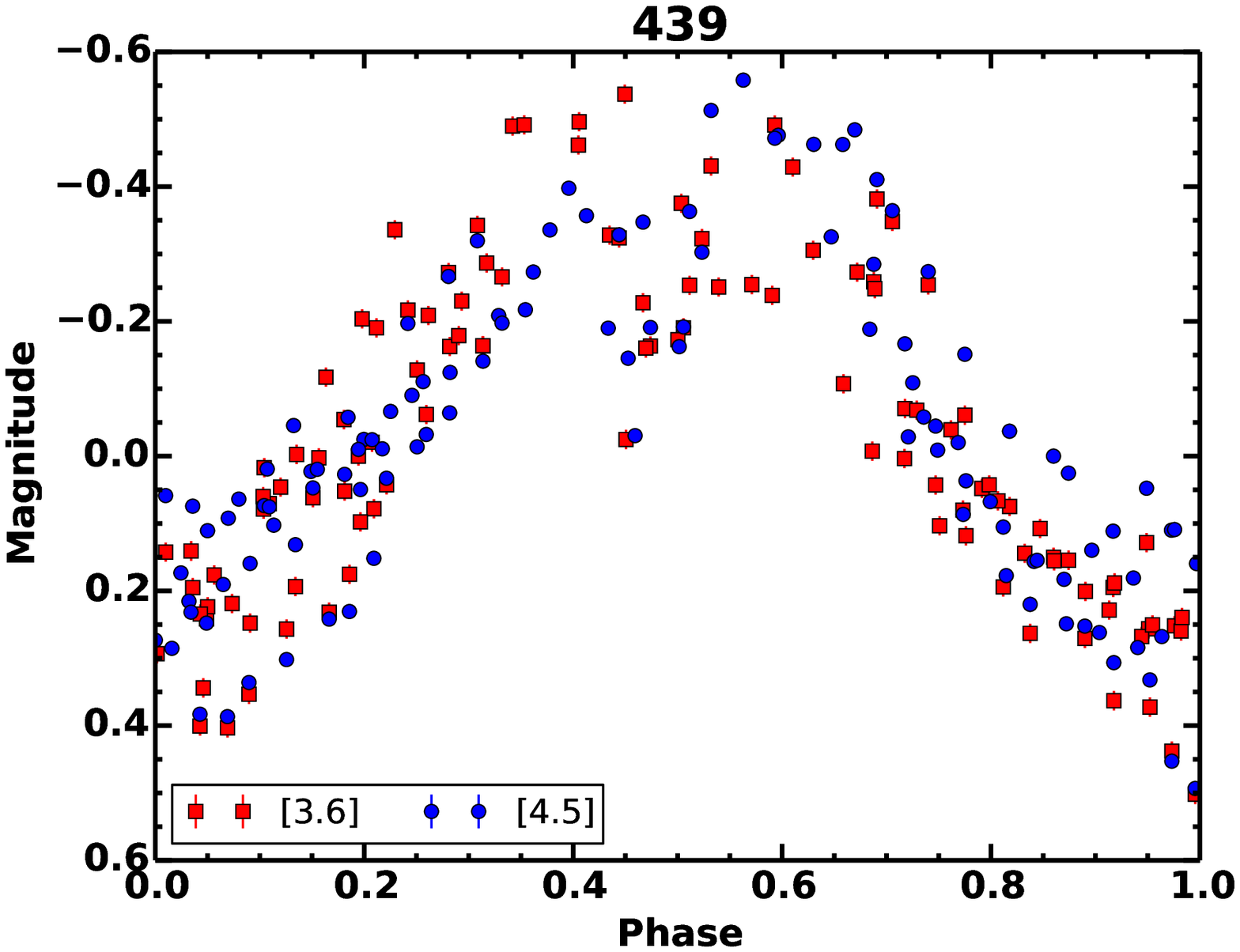}
\includegraphics[scale=.25]{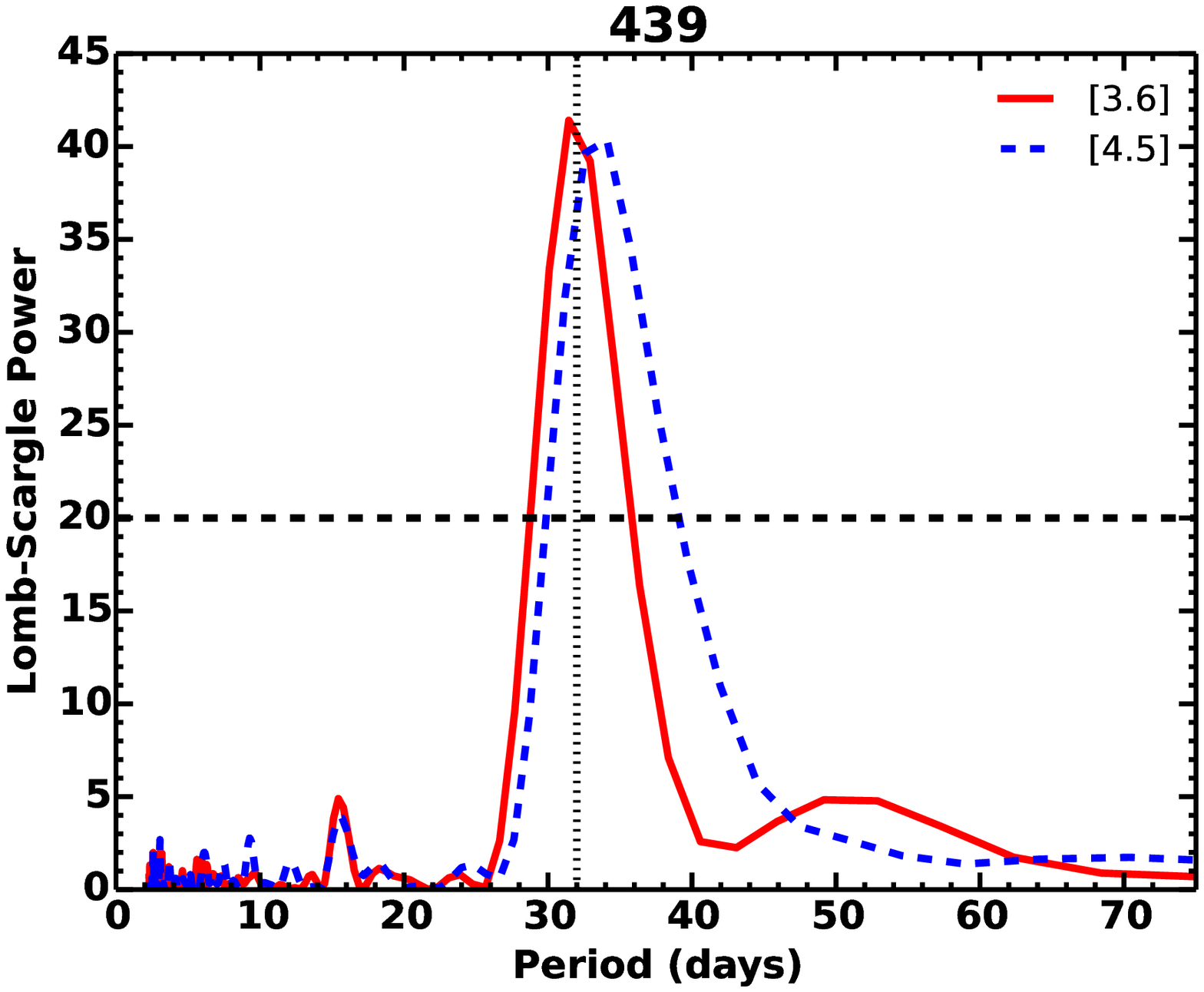}
\includegraphics[scale=.25]{lc530.eps}
\includegraphics[scale=.25]{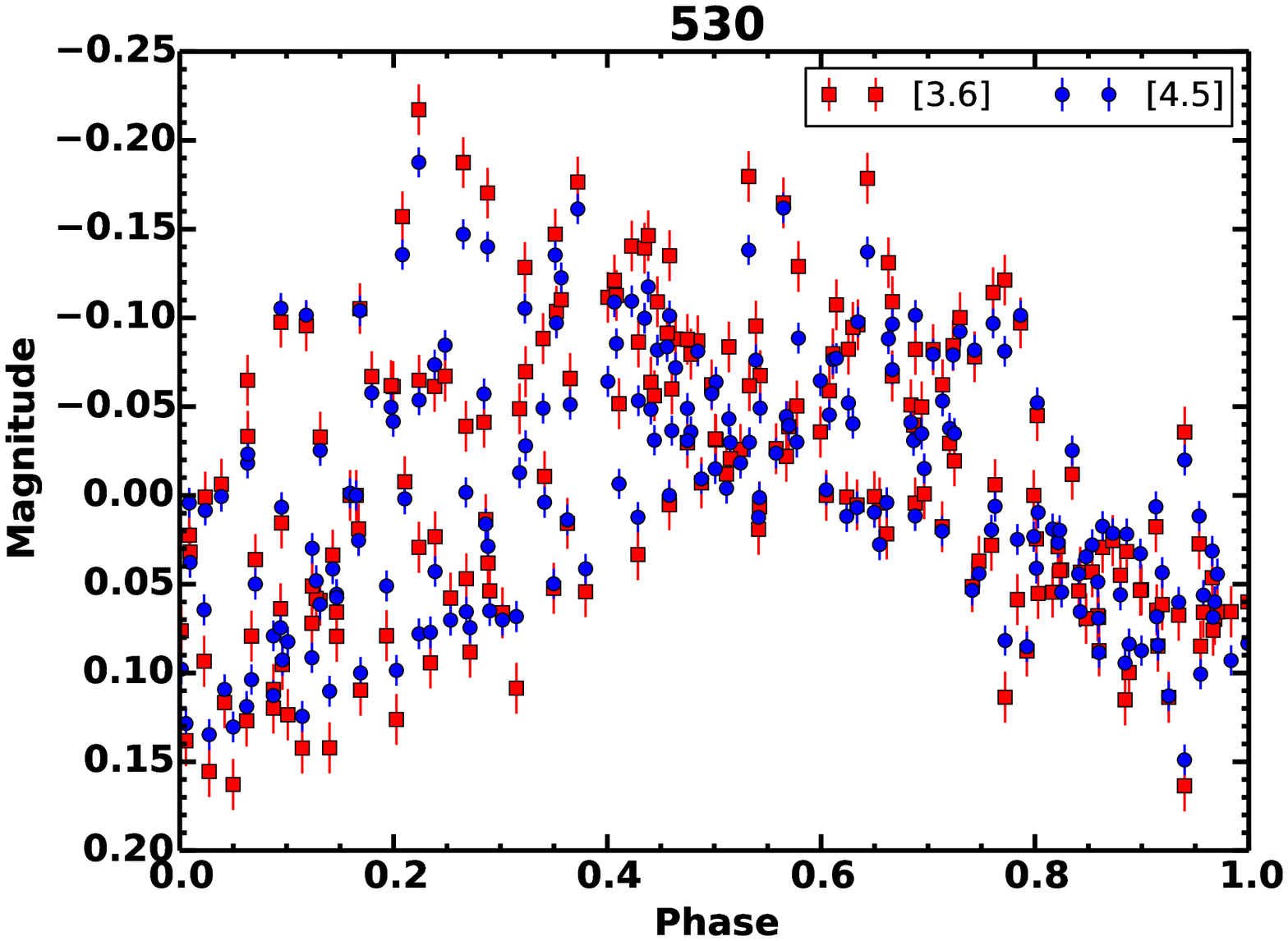}
\includegraphics[scale=.25]{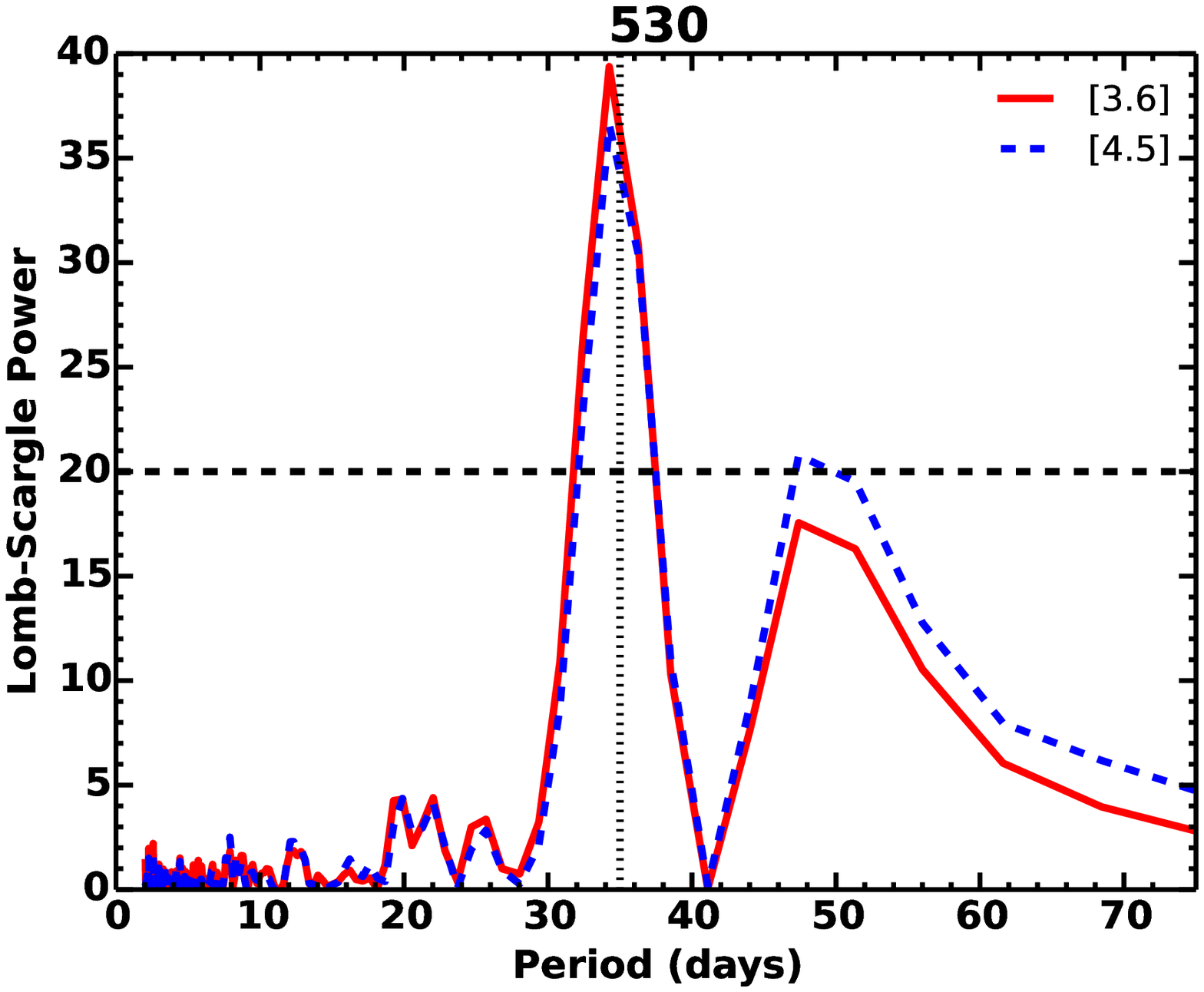}
\includegraphics[scale=.25]{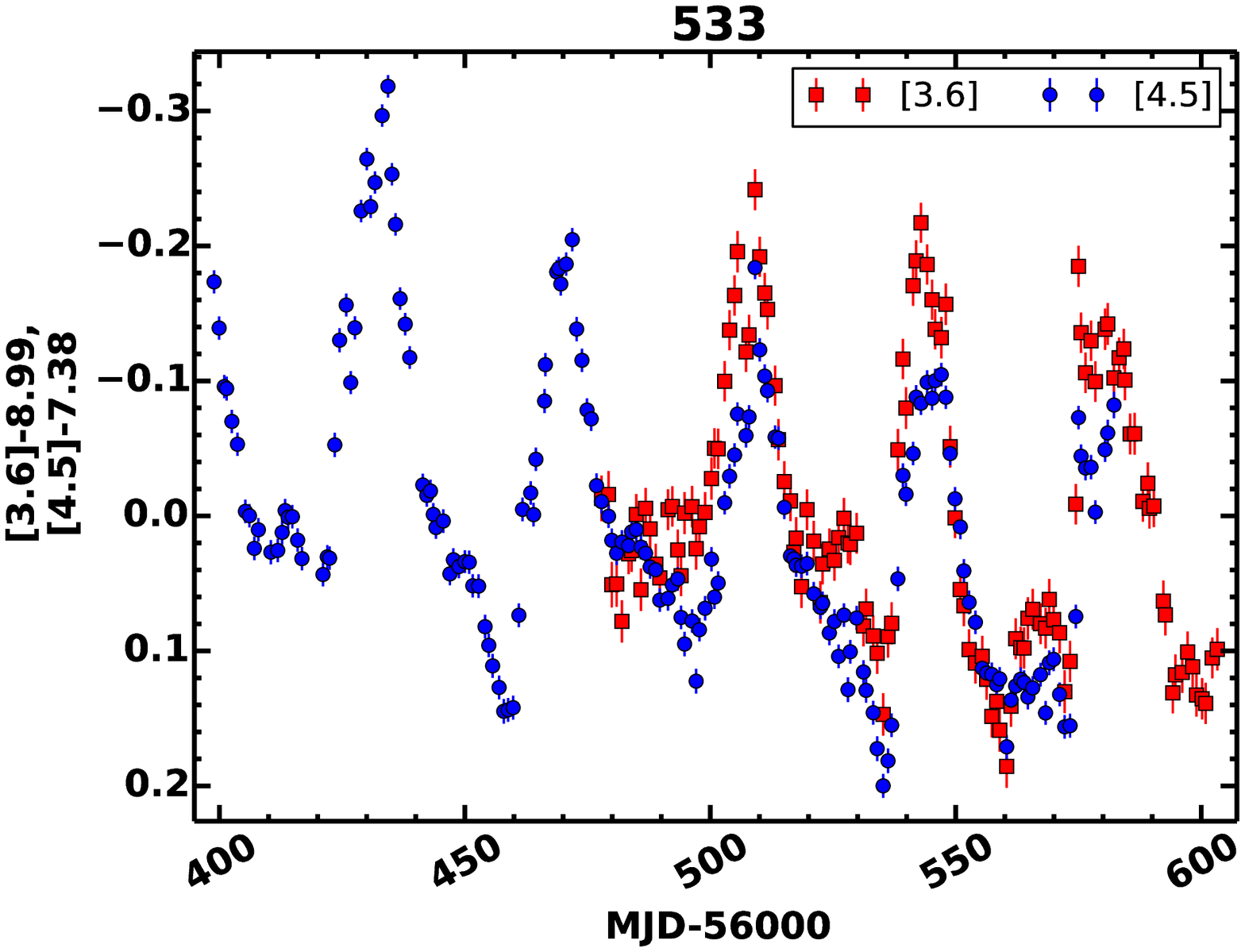}
\includegraphics[scale=.25]{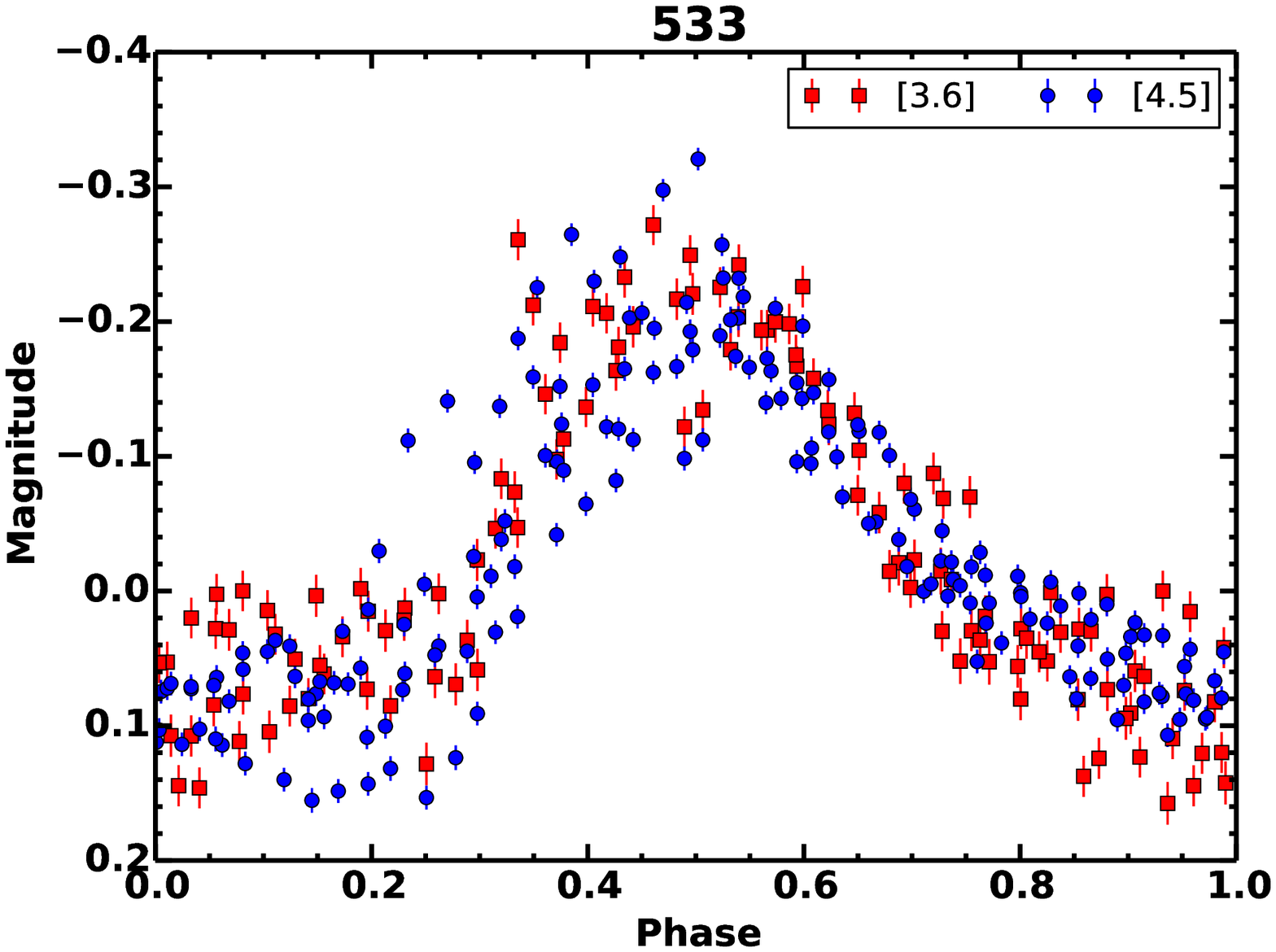}
\includegraphics[scale=.25]{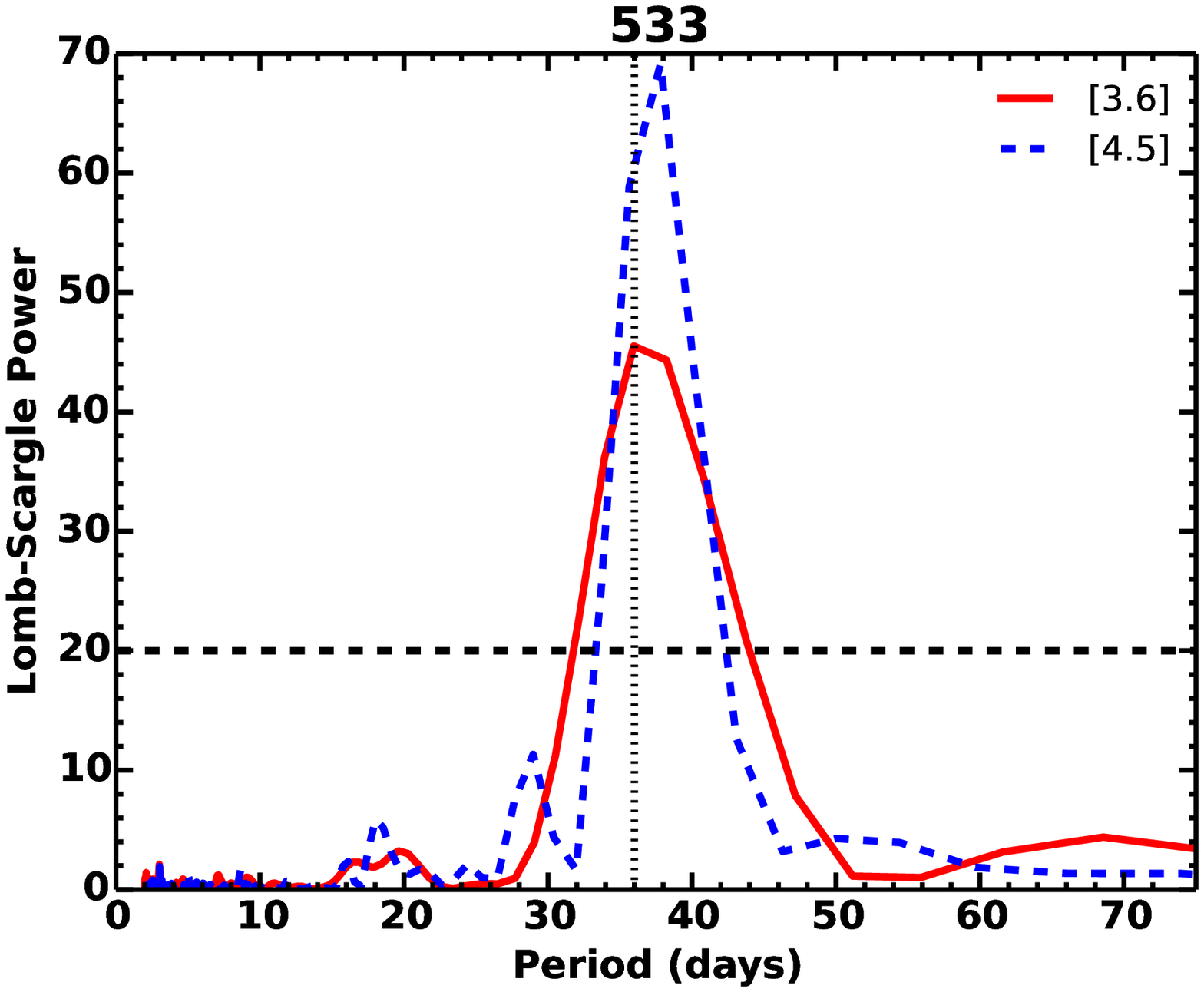}
\caption{Original (left) and phased (middle) light curves, along with the Lomb-Scargle periodogram (right) for the three sources that show slow periodic behavior (P=32, 35, and 36 days from top to bottom). In the periodograms the horizontal line marks the boundary for a significant signal and vertical dotted lines indicate the most significant period. \label{long_period}}
\end{figure*}

The other three periodic sources (\# 439, 530 and 533) have timescales (32, 35 and 36 days respectively) that are much longer than expected from stellar rotation, suggesting that the variability is not associated with motion close to the stellar surface. All three show strong fluctuations ($\Delta[3.6]$=0.53, 0.26, 0.27 respectively) and have a strong infrared excess; 439 and 530 are class II sources, while 533 is a class I source. The small number of sources prevent any strong interpretation, but we note that \citet{reb14} also found an infrared excess to be common among stars with longer periods. Light curves for our three long period sources are shown in Figure~\ref{long_period}. For stars 439 and 533 there is a consistent shape among the different cycles, while the phased light curve of star 530 highlights the fact that the size of the fluctuations, and possibly their period, vary from one cycle to the next. The periods correspond to Keplerian orbits at distances of $\sim$0.2 au. Previous surveys of near-infrared variability have found stars with periods over 100 days \citep{wol13,par14,pla08b}, indicating that long-period behavior is not uncommon among young stellar objects.

\subsubsection{Understanding the Timescale of Aperiodic Variability}
The majority of our cluster members are aperiodic, with fluctuations operating on a range of timescales. Even star 530, which shows significant periodicity, has a long term trend underlying the periodic behavior. The stochastic nature of much of the variability means that many cycles must be observed before the average size of the fluctuations can be accurately measured. Further complicating this analysis is that statistical tools designed to quantify the size of the fluctuations as a function of frequency, such as the structure function \citep{haw02,dev03} or Lomb-Scargle periodogram \citep{sca82}, suffer from biases due to the finite sampling and limited observing window \citep{emm10,fin15}. To accurately quantify the strength of the fluctuations as a function of frequency we utilize recent work in understanding AGN variability as laid out in \citet{kel09} and \citet{kel14}, and contained in the CARMA\_PACK software package\footnote{https://github.com/brandonckelly/carma\_pack}. In short, this code uses a Continuous time AutoRegressive Moving Average (CARMA) stochastic model to fit the lightcurve. The power spectrum is then derived analytically from the parameters of the light curve model, avoiding systematic distortions of the power spectrum shape by the finite sampling of the observed light curves. The Bayesian framework for the model fitting includes an accounting of the uncertainties in the data, which is translated into posterior distributions for the model parameters and statistical uncertainties in the shape of the power spectrum. With statistically robust estimates of the power spectrum that are unbiased by the observing cadence we can begin to characterize the relative strength of the stochastic fluctuations on different timescales and use this information to better understand the source of the variability. 

\begin{figure*}
\center
\includegraphics[scale=.3]{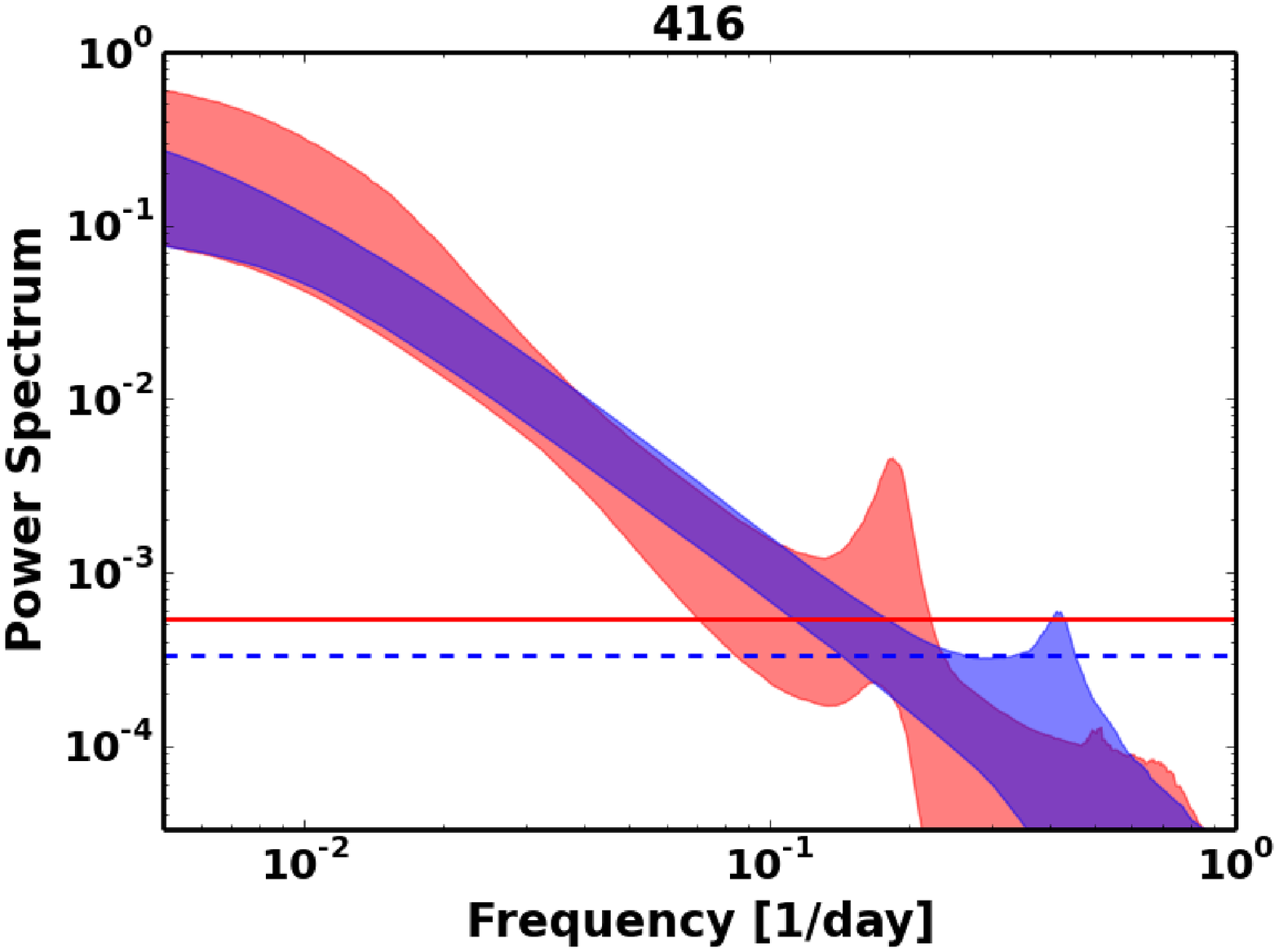}
\includegraphics[scale=.3]{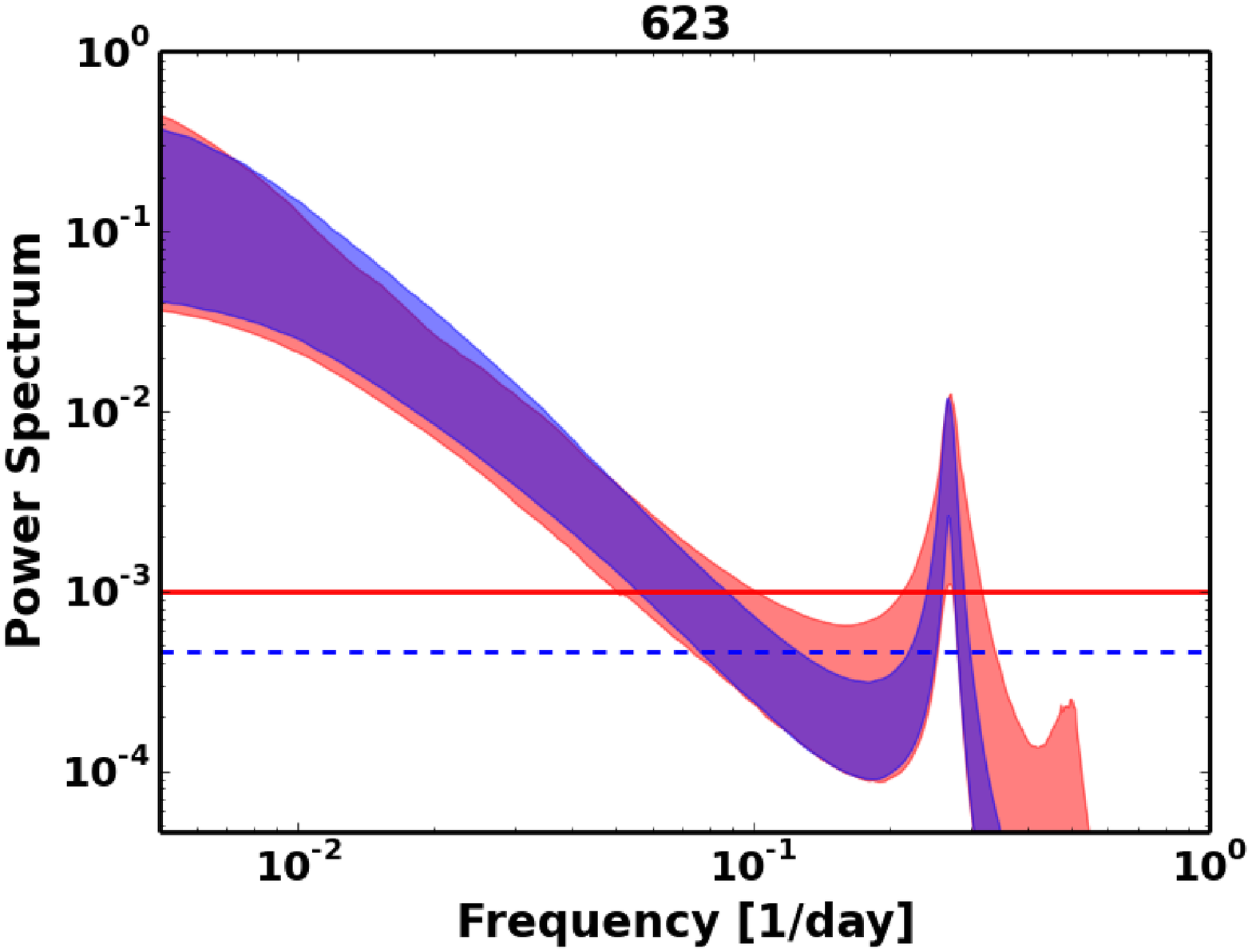}
\includegraphics[scale=.4]{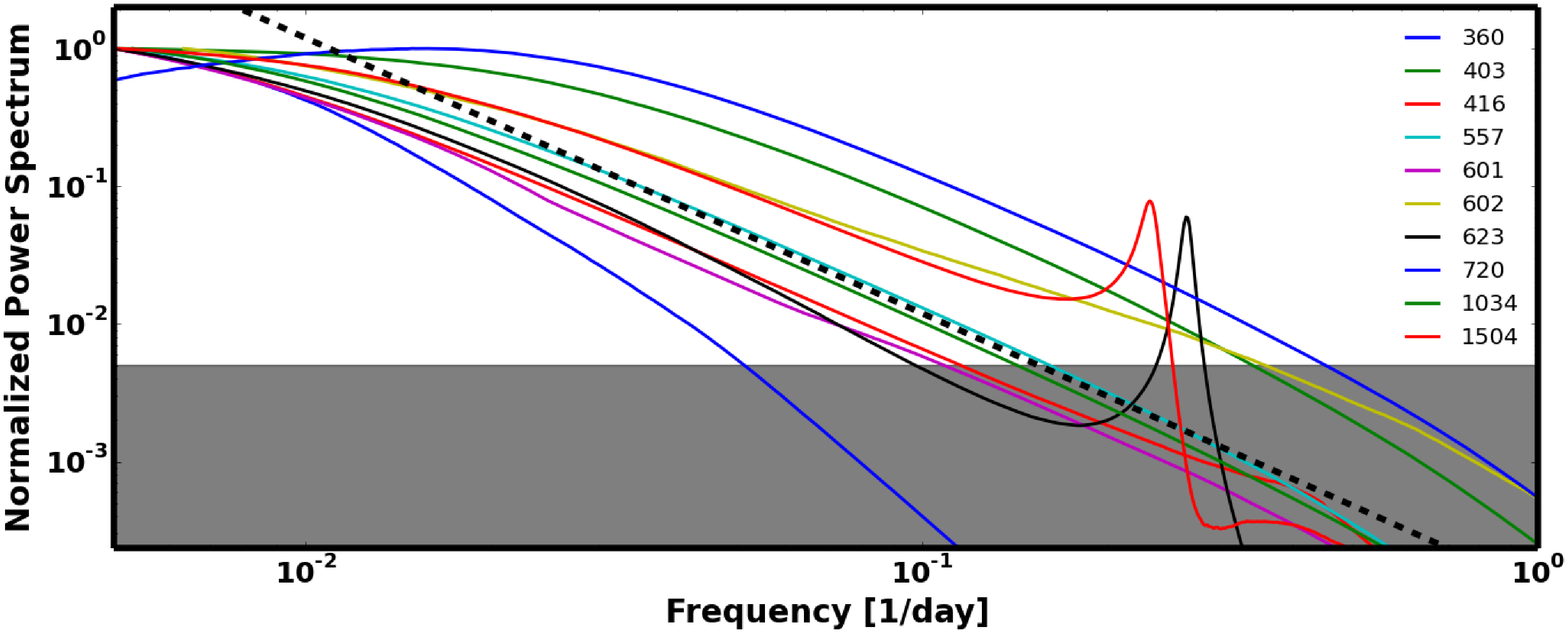}
\caption{(Top) Power spectral density for stars 416 (left) and 623 (right). The red band shows the results derived from the [3.6] lightcurve, while the blue band shows the results from [4.5]. The band represents the 95\%\ confidence interval of the power spectral density. Horizontal lines mark the uncertainties in the derived power spectra for [3.6] (red solid line) and [4.5] (blue dashed line). (Bottom) Power spectral density, normalized to their peak, derived from [4.5] for those stars with aperiodic variabilities and sufficiently sampled light curves. Only the median fit to the light curve is shown for each star. The grey band indicates the region where the uncertainties become substantial and the shape of the power spectra becomes highly uncertain. In general, the power spectra follow a $P(f)\propto f^{-2}$ shape, indicated by the black dashed line. This behavior is likely due to fluctuations arising from a range of radii within the disk. \label{psd}}
\end{figure*}

In Figure~\ref{psd} we highlight power spectral density curves that display some of the common behaviors seen among our sample. The bands in these figures represent the 95\%\ confidence interval on the shape of the power spectrum, while the horizontal lines represent the noise level of the power spectrum. We find that almost every source exhibits an increase in power toward low frequencies that is roughly consistent with $P(f)\propto f^{-2}$. The only source whose power spectrum is not dominated by a power law decline is 691. In this case the lightcurve is deterministic and not intrinsically stochastic, thus violating one of the assumptions of the CARMA model, and we exclude it from our analysis in the context of the CARMA model. The power law shape seen in the power spectral density indicates that the fluctuations are not dominated by day to week long timescales, but instead exhibit behavior covering a large range of timescales. Within each object, as the timescale increases the size of the fluctuations increases. Previous {\it Spitzer} surveys predominately sample frequencies down to 0.03 day$^{-1}$, while our data is able to extend down to 0.005 day$^{-1}$. The 200 day day span of our survey prevents us from determining if the power-law shape extends to lower frequencies, although there is evidence for fluctuations over years, as discussed below. The combination of high cadence and long baseline observations within our survey allows us to derive the smooth connection between daily and months-long infrared fluctuations as seen in the approximately power-law shape of the power spectrum. 

The presence of long-timescale fluctuations is consistent with recent Spitzer and ground-based near-infrared surveys, although with less dense time coverage. \citet{reb14} identify a handful of objects with variations between Spitzer observations separated by 6-7 years. Interestingly they find that these long-term variables are more common among clusters with a higher fraction of Class I protostars, and hence are likely younger. \citet{par14} observe $\rho$ Oph over the course of 2.5 years and find long-term variations in 31\%\ of the sample, with timescales ranging from 64 to 790 days, with 68\%\ of the variable systems showing a long term trend over the entire 2.5 year period. \citet{wol13} in their near-infrared study of Cyg OB7 also find variability on timescales of years, consistent with the previous surveys. In their study of four young clusters, \citet{sch12} find that the frequency of highly variable objects ($\Delta$m$>$0.5) increases with the length of the time window (from days to years). \citet{meg12} find variability to be common in Spitzer observations of Orion that cover a six-month baseline. In near-infrared observations spanning almost 900 days, \citet{ric15} find variability to be common, with fluctuations consistent with accretion/disk variability to have a longer timescale than those associated with variable obscuration. These results indicate that even our 200-day time coverage may not span the entire range of the variability, as discussed in more detail below. 

On top of the power-law behavior, we also occasionally find sharp peaks at high frequencies. This is best exemplified in star 623, where this peak is detected at the $>$3$\sigma$ level, although 1-2$\sigma$ peaks are seen in other sources. These features do show up in sources for which the Lomb-Scargle periodogram indicates periodic behavior (e.g. 330) but, as in the case of star 623 (discussed in more detail in Appendix~\ref{ots44}), this feature is not always picked up in our period search. This may be due to the stochastic nature of the periodic behavior. The Lomb-Scargle periodogram assumes that the amplitude and frequency of any periodic behavior is constant throughout the light-curve, while the CARMA model is designed for more stochastic behavior. Additional data at a higher observing cadence will help to confirm many of these features.

\subsubsection{Very Long Timescale variability}
While our survey only covered 200 days, we can use these data in combination with previous infrared surveys to help characterize the very-long term variability. Previous studies have found year-to-year fluctuations \citep{reb14,par14,wol15} suggesting that infrared variability is not limited to the timescales studied here. {\it Spitzer} was used to survey the Chameleon star-forming region in 2004 and 2005 \citep{luh08a}, covering our field of view, at a similar depth, allowing a direct comparison of measured photometry. There are 16 sources within our sample for which either the [3.6] or [4.5] flux differs significantly between the previous measurements and our survey (Table~\ref{other_phot}), suggesting fluctuations on decade long timescales. The prior photometry differs by 0.1-0.5 magnitudes from the average of our observations, which corresponds to fluctuations 3-5 times larger than we observe. These larger fluctuations on longer timescales are roughly consistent with an extension of the $P(f)\propto f^{-2}$ seen in our derived power spectral density functions, which would predict fluctuations $\sim$4 times larger on a nine year timescale than on a 200 day timescale, although more work is needed to confirm this behavior.

Year-long fluctuations can also be probed by comparing data taken by the Wide-field Infrared Survey Explorer (WISE) \citep{wri10} in the cryo and post-cryo \citep[NEOWISE:][]{mai11} campaigns. WISE surveyed the entire sky in four infrared bands, of which the shortest two have central wavelengths of 3.4$\micron$ and 4.6$\micron$, similar to that of {\it Spitzer}. During its cryogenic campaign WISE observed Cha I in February and August 2010, while the post-cryo NEOWISE campaign observed Cha I over four blocks in February and August of 2014 and 2015, providing a 5 year total baseline. Of the 19 sources in our sample detected in both bands in both campaigns, two show a substantial change in flux between the WISE and NEOWISE campaigns, of a few tenths of a magnitude. The smaller number of sources with long-term fluctuations detected by WISE as opposed to with {\it Spitzer} is likely not due to sensitivity, both the WISE and {\it Spitzer} photometry have uncertainties of 0.01-0.02 magnitudes, but because the NEOWISE data covers a year-long span, and likely already captures much of the slow fluctuations. 

Another form of long-term variability is when the size of the fluctuations change over time. WISE can help to address this since it observed Cha I during its cryogenic phase in two short blocks, one covering Feb 14-16, 2010 and the other covering Aug 12-17, 2010, with a total time coverage similar to our survey. By looking for sources that are variable in the WISE data but not in our survey, or vice-versa, we can begin to characterize the frequency of large changes in the amplitude of infrared variability. Of the 11 non-variable sources in our sample, one (\# 2712) shows significant variability ($\sim0.1$ mag) in the WISE data that if still ongoing would have been detectable in our survey. The majority of the Spitzer variable sources are also variable in the WISE data, with the exception of 602, 523, 1034 and 1504. While the fluctuations in 602 are small ($\sim$0.05) and may not have been detectable with WISE, the fluctuations in the other three stars are $\sim$0.1 mag and should have been detectable with WISE. Similarly, comparing the WISE and NEOWISE observations there are seven sources that show evidence of variability in NEOWISE, but are not detected as variable in the cryogenic WISE campaign. This suggests that some disks do change the strength of their variability on timescales of years. As with the slow modulations seen between the Spitzer epochs, this indicates that there are further long-term changes in circumstellar disks that are not fully probed by our survey. Whether or not these long-term fluctuations are physically related to the months-long variability, or a sign of distinct processes, is discussed in more detail below. 

\subsubsection{Color Variations with Time}
So far in analyzing the timescale of the fluctuations we have focused on changes in the [3.6] and [4.5] individually, but as noted in Section~\ref{color} the [3.6]-[4.5] color is not constant in all of our targets. The behavior of the color with time may also probe the physical cause of the variability (ie. stable spots cause periodic variations in both flux and color). 

Inspecting the light curves of the stars with significant color variations (Fig~\ref{lc_redder},\ref{lc_bluer}) reveals additional trends. In star 1504 there are distinct reddening events at the moments when the [3.6] and [4.5] flux drops, while in star 533 the color shows the same oscillations as the total flux, with the bluest color at the beginning of the outburst. Even among the stars that show no overall change in color, there is structure in their color lightcurve (Figure~\ref{lc_nocolor}). 

\begin{figure*}
\center
\includegraphics[scale=.3]{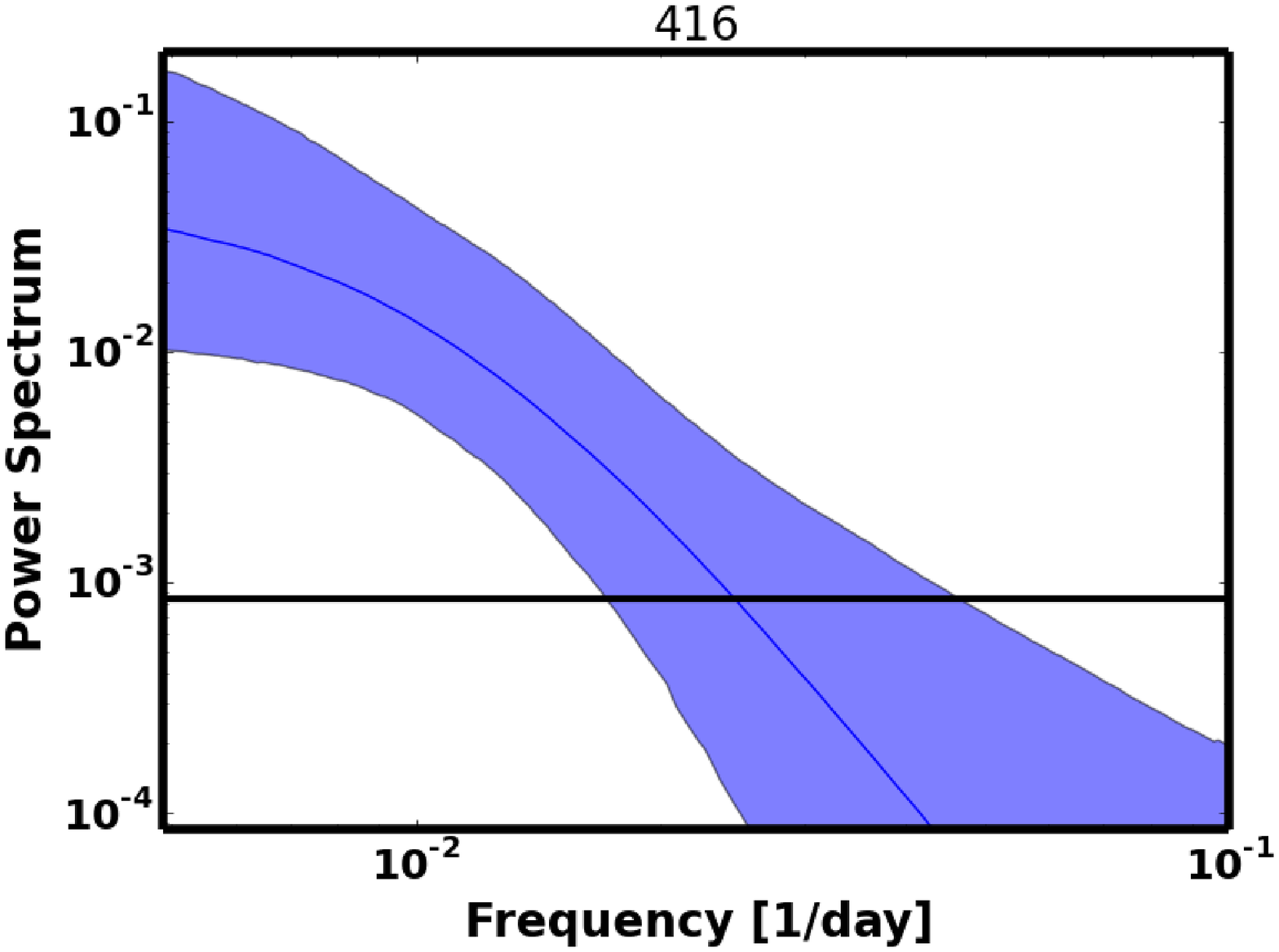}
\includegraphics[scale=.3]{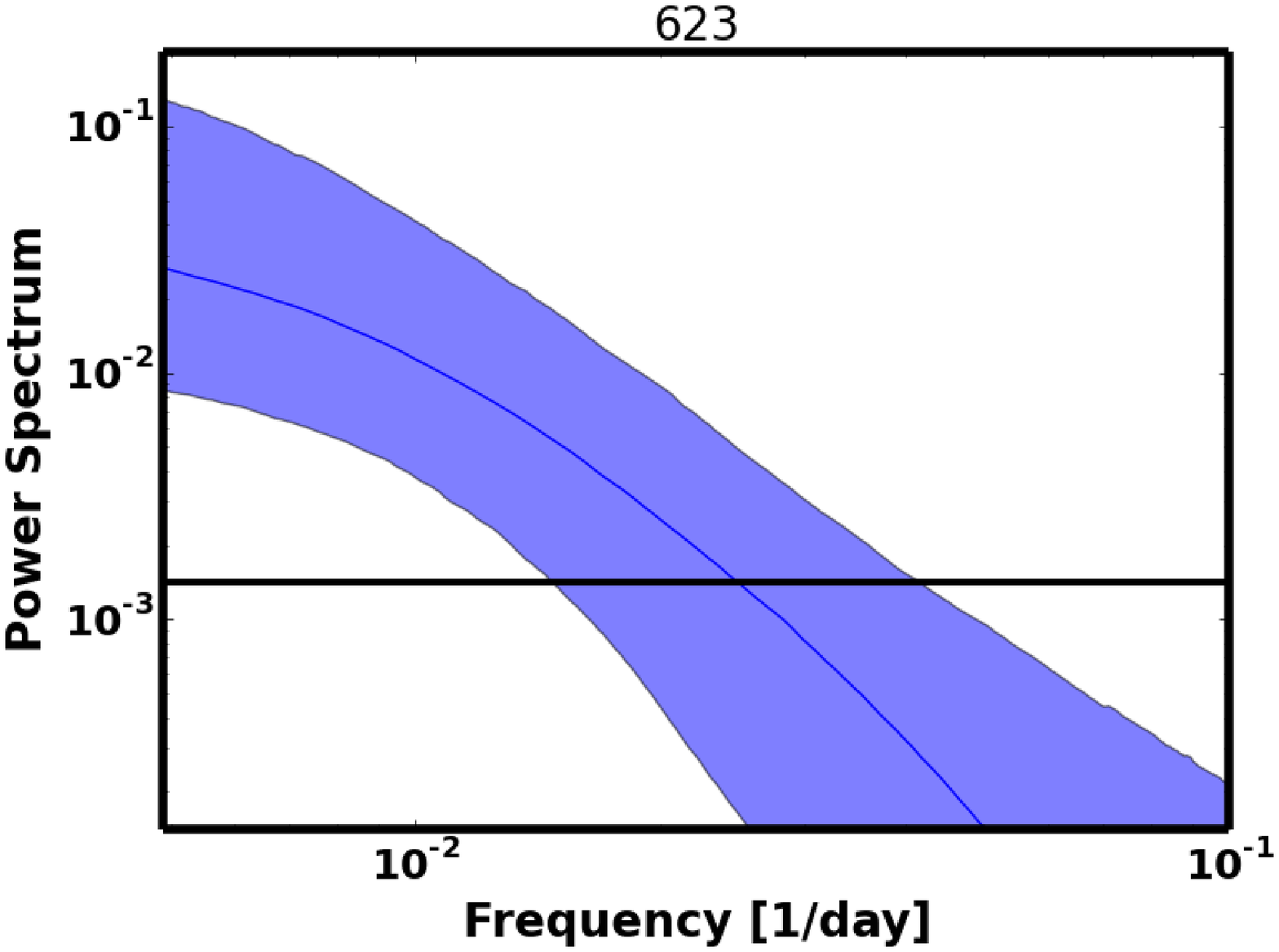}
\includegraphics[scale=.4]{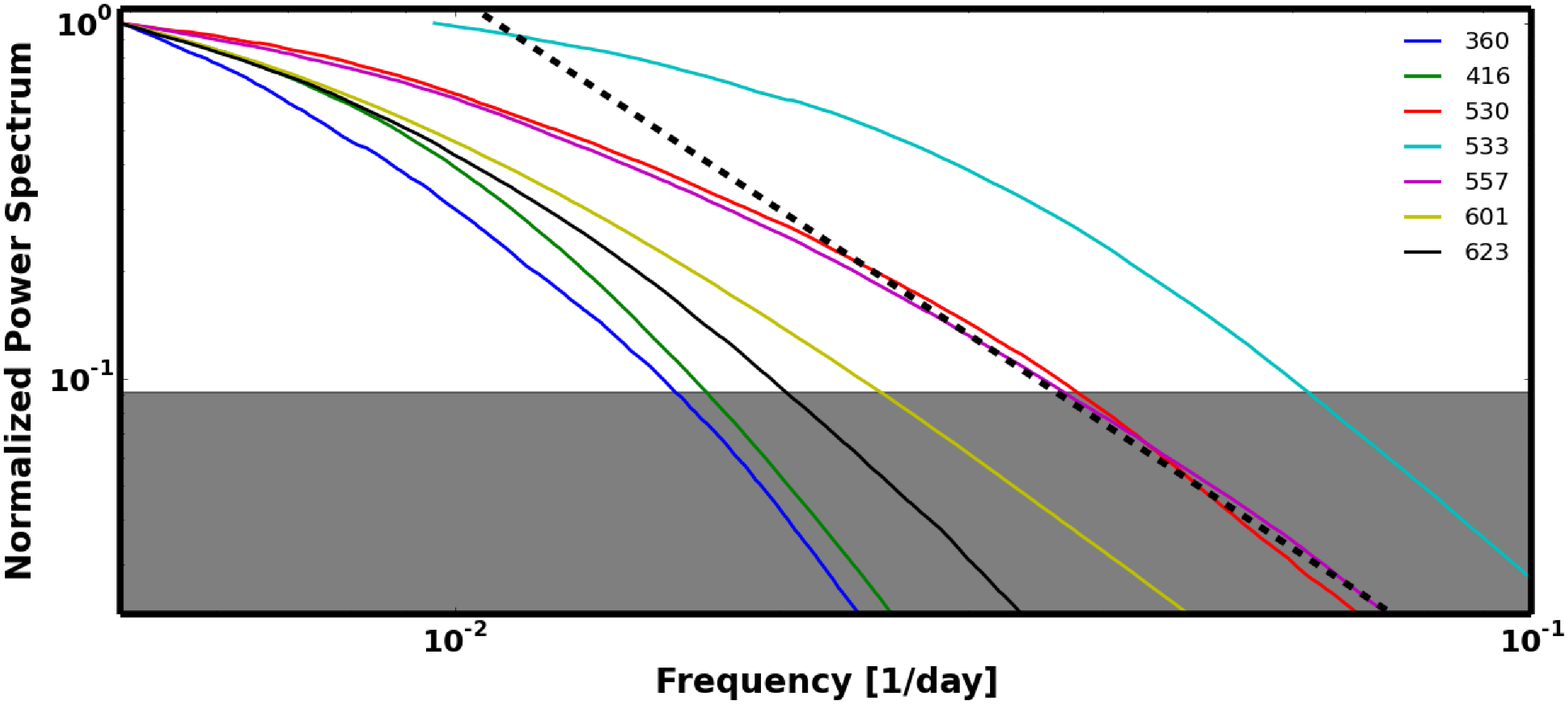}
\caption{(Top) Power spectral density derived from the [3.6]-[4.5] color for stars 416 (left) and 623 (right). As compared to the [3.6] and [4.5] fluctuations, the variations in [3.6]-[4.5] are smaller with noise limiting our ability to probe fluctuations more rapid that $\sim$50 days. (Bottom) Power spectral density, normalized to their peak, derived from [3.6]-[4.5]. Only the median fit to the light curve is shown for each star. The grey band indicates the region where the uncertainties become substantial and the shape of the power spectra becomes highly uncertain. A $P(f)\propto f^{-2}$ shape is indicated by the black dashed line. As with the flux, the color exhibits a roughly power law shape to its power spectrum, although the shape of the high frequency end is unknown. \label{psd_color}}
\end{figure*}

Applying the CARMA\_PACK tool to the [3.6]-[4.5] light curves, we find that the power spectral density has a roughly power law behavior (Figure~\ref{psd_color}), similar to the [3.6] and [4.5] magnitudes. Only star 330 has color fluctuations consistent with the noise at all timescales. Significant fluctuations are only detected on timescales $\gtrsim$50 days due to the small size of the variability relative to the noise, making it difficult to search for the narrow high frequency features seen in the [3.6] and [4.5] power spectral densities. Despite these limitations it appears that the change in color increases with timescale. This is seen even among sources that do not have a significant average change in color (stars 360, 416, 530, 601, and 623). This can occur if the direction of the change in color, which is not captured by the CARMA model, varies with timescale (e.g. extinction dominates on short timescales while disk changes dominate on longer timescales).

The derived power spectral densities only apply to timescales up to 200 days, but we can constrain slower effects by comparing our measured color with that of \citet{luh08a}. We find that eleven stars (100, 330, 530, 533, 602, 623, 691, 966, 1103, 2712, and 4342) show a significant variation of color (0.1-0.6 mag) over the $\sim$decade between these observations. Half of these very long timescale variations are consistent with disk fluctuations while half are consistent with extinction. In none of these sources does the color change enough to affect the SED classification. Of these eleven sources, four (530, 533, 602, and 623) show an increasing color power spectrum in our data, one (330) has a color power spectrum below the noise, while for the remaining six (100, 691, 966, 1103, 2712, and 4342) we do not have enough data to constrain the color variations in our survey. The presence of color variations over such long timescales is consistent with increased color fluctuations with time, as was seen with the power spectral density curves.

\section{Discussion\label{discussion}}
We find that long-timescale (20-200 day) variability is common among young stellar objects and the color changes observed in our sample are consistent with a mix of extinction and disk emission fluctuations. This behavior is an extension of the short-timescale variability seen in previous surveys, and may continue to even longer timescales. Here we consider a handful of possible models for infrared variability that can be constrained by our observations. 

{\bf Star Spots:} Optical variability is often associated with the rotation of hot or cold spots across the stellar surface \citep{her94}. Spots can be seen in the mid-infrared in systems with very little dust emission \citep{pop15}, as well as in the near-infrared \citep{car01,par14,wol15} where the contribution to the observed flux from the dust is weaker and the stellar flux is stronger. Since the spots are affixed to the stellar surface, they generate periodic signatures in the light curve at the stellar rotation period. This periodicity is more often seen with cold spots, which tend to be long-lived, than with hot spots, which tend to vary substantially in size from one period to the next \citep{her94}. We do find two sources with stable periods consistent with stellar rotation (P=7.9 days for stars 330 and 691), but the size of the observed fluctuations ($\Delta$[3.6]=0.19,0.44) is much larger than expected for spots ($\Delta$[3.6]$<$0.15). Also, both sources have a strong infrared excess, 330 is a class II source while 691 is a class I source, making the spot explanation less likely. 

Spots are more likely an explanation for those sources with little to no infrared excess. While class I and II sources are the main focus of our survey, there are two variable class III sources within our sample (2712 and 5073). In both cases the fluctuations are small ($\Delta$mag$\sim$0.03-0.06) and consistent with that expected from cold spots. There is no strong evidence for periodicity in the light curves, although the fluctuations are at the edge of our detection limit making any periodic signal difficult to detect. 

{\bf Perturbation from a Companion:} Some known tight binaries show strong variability due to changes in the accretion flow associated with the eccentric motion of the companion \citep{bas97,mat97,jen07}. Wider binaries can experience periodic dimming as components of the binary pass behind a circumbinary disk, as is the case in KH 15D \citep[e.g.][]{win14,aru16}. The 131 day period WL 4 system \citep{pla08b} and the 93 day period YLW 16A system \citep{pla13} show similar dimming behavior, suggesting a similar origin for their variability. A companion misaligned with the disk can also drag material out of the midplane \citep{fra10}, leading to periodic occultations of the primary star. Periodic variability has been used to infer the presence of a tight binary in sources in which the central objects cannot be directly observed \citep{muz13}. The motion of the binary can result in stable periodic fluctuations on timescales extending from days to thousands of years, depending on the separation of the two stars. Previous near-infrared surveys have found stars with up to 90 day periodic variations \citep{wol13,par14} and binarity is known to be common among young stellar objects \citep[e.g.][]{ngu12}. While the majority of the observed sources in Cha I exhibit aperiodic behavior, a small handful (\# 439, 530, 533) are periodic with timescales ($\sim$30 days) that are longer than typical stellar rotation periods, and are consistent with a companion, such as a massive planet or low-mass star, at a separation of $\sim$0.2 au. 

Star 533 exhibits outburst behavior reminiscent of that seen in LRLL 54361, which is characterized by $\sim$3 mag outbursts every 25 days \citep{muz13}. More recently, \citet{hod15} discovered another outbursting class 0/I object in Orion with similar properties ($\Delta$K=2 mag, P=30 days). Such behavior can be explained by material being dragged inward from the circumbinary disk during apastron, which eventually causes a large outburst in accretion luminosity as it falls onto the stars during periastron. Star 533 had been noted as a possible FUOr outbursting star by \citet{gra14}, with K-band outbursts of 2 magnitudes. While the period of star 533 in our data (P=37 days) is slightly longer and the size of the fluctuations ($\sim$0.4 mag) are smaller than these other objects, all three are deeply embedded class I objects with stable periods. Star 533 also gets bluer as its becomes brighter (Figure~\ref{lc_redder}), similar to the behavior seen in LRLL 54361, which is consistent with an enhanced contribution from high-temperature accretion luminosity during outbursts. In this way, 533's variability is consistent with the presence of an unseen, tight, eccentric binary driving periodic accretion bursts. 

 The source of the periodicity in stars 439 and 530 is less clear, but may still be connected with a binary. Star 439 exhibits a strongly sinusoidal pattern with a period of 33 days, and is one of the strongest variables in the sample, with $\Delta[3.6],[4.5]$ = 0.53,0.54. Phase folding the light curves reveals a dip in flux near its peak of $\sim$0.2 mag (Figure~\ref{long_period}), possibly associated with obscuration of the central source. \citet{luh08a} observe significant variability at 24$\micron$ with MIPS, with the [24] magnitude varying from 1.46$\pm$0.04 in 2004 to 1.80$\pm$0.04 in 2005. It is also has a high accretion rate; at $\dot{M}=4.7\times10^{-8}$M$_{\odot}$ yr$^{-1}$ its accretion luminosity is $\sim$50\%\ of the stellar luminosity \citep{ant11}. Star 530 exhibits an irregular sinusoidal pattern on top of a longer term trend, with modest variations in flux ($\Delta$[3.6],[4.5]=0.26,0.26). It shows no significant average change in color, although there is evidence for small fluctuations in color that follow the sinusoidal pattern (Figure~\ref{lc_nocolor}) and an increase in color fluctuations toward long timescales (Figure~\ref{psd_color}). This source was observed to be variable in WISE observations in 2010 (Table~\ref{other_phot}) suggesting that the fluctuations observed here are not a recent phenomenon. The phased light curve highlights the fact that the size of the fluctuations, and possibly their period, vary from one cycle to the next. While no other significant periods are detected consistently in both bands, they may still exist at a weak level. 

While the presence of an unseen binary is difficult to constrain from the infrared variability alone, if all three of these systems were binaries it would be consistent with the expected fraction of tight binaries. \citet{ngu12}, in their spectroscopic binary survey of Cha I with sensitivity to $\sim$0.1 M$_{\odot}$ companions and periods less than 60 days, find that 7\% of cluster members in Cha I are tight binaries, similar to surveys of other clusters \citep{tob09,rei14,kou16}. Further work is needed to confirm the binary nature of these three sources and its role in driving the infrared variability.

{\bf Heating by variable accretion:} Gas flows inward through the disk, and along stellar magnetic fields lines where it free falls onto the stellar surface. This is a highly variable process, with emission line measurements finding variations up to a factor of three on timescales of days \citep{ngu09,cos14}. Recent analysis of well-sampled optical light curves of young stellar objects in the NGC 2264 cluster find that some of the optical variability can be explained by a mix of strong \citep{sta14} and weak \citep{sta16} accretion bursts. Shocks created as gas strikes the stellar surface are associated with high temperature emission and while this emission does not directly contribute much to the mid-infrared flux, it may heat the dust in the inner disk leading to potentially observable infrared fluctuations. Large, orders of magnitude, outbursts in accretion certainly lead to variable infrared emission \citep{aud14}, and may apply in some cases here (e.g. star 533), but in this section we focus on the more modest fluctuations that occur more frequently among young stellar objects.

To affect the heating of the inner disk, the variable accretion luminosity must be substantial compared to the more stable stellar luminosity. Among the sources within our sample with measured accretion rates \citep{har98,ant11,rob12,fra15,man16}, typical L$_{\rm acc}$/L$_*\sim$0.1-0.4, consistent with surveys of other low-mass star-forming regions \citep[e.g.][]{ant14}. Some variation is seen in the accretion rate as measured by different surveys, but part of this may be due to differences in methodologies for deriving the accretion rate rather than real fluctuations. \citet{nag15} use detailed radiative transfer models of the inner disk and find that starting with such modest accretion luminosities, they can reach $\Delta$[3.6]$\sim$0.1-0.2 with changes in accretion rate of 2-3 orders of magnitude. The required accretion rate variations are much larger than typically observed except in the most extreme cases, suggesting that variable heating is unlikely unless an asymmetric feature within the disk, such as a warp lined up with the accretion shock \citep{mor11}, amplifies the effect.

Accretion variability also appears to operate on days to week long timescales \citep{cos14,sta16}, corresponding to the dynamical timescale at such small distances from the star \citep{rom08}. This suggests that variable accretion heating of the inner disk is most viable in systems with not only high accretion luminosities, but also variability timescales on the order of days. Star 439 has the highest L$_{\rm acc}$/L$_{*}$, but its stable period is inconsistent with the generally stochastic nature of variability associated with accretion variations \citep{her94,sta14,sta16}. As with star spots, variations in accretion heating of the disk may still play a role, but it is unlikely to be responsible for the majority of our observed infrared variability. 

{\bf Interactions between the stellar magnetosphere and the disk:} The innermost extent of the gas disk is set by the stellar magnetic field, while the dust boundary is defined by where the temperature is high enough to sublimate the dust. Typically dust is truncated farther from the star than the gas, but if dust extends inward far enough, then any instabilities in the disk/magnetosphere interface could lead to infrared fluctuations. Evidence for complex structure in this region has been revealed by optical variability studies, which are sensitive to obscuration of the central star by these features. AA Tau experiences periodic occultations of the central star by a warp that is generated as the gas and dust flows onto a stellar magnetic field that is misaligned with the disk rotation axis \citep{bou03}. Similar occultation events, also called 'dippers' or 'AA Tau-like', have been seen in NGC 2264 \citep{ale10} and other clusters \citep[e.g.][]{ans16}, with rapid extinction events associated with material within the accretion flow \citep{sta14} and longer, quasi-periodic events associated with warps in the disk \citep{mcg15}. Inner warps can also be generated by massive companions mis-aligned with the disk, as discussed earlier, although such a model has difficulty explaining the quasi-periodic nature of many of the dippers.

These structural effects can lead to changes in the infrared emission either through the increase in emitting area associated with a variable warp, or through the obscuration of the stellar flux. The observed changes in [3.6]-[4.5] color fall between these two models for much of our sample. Signs of this interaction can be also searched for in the timescale of the variability. The stellar magnetic field intersects the disk at the point where the disk co-rotates with the star, which implies that the timescale for these interactions is the stellar rotation period. While the rotation period has not been measured for much of our sample, young stellar objects typically display periods of 1-15 days \citep{reb04,reb06,cie06,cod10}, corresponding to corotation radii a few times the stellar radius. As a result we expect disk/magnetosphere interactions to lead to variability that peaks on timescales of days. The exact range of timescales over which this process operates depends on the range of radii over which the disk interacts with the stellar magnetic field; a narrow (wide) range in timescales would arise if this interaction occurs over a narrow (wide) range of radii. While some of the warps can be long-lived, many are highly stochastic \citep{mcg15} leading to variations that are quasi-periodic.

This model is a promising explanation for the days to week long variability seen in some of our sources. As discussed earlier, stars 330 and 691 haver periods consistent with stellar rotation, but amplitudes inconsistent with star spots. Extinction can produce arbitrarily large fluctuations, making the amplitudes seen in these sources consistent with periodic obscuration by a stable warp. Star 691 also has significant color variations that are consistent with reddening, supporting this scenario. While the full light curve of star 330 does not show any significant color change, by subtracting off a 50-day median smoothed version of its light curve we better isolate the short timescale variability. Within this median-smoothed light curve we find a significant change in color ($\Delta$[4.5]/$\Delta$[3.6]=0.80$\pm$0.05) that is consistent with reddening. In addition, star 623 has a sharp peak in its power spectrum at 3.7 days and when phased at this period (Figure~\ref{star623}), the light curve exhibits a dimming feature that could be a sign of obscuration. The quasi-periodic nature of this feature may be a result of instabilities in the size of the warp. Star 1504 exhibits rapid dimming events separated by $\sim$20 days during which its color becomes redder (Figure~\ref{lc_redder}). All four of these sources are consistent with quasi-periodic extinction by a disk warp close to the co-rotation radius.

Variations in disk/magnetosphere structure and orientation from one star to the next will influence the observed strength of this processes, and may explain why disk/magnetosphere interactions are not seen more frequently in our sample. In particular, this process requires dust very close to the star. Dust sublimation radii are $\sim$0.1 au, while typical stellar rotation periods correspond to corotation radii of $\sim$0.05 au, indicating that for most stars dust/magnetosphere interactions will not produce observable effects in the infrared. Large dust grains, or a low stellar luminosity, could allow the dust to encroach on the star without sublimating, making this scenario more likely. This model also relies on a strong dipole magnetic field to create a warp, which may not be the case in all pre-main sequence stars \citep{don09} and may be less common among high mass stars \citep{wad07}. \citet{ans16}, using Kepler observations of the Upper Scorpius and $\rho$ Ophiucus star forming regions, primarily find 'dippers' among low mass stars. While our sample is not large enough to search for a mass dependence, we note that stars 330, 623, 691, and 1504 are all low mass M-type stars (Table~\ref{stellar_params}). Orientation will further complicate our ability to observe a stable warp, since high inclinations are needed to generate occultation events. The average accretion rate through the disk may also play a role since the structure of the inner disk appears to vary with accretion rate \citep{mcc13b} and the form of the optical variability, bursts vs occultations vs spots, appears to be broadly correlated with accretion rate \citep{sou16,bod16}. This connection between disk/magnetosphere interactions and average accretion rate through the disk is independent of any of the infrared fluctuations induced by rapid accretion rate variability discussed earlier. 

{\bf In-situ perturbations of the disk:} While disk/magnetosphere interactions can explain days to week long fluctuations, they have trouble explaining months to year long timescales. Accounting for the power seen at low frequencies would require a stellar magnetic field that stretches out to $\sim$0.5 au, much beyond the corotation radius. Another possibility is that there are perturbations within the disk, operating at the local dynamical timescale, that lead to observable changes in the infrared emission. This could include structural perturbations, such as those generated by turbulence lifting dust away from the midplane \citep{tur10}, or temperature perturbations, such as collisions between massive bodies that heat up a substantial fraction of the dust, which have been seen to create significant variability among debris disks \citep{men15}. 

The key property in this class of models, which distinguishes it from interactions with the stellar magnetosphere, is that they can operate on a wide range of radii, and hence create fluctuations in emission on a wide range of timescales. Their contribution to the power spectrum is only limited by the radii over which a particular process operates, and the range of radii that generate significant [3.6] and [4.5] emission. The power law form of the power spectral densities seen among many of the cluster members is consistent with fluctuations arising from a wide range of radii. With our well-sampled light curves we can see that this behavior is common, suggesting that in-situ perturbations are not restricted to a small sample of young stellar objects.

The color variations in much of our sample are consistent with changes in disk emission, which can arise as the structure of the hot small grains in the upper disk atmosphere rapidly varies. \citet{pop15} and \citet{ric15} in their surveys of IRAS:20050+2720 and Orion find that long timescale behavior exhibits variations in color more consistent with disk perturbations. The increase in color power spectral density towards long timescales in our sample (Figure~\ref{psd_color}) is also consistent with the increased role of disk fluctuations on longer timescales. Evidence for changes in the structure of the inner disk have also been observed in pre-transition disks. These systems, with an optically thick inner disk that is separated from the optically thick outer disk by a large gap, have been observed to exhibit 'seesaw' behavior in which the short-wavelength flux increases while the long wavelength flux decreases, or vice-versa \citep{muz09,fla12}. This behavior can be explained by a variable inner disk height; as the inner disk puffs up its own flux increases, while increasing the shadowing of the outer disk leading to a decrease in the long wavelength flux \citep{esp11}. Similar scale-height fluctuations may be common in less evolved systems, and present at a wide range of radii, leading to the the infrared fluctuations that we observe here. 

In-situ perturbations to disk structure are not mutually exclusive from perturbations associated with the disk/magnetosphere interface. In fact, stars 330 and 623 both show low-frequency structure in their power spectrum in addition to the more rapid behavior. Extinction may explain the rapid dimming events in star 1504, but may not explain the brightening during the middle of the light curve. If extinction operates on short timescales, while disk fluctuations operate on long timescales, the end result can be zero average change in color, as is seen in a number of sources. Which process is more evident in the power spectrum likely depends on the parameters of the system (stellar mass, accretion rate, etc) and the timescales being studied in detail. Further work is needed to better understand the details of the perturbations, but our data are consistent with in-situ perturbations at a wide range of radii being common among young stellar objects.

{\bf Beyond the Dynamical Timescale:} In addition to the months-long variability observed during our monitoring campaign, we find evidence for longer timescale variability, stretching out to years and decades, consistent with previous infrared variability studies covering similar timescales \citep{meg12,reb14,par14,wol15}. The models discussed above all rely on perturbations on the local dynamical timescale; to reach years would require perturbations out at 1-10 au, a region that does not contribute significantly to the [3.6] and [4.5] flux \citep{dal06}. Instead, perturbations may be operating at the local thermal or viscous timescales. At the sublimation radius ($\sim$0.1 au), where the dynamical timescale is days, the thermal timescale is 1-2 years \citep{cg97} and previous models have found that thermal instabilities can generate significant changes in the physical structure of the disk \citep{wat08}. The viscous timescale, the timescale on which significant perturbations to the surface density and accretion rate through the disk can propagate inward \citep{arm11}, is hundreds of years at $\sim$0.1 au. These features are more likely to influence the very longest timescales probed, over decades of observations. More detailed characterization of the years to decades long fluctuations can help to probe these additional mechanisms.

\section{Conclusions}
Based on {\it Spitzer} observations of the Cha I cluster taken roughly once per day over 200 days, we find that large (0.05-0.3) infrared fluctuations are common, occurring in $\sim$80\% of the observed young stellar objects. We find that the behavior seen in previous {\it Spitzer} surveys, which focused on well-sampled observations over $\sim$40 day observing windows, continues to much longer timescales. In examining the properties of the variability, we find that changes in color ($\Delta$[4.5]/$\Delta$[3.6]=0.75-1.61) occur in some of the young stellar objects, and that these fluctuations fall between the behavior expected of extinction and fluctuations in disk emission ($\Delta$[4.5]/$\Delta$[3.6]=0.84,1.25 respectively). We find a small handful of the cluster members to be periodic (5/30), including three with periods of 30-40 days, longer than the typical rotational periods of young stellar objects. With the combination of high cadence and long observing window, we are able to simultaneously sample both daily and months-long fluctuations. With these data, we find a smooth increase in power across two orders of magnitude in timescale in almost every young stellar object, with this behavior seen in the [3.6] and [4.5] fluxes as well as the [3.6]-[4.5] color. The shape of the power spectrum indicates that the infrared variability is largest on the longest timescales, with fluctuations $\sim$2 times larger over 200 days than over the typical 40 day windows sampled by previous {\it Spitzer} variability surveys. A comparison of our observations with prior {\it Spitzer} photometry of Cha I suggests that this trend extends out to decades-long timescales. 

Based on the size of the fluctuations we can rule out the direct contribution of star spots, while perturbations from a companion are possible for the three sources with long ($\sim$30 day) periods. Interactions at the disk/magnetosphere boundary can explain some of the days to week long fluctuations, while longer months to year long variability is likely a result of structural perturbations to the disk out as far as $\sim$0.5 au. These processes are not mutually exclusive, and can operate simultaneously within an individual system. Whichever process appears more prominent in the light curves depends on the timescales being probed, as well as the intrinsic properties of the system (e.g. accretion rate, stellar magnetic field strength, stellar luminosity). More detailed measurements of accretion rate, stellar rotation period and inclination could further guide our understanding of which processes are strongest within a given system.

\acknowledgements
We thank the referee for the detailed report that greatly improved the manuscript. This work is based on observations made with the {\it Spitzer Space Telescope}, which is operated by the Jet Propulsion Laboratory, California Institute of Technology under a contract with NASA. Support for this work was provided by NASA through an award issued by JPL/Caltech. The research made use of Astropy, a community-developed core Python package for Astronomy \citep{astropy13}. And thanks to Mary Drennan; I have not found any new stars, but there might be some new planets in here.

\appendix
\section{OTS 44\label{ots44}}
While many of our targets are $\sim$0.5 M$_{\odot}$, our survey also includes one of the least massive known young (sub)stellar objects. OTS 44 \citep{luh04}, star 623 in our sample, has a spectral type of M9.5 and is estimated to have a mass of only $\sim$15 M$_{\rm jup}$ \citep{bon14}. It has a strong infrared excess \citep{luh05,har12,liu15} and is actively accreting \citep{jor13}. 

\begin{figure}
\center
\includegraphics[scale=.4]{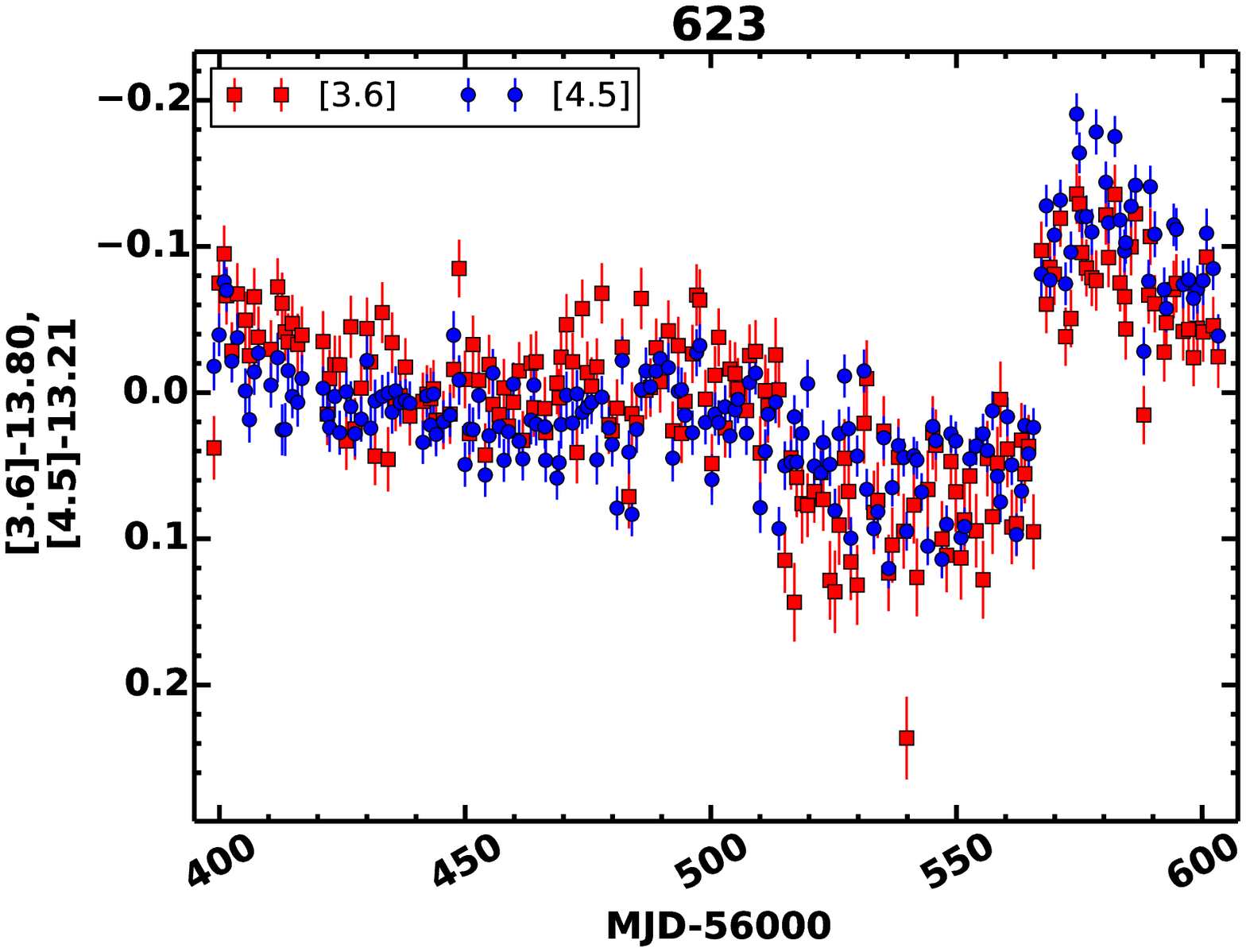}
\includegraphics[scale=.4]{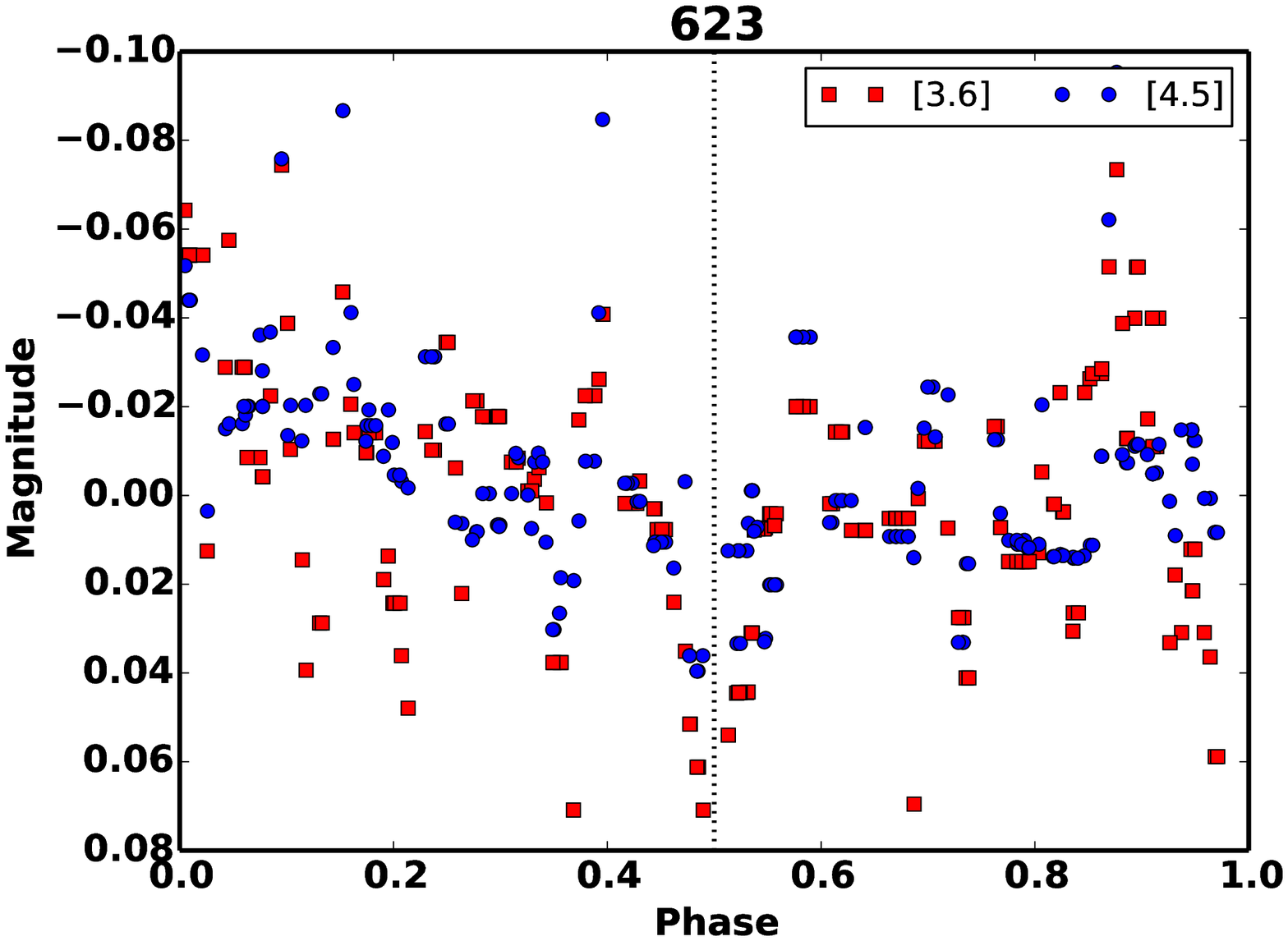}
\caption{[3.6] (red squares) and [4.5] (blue circles) light curves for the low mass brown dwarf OTS 44 (star 623). With a mass of $\sim$15 M$_{\rm jup}$ it is one of the lowest mass objects with a disk, and the lowest mass object with known variability in its disk. The right panel shows the light curve phased to a period of 3.73 days, as indicated by analysis of the power spectrum. This phased curve has been median smoothed over 0.1 in phase to reduce the noise and highlight the absorption feature at phase=0.5.\label{star623}}
\end{figure}

{\bf OTS 44} is also variable in the infrared. In our data, the [3.6] and [4.5] emission varies by 0.15 magnitudes over the length of our observations (Figure~\ref{star623}), while \citet{luh08a} find that the [24] flux changes from 8.81$\pm$0.11 in 2004 to 9.65$\pm$0.20 in 2005. While this variability is much different than the typical infrared fluctuations among field brown dwarfs that are attributed to clouds, which have timescales of a few hours and amplitudes of a few percent \citep{met15}, it is not unique among young sub-stellar objects \citep{reb15}. 

The mid-infrared flux of OTS 44 slowly declines over the first 150 days during which the color changes substantially ($\Delta$[4.5]/$\Delta$[3.6]=0.63$\pm$0.05), with a stronger reddening than even that predicted for extinction. This decline is interrupted by an abrupt increase in flux that occurs between two observations, limiting its length to less than 1.6 days. Given the very low luminosity of this source the dust sublimation radius is at 0.002 au, which has a Keplerian period of 0.2 days; such a short dynamical timescale allows for a substantial change in structure between our individual epochs, and may explain the sharpness of the jump. After this abrupt increase there is a steep decline, with fluxes almost reaching pre-outburst levels by the end of our observing campaign. The power spectrum also indicates a significant quasi-periodic feature at 3.7 days, similar to periods observed in the optical for other low mass objects in the Cha I cluster \citep{cod14bd}. In the smoothed light curve phased to a period of 3.7 days (Figure~\ref{star623}) there is evidence for a dimming feature at phase $\sim$0.5, consistent with obscuration by a clump of material at 0.016 AU. In their survey of brown dwarfs in NGC 1333, \citet{reb15} find one system with obscuration features that occur every 4.44 days, similar to the behavior seen here in OTS 44.

In the other observed cluster members our 200 day coverage encompasses the dynamical timescale of the inner disk, but given the small disk around this low-luminosity source our observations extend much beyond the dynamical timescale, opening up the thermal and even viscous timescales of the disk as relevant for the gradual behavior. This allows for thermal instabilities related to variable heating of the midplane \citep{wat08} as well as perturbations to the density of inflowing material \citep[e.g.][]{kel14} as possible explanations for the variability. Given the complexity of the light curves, it is likely that multiple processes are at play in this system.

\begin{turnpage}
\begin{deluxetable}{cccccccccccc}
\tabletypesize{\tiny}
\tablewidth{0pt}
\tablecaption{Stellar Parameters\label{stellar_params}}
\tablehead{\colhead{Index} & \colhead{Name\tablenotemark{a}} & \colhead {RA} & \colhead{Dec} & \colhead{Spectral Type} & \colhead{T$_{\rm eff}$ (K)} & \colhead{L$_{*}$ (L$_{\odot}$)} & \colhead{M$_{*}$ (M$_{\odot}$)} & \colhead{$\log$($\dot{M}$ (M$_{\odot}$ yr$^{-1}$))} & \colhead{IR Class} & \colhead{n$_{3.6-4.5}$\tablenotemark{b}} & \colhead{n$_{4.5-24}$\tablenotemark{b}}}
\startdata
75 & J11095215-7639128 &  11:09:52.14 & -76:39:12.8 & M6.25 & 2962 & 0.023 & 0.09 & ... & II & -0.85$\pm$0.11 & -0.98$\pm$0.02\\
100 & J11084952-7638443 & 11:08:49.51 & -76:38:44.3 & M8.75 & 2478 & 0.0014 & 0.07 & ... & II & -1.43$\pm$0.11 & -1.07$\pm$0.05\\
239\tablenotemark{c} & J11095003-7636476 & 11:09:50.01 & -76:36:47.7 & B9 & 10500 & 57 & 2.6\tablenotemark{d} & ... & I & -2.88$\pm$ 0.14 & \ldots\\
299\tablenotemark{e} & J11084069-763078 & 11:08:40.68 & -76:36:07.8 & M2.5 & 3488 & 0.74 & 0.4 & ... & III & -2.71$\pm$0.15 & -2.84$\pm$0.04\\
330 & J11100469-7635452 &  11:10:04.67 & -76:35:45.3 & M1 & 3705 & 0.43 & 0.46 & $<$-9.91\tablenotemark{f},-8.87$\pm$0.5\tablenotemark{g}, -8.9$\pm$0.2\tablenotemark{h} & II & -1.93$\pm$0.11 & -0.87$\pm$0.02\\
360 & J11101141-7635292 & 11:10:11.39 & -76:35:29.3 & K5.5 & 4278 & 1.2 & 0.9 & -7.98\tablenotemark{i},-9.79\tablenotemark{f}, -9.18$\pm$0.5\tablenotemark{g} & II & -1.23$\pm$0.11 & -0.04$\pm$0.01\\
383 & J11095493-7635101 & 11:09:54.92 & -76:35:10.3 & M5.75 & 3024 & 0.0014 & 0.17 & ... & II & -0.7$\pm$1.1 & -2.1$\pm$0.2\\
402\tablenotemark{e} & J11100010-7634578 & 11:10:00.08 & -76:34:57.9 & K5 & 4350 & 5.5 & 1.0 & -6.60\tablenotemark{f}, -6.6$\pm$0.1\tablenotemark{h} & II & 0.05$\pm$0.11 & -0.54$\pm$0.01\\
403 & J11094192-7634584 & 11:09:41.91 & -76:34:58.5 & $<$M0 & $>$3850 & ... & ... & ... & II & -0.69$\pm$0.11 & -0.39$\pm$0.02\\
416\tablenotemark{e} & J11094621-7634463 & 11:09:46.19 & -76:34:46.4 & M3.25 & 3379 & 0.15 & 0.29 & -8.11\tablenotemark{i},-8.3$\pm$0.1\tablenotemark{h} & II & -0.53$\pm$0.11 & -0.21$\pm$0.03\\
439 & J11092266-7634320 & 11:09:22.65 & -76:34:32.0 & M1.25 & 3669 & 0.8 & 0.4 & -7.32\tablenotemark{i} & II & -0.20$\pm$0.11 & -0.15$\pm$0.01\\
530 & J11100369-7633291 & 11:10:03.69 & -76:33:29.2 & K8 & 3955 & 0.64 & 0.63 & -8.7$\pm$0.3\tablenotemark{h} & II & -0.24$\pm$0.11 & -0.57$\pm$0.01\\
533 & J11092855-7633281 & 11:09:28.54 & -76:33:28.1 & M3.5-M6.5\tablenotemark{j} & 3100 & 1.5 & 0.1 & \ldots & I & 3.19$\pm$0.15 & 0.70$\pm$0.03\\
557 & J11100336-7633111 & 11:10:03.35 & -76:33:11.1 & M4 & 3270 & 0.0025 & 0.25 & ... & II & -0.07$\pm$0.11 & 0.82$\pm$0.02\\
593 & J11085176-7632502 & 11:08:51.75 & -76:32:50.3 & M7.25 & 2838 & 0.011 & 0.09 & ... & II & -2.42$\pm$0.15 & \ldots\\
601 & J11095505-7632409 & 11:09:55.04 & -76:32:41.0 & $<$M0 & $>$3850 & ... & ... & ... & II & 0.05$\pm$0.11 & -0.18$\pm$0.02\\
602 & J11085497-7632410 & 11:08:54.96 & -76:32:41.1 & M5.5 & 3058 & 0.035 & 0.12 & ... & II & -1.81$\pm$0.11 & -1.14$\pm$0.02\\
623 & J11100934-7632178 & 11:10:09.31 & -76:32:17.9 & M9.5 & 2300 & 0.00077 & 0.075 & ... & II & -1.35$\pm$0.11 & -0.59$\pm$0.08\\
691 & J11095437-7631113 & 11:09:54.37 & -76:31:11.4 & M1.75 & 3596 & 0.013 & 0.35 & -10.8$\pm$0.4\tablenotemark{h} & I & 0.54$\pm$0.15 & 0.10$\pm$0.03\\
720\tablenotemark{e,k} & J11091812-7630292 & 11:09:18.12 & -76:30:29.3 & M1.25 & 3669 & 0.55 & 0.43 & $-7.80$\tablenotemark{i} & II & -0.94$\pm$0.11 & -0.78$\pm$0.01\\
791\tablenotemark{e} & J11120984-7634366 & 11:12:09.83 & -76:34:36.6 & M5 & 3125 & 0.15 & 0.16 & ... & II & -1.72$\pm$0.11 & -0.79$\pm$0.01\\
793 & J11105333-7634319 & 11:10:53.32 & -76:34:32.0 & M3.75 & 3306 & 0.13 & 0.26 & -8.46$\pm$0.5\tablenotemark{g}, -7.90$\pm$0.6\tablenotemark{l} & II & -0.73$\pm$0.11 & -0.67$\pm$0.02\\
966 & J11104006-7630547 & 11:10:40.05 & -76:30:54.7 & M7.25 & 2838 & 0.0082 & 0.073 & ... & II & -2.51$\pm$0.15 & \ldots\\
1034\tablenotemark{e,m} & J11100704-7629376 & 11:10:07.03 & -76:29:37.7 & M0 & 3850 & 1.4 & 0.6 & -8.24$\pm$0.5\tablenotemark{g},-8.44$\pm$0.6\tablenotemark{l} & II & -1.35$\pm$0.11 & -0.83$\pm$0.01\\
1103\tablenotemark{e,n} & J11094006-7628391 & 11:09:40.06 & -76:28:39.2 & M1.25 & 3669 & 0.68 & 0.41 & ... & II & -3.00$\pm$0.15 & -2.74$\pm$0.03\\
1504\tablenotemark{e,o} & J11095407-7629253 & 11:09:54.07 & -76:29:25.3 & M2 & 3560 & 0.48 & 0.36 & -7.79\tablenotemark{i} & II & -1.23$\pm$0.11 & -0.54$\pm$0.02\\
2712\tablenotemark{e} & J11113474-7636211 & 11:11:34.73 & -76:36:21.2 & M2.5 & 3488 & 0.21 & 0.36 & ... & III & -2.87$\pm$0.15 & -2.74$\pm$0.04\\
4342 & J11114533-7636505 & 11:11:45.32 & -76:36:50.5 & M8 & 2710 & 0.0033 & 0.073 & ... & II & -1.43$\pm$0.11 & -1.33$\pm$0.05\\
5073 & J11120327-7637034 & 11:12:3.26 & -76:37:03.4 & M5.5 & 3058 & 0.082 & 0.13 & \ldots & III & -2.50$\pm$0.15 & -2.67$\pm$0.08\\
6555 & J11091380-7628396 & 11:09:13.80 & -76:28:39.7 & M4.75 & 3161 & 0.072 & 0.18 & ... & II & -2.55$\pm$0.15 & -2.74$\pm$0.11\\
8665\tablenotemark{e,p} & J11091769-7627578 & 11:09:17.69 & -76:27:57.8 & K7 & 4060 & 1.1 & 0.68 & ... & III & -3.04$\pm$0.14 & -2.78$\pm$0.04\\
14260 & J11102226-7625138 & 11:10:22.25 & -76:25:13.8 & M8 & 2710 & 0.011 & 0.076 & \ldots & III & -2.59$\pm$0.15 & \ldots\\
\enddata
\tablecomments{Information on the detected cluster members within our sample. Spectral types are taken from \citet{luh08b} unless otherwise specified. Masses are estimated using the \citet{bar15} isochrones unless otherwise noted. Infrared class is taken from \citet{gut09} and their classification scheme.}
\tablenotetext{a}{Numerous naming conventions have been applied to members of Cha I. Here we list the 2MASS identifier, and refer to \citet{luh08b} for a more complete list of alternate names}
\tablenotetext{b}{$n_{\lambda 1-\lambda 2}=(\log(\lambda 2F_{\lambda 2})-\log(\lambda 1F_{\lambda 1}))/(\log(\lambda 2)-\log(\lambda 1))$. Photometry taken from \citet{gut09} and \citet{luh08a}.}
\tablenotetext{d}{Mass estimate from \citet{sie00} isochrones.}
\tablenotetext{c}{T41: 0.78'' (125 au) binary with B9, M3.5 components \citep{dae13}}
\tablenotetext{e}{No spectroscopic binary companion detected by \citet{ngu12}, with sensitivity down to $\sim$0.1M$_{\odot}$ over 60 day periods}
\tablenotetext{f}{\citet{man16}}
\tablenotetext{g}{\citet{fra15}}
\tablenotetext{h}{\citet{rob12}}
\tablenotetext{i}{\citet{ant11}}
\tablenotetext{j}{\citet{gra14}}
\tablenotetext{k}{CHXR 79: 0.88'' (140 au) binary with M2, M4 components \citep{dae13}}
\tablenotetext{l}{\citet{har98}}
\tablenotetext{m}{T46: 0.123'' (20 au) binary \citep{laf08}}
\tablenotetext{n}{CHXR 40: 0.151'' (24 au) binary \citep{laf08}}
\tablenotetext{o}{T43: 0.78'' (126 au) binary with M3.5, M7 components \citep{dae13}}
\tablenotetext{p}{CHXR 37: 0.079'' (13 au) binary \citep{laf08}}
\end{deluxetable}
\end{turnpage}
\clearpage

\begin{deluxetable}{cccccccccccc}
\tabletypesize{\footnotesize}
\tablewidth{0pt}
\tablecolumns{11}
\tablecaption{Infrared Variability\label{invar_table}}
\tablehead{\colhead{} & \multicolumn{3}{c}{[3.6]} & \colhead{} & \multicolumn{3}{c}{[4.5]} & \colhead{} & \colhead{} & \colhead{}\\
\cline{2-4} \cline{6-8}\\
 \colhead{ID} & \colhead {N$_{\rm epochs}$} & \colhead{$\chi^2_{\nu}$} & \colhead{$\Delta$mag} & \colhead{} & \colhead {N$_{\rm epochs}$} & \colhead{$\chi^2_{\nu}$} & \colhead{$\Delta$mag} & \colhead{} & \colhead{N$_{\rm pair}$} & \colhead{Stetson}}
\startdata
75 & 121 & 23.6 & 0.22 & & 6 & 30.7 & 0.13 & & 0 & \ldots\\
100 & 91 & 1.8 & 0.07 & & 26 & 1.7 & 0.06 & & 0 & \ldots\\
239 & 185 & 1.5 & 0.05 & & 73 & 1.6 & 0.03 & & 58 & 0.01\\
299 & 86 & 2.0 & 0.06 & & 71 & 1.4 & 0.04 & & 0 & \ldots\\
330 & 200 & 22.9 & 0.19 & & 111 & 56.1 & 0.20 & & 111 & 5.1\\
360 & 200 & 6.6 & 0.10 & & 122 & 25.7 & 0.12 & & 122 & 2.3\\
383 & 47 & 4.1 & 0.41 & & 49 & 44.9 & 1.05 & & 31 & 0.4\\
402 & 0 & \ldots & \ldots & & 25 & 5.6 & 0.15 & & 0 & \ldots\\
403 & 188 & 36.2 & 0.21 & & 137 & 61.2 & 0.18 & & 125 & 4.3\\
416 & 200 & 25.1 & 0.21 & & 200 & 45.4 & 0.22 & & 200 & 4.1\\
439 & 117 & 214 & 0.53 & & 114 & 566 & 0.54 & & 59 & 15.0\\
530 & 200 & 46.9 & 0.26 & & 200 & 129.1 & 0.26 & & 200 & 6.8\\
533 & 126 & 43.8 & 0.27 & & 179 & 147.7 & 0.27 & & 105 & 5.6\\
557 & 200 & 25.5 & 0.23 & & 200 & 67.9 & 0.27 & & 200 & 4.8\\
593 & 92 & 1.5 & 0.04 & & 150 & 1.1 & 0.03 & & 42 & 0.06\\
601 & 200 & 36.5 & 0.22 & & 200 & 88.6 & 0.21 & & 200 & 5.9\\
602 & 91 & 6.0 & 0.08 & & 153 & 30.9 & 0.12 & & 44 & 1.5\\
623 & 199 & 7.7 & 0.16 & & 200 & 18.3 & 0.17 & & 199 & 1.9\\
691 & 88 & 129 & 0.44 & & 200 & 207 & 0.32 & & 88 & 10.5\\
720 & 62 & 26.0 & 0.19 & & 197 & 61.1 & 0.16 & & 59 & 4.7\\
791 & 79 & 7.0 & 0.10 & & 20 & 5.7 & 0.06 & & 0 & \ldots\\
793 & 158 & 11.4 & 0.14 & & 94 & 16.6 & 0.11 & & 52 & 2.1\\
966 & 90 & 1.3 & 0.05 & & 155 & 1.2 & 0.04 & & 45 & -0.04\\
1034 & 56 & 43.2 & 0.25 & & 188 & 256 & 0.37 & & 44 & 5.8\\
1103 & 2 & \ldots & 0.04 & & 148 & 1.9 & 0.04 & & 0 & \ldots\\
1504 & 55 & 18.0 & 0.19 & & 196 & 50.0 & 0.16 & & 51 & 3.1\\
2712 & 102 & 1.9 & 0.03 & & 12 & 3.1 & 0.02 & & 0 & \ldots\\
4342 & 87 & 1.3 & 0.09 & & 2 & \ldots & 0.04 & & 0 & \ldots\\
5073 & 76 & 3.4 & 0.06 & & 0 & \ldots & \ldots & & 0 & \ldots\\
6555 & 0 & \ldots & \ldots & & 118 & 1.4 & 0.03 & & 0 & \ldots\\
8665 & 0 & \ldots & \ldots & & 100 & 1.8 & 0.03 & & 0 & \ldots\\
14260 & 0 & \ldots & \ldots & & 39 & 2.1 & 0.05 & & 0 & \ldots\\
\enddata
\tablecomments{Infrared variability statistics for our sample. Included for each band are the number of epochs at which the YSO was detected with $\sigma<0.1$ (N$_{\rm epochs}$), the reduced chi-squared ($\chi^2_{\nu}$, with $\chi^2_{\nu}>3$ indicating significant variability) and the difference between the magnitudes at the 10th and 90th percentile. Also listed is the number of epochs with simultaneous [3.6] and [4.5] data (N$_{\rm pair}$) and the Stetson index for those stars with N$_{\rm pair}>$0 (S$>$0.45 indicates significant variability).} 
\end{deluxetable}

\begin{deluxetable}{cccc}
\tablewidth{0pt}
\tablecaption{Star Forming Region Comparison}\label{cluster_comparison}
\tablehead{\colhead{Region} & \colhead{Variable Fraction} & \colhead{$\Delta$mag} & \colhead{Reference}}
\startdata
Cha I &  0.80 & 0.05-0.3 & this paper\\
\hline
IRAS:20050+2720  & 0.70 & 0.17 & 1\\
NGC 1333  & 0.62 & $\sim$0.1 & 2\\
ONC  & 0.70 & $\sim$0.2 & 3\\ 
L1688 & 0.85 & 0.1-0.26 & 4\\
GGD 12-15 & 0.70 & $\sim$0.15 & 5\\
IC 348 & 0.60 & $\sim$0.1 & 6\\
NGC 2264 & $\sim$0.90 & 0.1-0.3 & 7\\
\enddata
\tablecomments{Fraction of YSOs with an infrared excess that are variable in the infrared, as measured by {\it Spitzer}, and the typical range in $\Delta$[3.6],$\Delta$[4.5] as reported by these studies. Refs: 1- \citet{pop15}, 2 - \citet{reb15}, 3 - \citet{mor11}, 4 - \citet{gun14}, 5 - \citet{wol15}, 6 - \citet{fla13}, 7 - \citet{cod14}}
\end{deluxetable}

\begin{deluxetable}{cccc}
\tablewidth{0pt}
\tablecaption{Color Changes\label{color_table}}
\tablehead{\colhead{ID} & \colhead{N$_{\rm pair}$} & \colhead{$\Delta$[4.5]/$\Delta$[3.6]}}
\startdata
330 & 111 & 0.99 $\pm$ 0.03\\
360 & 122 & 1.61 $\pm$ 0.12\\
403 & 125 & 0.92 $\pm$ 0.03\\
416 & 200 & 1.01 $\pm$ 0.03\\
439 & 59 & 0.97 $\pm$ 0.02\\
530 & 200 & 1.01 $\pm$ 0.02\\
533 & 105 & 0.85 $\pm$ 0.04\\
557 & 200 & 1.13 $\pm$ 0.02\\
601 & 200 & 0.96 $\pm$ 0.03 \\
602 & 44 & 0.96 $\pm$ 0.10\\
623 & 199 & 1.15 $\pm$ 0.06\\
691 & 88 & 0.75 $\pm$ 0.01\\
720 & 59 & 0.81 $\pm$ 0.03\\
793 & 52 & 1.02 $\pm$ 0.08\\
1034 & 44 & 1.13 $\pm$ 0.05\\
1504 & 51 & 0.80 $\pm$ 0.04\\
\hline
Hot spot & - & 1.1\\
Cold spot & - & 0.9\\
Extinction & - & 0.84\\
Disk changes & - & 1.25\\
\enddata
\end{deluxetable}

\begin{turnpage}
\begin{deluxetable}{cccccccc}
\tabletypesize{\footnotesize}
\tablewidth{0pt}
\tablecolumns{8}
\tablecaption{Other photometry\label{other_phot}}
\tablehead{\colhead{} & \multicolumn{2}{c}{This Work} & \multicolumn{3}{c}{Luhman et al. 2008} & \colhead{} & \colhead{}\\
\colhead{ID} & \colhead{$\overline{\rm [3.6]}$} & \colhead{$\overline{\rm [4.5]}$} & \colhead{$\overline{\rm [3.6]}$} & \colhead{$\overline{\rm [4.5]}$} & \colhead{Epoch} & \colhead{WISE Var?} & \colhead{NEOWISE Var?}}
\startdata
75 & 10.94 $\pm$ 0.11 & 10.43 $\pm$ 0.06 & 10.80 $\pm$ 0.02 & 10.29 $\pm$ 0.02 & 2004 july 4 & Y & Y\\
100 & 13.76 $\pm$ 0.03 & 13.27 $\pm$ 0.03 & 13.61\tablenotemark{a} $\pm$ 0.02 & 12.86\tablenotemark{a} $\pm$ 0.05 & 2004 july 4 & N & N\\
 & & & 13.70 $\pm$ 0.02 & 12.88\tablenotemark{a} $\pm$ 0.02 & 2005 may 9 & N\\ 
239 & 7.21 $\pm$ 0.02 & 7.07 $\pm$ 0.01 & 7.08 $\pm$ 0.02 & 7.05 $\pm$ 0.02 & 2004 july 4 & N & Y\\
299 & 9.19 $\pm$ 0.03 & 9.06 $\pm$ 0.02 & 9.05 $\pm$ 0.02 & 8.94 $\pm$ 0.03 & 2004 july 4 & N & N\\
330 & 8.89 $\pm$ 0.09 & 8.51 $\pm$ 0.10 & 8.70 $\pm$ 0.02 & 8.10\tablenotemark{a} $\pm$ 0.03 & 2004 july 4 & Y & \ldots\\
360 & 8.22 $\pm$ 0.05 & 7.74 $\pm$ 0.06 & 8.07 $\pm$ 0.02 & 7.55\tablenotemark{a} $\pm$ 0.02 & 2004 july 4 & Y & \ldots\\
383 & 14.51 $\pm$ 0.21 & 13.44 $\pm$ 0.53 & 15.00 $\pm$ 0.13 & 14.43 $\pm$ 0.23 & \ldots\tablenotemark{b} & \ldots & \ldots\\
402 & \ldots & 7.46 $\pm$ 0.07 & saturated & saturated & 2004 july 4 & N & N\\
403 & 8.59 $\pm$ 0.11 & 7.85 $\pm$ 0.09 & 8.53 $\pm$ 0.02 & 7.95 $\pm$ 0.02 & 2004 july 4 & N & N\\
416 & 9.65 $\pm$ 0.11 & 9.07 $\pm$ 0.11 & 9.49 $\pm$ 0.02 & 8.92 $\pm$ 0.02 & 2005 july 4 & N & \ldots\\
439 & 7.22 $\pm$ 0.27 & 6.47 $\pm$ 0.27 & 7.22 $\pm$ 0.02 & 6.51 $\pm$ 0.02 & 2004 july 4 & Y & Y\\ 
530 & 8.30 $\pm$ 0.13 & 7.67 $\pm$ 0.13 & 8.18 $\pm$ 0.02 & 6.84\tablenotemark{a} $\pm$ 0.03 & 2004 july 4 & Y & Y\\
533 & 8.99 $\pm$ 0.14 & 7.38 $\pm$ 0.13 & 8.47\tablenotemark{a} $\pm$ 0.02 & 6.97\tablenotemark{a} $\pm$ 0.02 & 2005 july 4 & \ldots & Y\\
557 & 12.76 $\pm$ 0.11 & 11.94 $\pm$ 0.13 & \ldots & \ldots & \ldots & Y & \ldots\\
593 & 12.59 $\pm$ 0.02 & 12.36 $\pm$ 0.02 & 12.43\tablenotemark{a} $\pm$ 0.02 & 12.27 $\pm$ 0.05 & 2004 july 4 & N & N\\
 & & & 12.42\tablenotemark{a} $\pm$ 0.02 & 12.19 $\pm$ 0.03 & 2005 may 9 & N & \ldots\\
601 & 7.91 $\pm$ 0.11 & 7.14 $\pm$ 0.10 & 7.90 $\pm$ 0.02 & 7.15 $\pm$ 0.02 & 2004 july 4 & \ldots & Y\\
602 & 10.84 $\pm$ 0.04 & 10.41 $\pm$ 0.06 & 10.66\tablenotemark{a} $\pm$ 0.02 & 10.34 $\pm$ 0.02 & 2004 july 4 & N & Y\\
623 & 13.80 $\pm$ 0.08 & 13.21 $\pm$ 0.08 & 13.69 $\pm$ 0.02 & 13.22 $\pm$ 0.02 & 2004 july 4 & N & Y\\
 & & & 13.67 $\pm$ 0.02 & 13.17 $\pm$ 0.02 & 2005 may 9 & N & Y\\
691 & 10.90 $\pm$ 0.22 & 10.12 $\pm$ 0.16 & 11.22 $\pm$ 0.02 & 10.35 $\pm$ 0.02 & 2004 july 4 & Y & Y\\
720 & 8.25 $\pm$ 0.09 & 7.71 $\pm$ 0.08 & 7.92\tablenotemark{a} $\pm$ 0.02 & 7.43\tablenotemark{a} $\pm$ 0.02 & 2004 july 4 & Y & Y\\
791 & 9.47 $\pm$ 0.05 & 9.08 $\pm$ 0.03 & 9.35 $\pm$ 0.02 & 9.03 $\pm$ 0.02 & 2004 july 4 & Y & Y\\
793 & 9.37 $\pm$ 0.07 & 8.75 $\pm$ 0.06 & 9.14\tablenotemark{a} $\pm$ 0.02 & 8.56\tablenotemark{a} $\pm$ 0.02 & 2004 july 4 & Y & Y\\
966 & 13.00 $\pm$ 0.03 & 12.77 $\pm$ 0.02 & 12.84\tablenotemark{a} $\pm$ 0.02 & 12.72 $\pm$ 0.02 & 2004 july 4 & N & Y\\
 & & & 12.83\tablenotemark{a} $\pm$ 0.02 & 12.65\tablenotemark{a} $\pm$ 0.02 & 2005 may 9 & N & Y\\
1034 & 7.70 $\pm$ 0.13 & 7.32 $\pm$ 0.19 & 7.69 $\pm$ 0.02 & 7.29 $\pm$ 0.02 & 2004 july 4 & N & Y\\
1103 & 8.99 $\pm$ 0.02 & 8.83 $\pm$ 0.02 & 8.76\tablenotemark{a} $\pm$ 0.02 & 8.76 $\pm$ 0.02 & 2004 july 4 & N & \ldots\\
1504 & 8.51 $\pm$ 0.09 & 7.92 $\pm$ 0.08 & 8.55 $\pm$ 0.02 & 8.14\tablenotemark{a} $\pm$ 0.02 & 2004 july 4 & N & Y\\
2712 & 9.68 $\pm$ 0.03 & 9.56 $\pm$ 0.02 & 9.50\tablenotemark{a} $\pm$ 0.02 & 9.47 $\pm$ 0.02 & 2004 july 4 & Y & N\\
4342 & 13.50 $\pm$ 0.05 & 13.01 $\pm$ 0.02 & 13.31\tablenotemark{a} $\pm$ 0.02 & 12.92 $\pm$ 0.02 & 2004 july 4 & N & N\\
 & & & 13.32\tablenotemark{a} $\pm$ 0.02 & 12.98 $\pm$ 0.02 & 2005 may 9 & N & N\\
5073 & 10.58$\pm$0.03 & \ldots & 10.39$\pm$0.02\tablenotemark{a} & 10.23$\pm$0.03 & 2004 july 4 & N & Y\\
 & & & \ldots & 10.28$\pm$0.02 & 2004 july 21 & N & Y\\
6555 & \ldots & 10.52 $\pm$ 0.01 & 10.55 $\pm$ 0.02 & 10.44 $\pm$ 0.02 & 2004 july 4 & N & Y\\
8665 & \ldots & 8.50 $\pm$ 0.02 & 8.44 $\pm$ 0.02 & 8.45 $\pm$ 0.02 & 2004 july 4 & N & Y\\
14260 & \ldots & 11.91$\pm$0.02 & 11.97$\pm$0.02 & 11.87$\pm$0.02 & 2004 july 4 & N & Y\\
 & & & 11.95$\pm$0.02 & 11.84$\pm$0.02 & 2005 may 9 & N & Y\\
\enddata
\tablecomments{Comparison of the fluxes observed in our survey, and those from the original \citet{luh08a} survey of Cha I. For  data from this work we list the mean magnitude and the size of the fluctuations, as defined by the difference between the 10th and 90th percentile. For the \citet{luh08a} sample we report mean magnitudes and uncertainties, as well as the date of observations, from their survey. The last two columns indicate if the source was variable in the two short-wavelength bands of WISE as observed during the cryogenic WISE and the post-cryogenic NEOWISE missions. The WISE observations of Cha I spanned Feb-Aug 2010 and Feb 2014-Aug 2015 respectively.}
\tablenotetext{a}{Differs significantly from the range of [3.6] or [4.5] fluxes seen in our survey}
\tablenotetext{b}{No exact date for the photometry is reported.}
\end{deluxetable}
\end{turnpage}
\clearpage

\begin{deluxetable}{cc}
\tablewidth{0pt}
\tablecaption{Periodic Stars\label{periodic_table}}
\tablehead{\colhead{ID} & \colhead{Period (days)}}
\startdata
330 & 7.9 \\
439 & 32\\
530 & 35\\
533 & 36\\
691 & 7.9\\
\enddata
\tablecomments{Periods derived from the mid-infrared light curves.}
\end{deluxetable}


\clearpage

\end{document}